\documentclass[amssymb,aps,prd,two column,nofootinbib,superscriptaddress]{revtex4-2}
\usepackage{graphicx}
\usepackage{subcaption} 
\usepackage{mathtools}
\usepackage{physics}
\usepackage{url}
\usepackage[normalem]{ulem}
\usepackage{caption}
\usepackage{enumitem}
\usepackage[colorlinks=true, linkcolor=blue, citecolor=magenta, urlcolor=red]{hyperref}    
\usepackage[percent]{overpic}
\usepackage{float}
\usepackage{amsmath, amsfonts, mathrsfs}
\usepackage{multirow}
\usepackage{makecell}
\usepackage{tikz}
\usetikzlibrary{tikzmark}
\newcommand{\be}{\begin{eqnarray}}
	\newcommand{\ee}{\end{eqnarray}}

\newcommand{\bfk}{{\bf k}_{\perp}}

\usepackage{xcolor}
\newcommand{\orcid}[1]{\href{https://orcid.org/#1}{\includegraphics[width=8pt]
{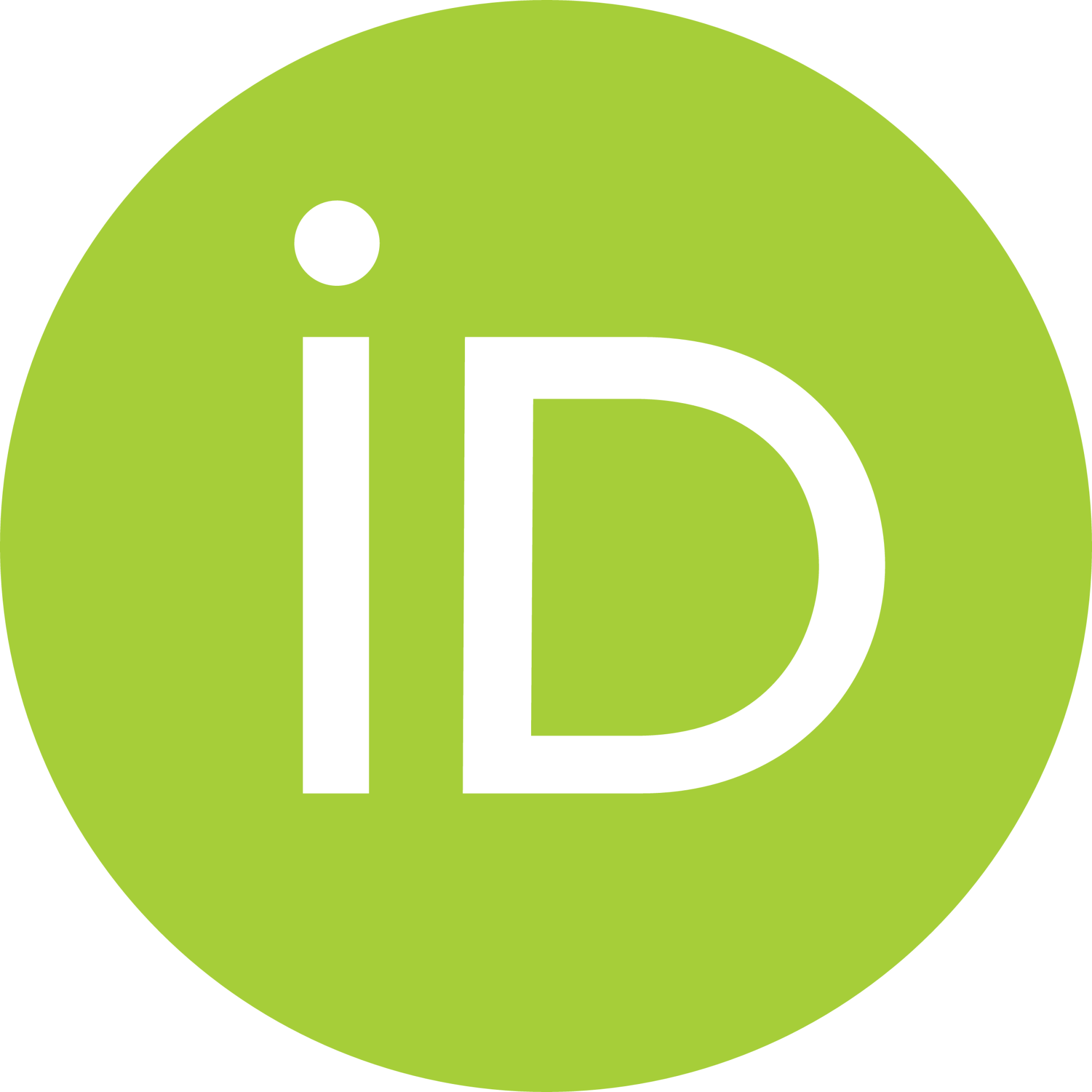}}}
\begin{document}
\title{Mechanical distribution of the pseudoscalar charmonium and bottomonium on the light-front}

\author{Ashutosh Dwibedi\orcid{https://orcid.org/0009-0004-1568-2806}}\email{ashutoshdwibedi92@gmail.com}\affiliation{Department of Physics, Indian Institute of Technology Bhilai, Kutelabhata, Durg, 491002, Chhattisgarh, India}
\author{Satyajit Puhan\orcid{https://orcid.org/0009-0004-9766-5005}}\email{puhansatyajit@gmail.com}\affiliation{Institute of Physics, Academia Sinica, Taipei 11529, Taiwan}
\author{Sabyasachi Ghosh\orcid{https://orcid.org/0000-0003-1212-824X}}\email{sabya@iitbhilai.ac.in}\affiliation{Department of Physics, Indian Institute of Technology Bhilai, Kutelabhata, Durg, 491002, Chhattisgarh, India}

\begin{abstract}
We investigate the energy-momentum tensor of pseudoscalar charmonium and bottomonium within the framework of the self-consistent light-front quark model. The gravitational form factors (GFFs), namely the $A$ and $D$-terms, are evaluated in terms of the light-front wave functions. The corresponding spatial mechanical distributions in the transverse plane are obtained through the Fourier transform of these GFFs. To examine the sensitivity of the results to the internal quark-antiquark distribution inside the meson, two distinct Gaussian forms are employed for the spatial part of the wave function. We analyze several mechanical properties in the transverse plane, including the momentum density, pressure distribution, shear stress, force density, and internal energy density. The pressure distribution exhibits a node where it changes sign from positive (repulsive) to negative (attractive) with increasing transverse distance. The force distribution remains positive throughout the transverse plane, supporting the stability condition proposed in earlier studies. Most of the spatial distributions, except for the shear stress, are found to be sensitive to the choice of the spatial wave function near the center of the meson, while they become nearly insensitive toward the periphery. In contrast, the shear stress distribution exhibits noticeable sensitivity to the choice of wave function in the intermediate transverse region.    
\end{abstract}
\maketitle
\section{Introduction}
The study of mechanical distributions inside hadrons has recently gained significant attention, particularly following the extraction of the proton pressure from experimental data~\cite{Burkert:2018bqq}. The internal pressure and force distributions of hadrons are closely related to the so-called $D$-term (or $D$-form factor), also referred to as the Druck term~\cite{Polyakov:2002yz,Polyakov:2018zvc}, which has been described as the “last global unknown” of hadron structure~\cite{Polyakov:2018zvc}. The electromagnetic form factors (EMFFs) are well established. They are obtained from matrix elements of the electromagnetic current between hadronic states. These form factors encode information about the spatial distributions of charge and current inside hadrons. In contrast, the mechanical properties of hadrons are determined from matrix elements of the energy–momentum tensor (EMT). These matrix elements give rise to gravitational form factors (GFFs), which remain comparatively less explored. The GFFs provide access to fundamental properties of hadrons, including their mass, energy, angular momentum, and internal stress distributions~\cite{Polyakov:2018zvc,Burkert:2023wzr}. From the experimental perspective, the expected tools to measure GFFs are the hard exclusive processes like the deeply virtual Compton scattering (DVCS)~\cite{Ji:1996nm}, which probes the generalized parton distributions (GPDs)~\cite{Burkardt:2002hr,Diehl:2003ny,Belitsky:2005qn,Lorce:2025aqp}. The future Electron-Ion Collider (EIC)~\cite{Accardi:2012qut,Abir:2023fpo,AbdulKhalek:2021gbh} with its larger kinematic range and high luminosity is expected to significantly advance the experimental determination of GPDs. 

While considerable progress has been achieved in understanding the gravitational structure of the nucleons~\cite{PhysRevLett.74.1071,Polyakov:2002yz,Goeke:2007fp,Cebulla:2007ei,Kim:2012ts,Shanahan:2018nnv,Lorce:2018egm,Chakrabarti:2020kdc,Ji:2021mfb,Freese:2021qtb,Fujita:2022jus,Choudhary:2022den,GarciaMartin-Caro:2023klo,GarciaMartin-Caro:2023toa,Hackett:2023rif,Cao:2024zlf,Nair:2024fit,Guo:2025jiz,Goharipour:2025lep,Ji:2025gsq,Sugimoto:2025btn,Tanaka:2025pny,Ji:2025qax,Deng:2025azp,Fukushima:2025jah,Fukushima:2026wwc,Deng:2026gik,Stegeman:2025sca,Cao:2025dkv,Won:2025dgc,Won:2026ljg,KAUR2026140686,Nair:2025sfr} and pions~\cite{Broniowski:2008hx,Freese:2019bhb,Stegeman:2025sca,Cao:2025dkv,Liu:2024vkj,Liu:2024jno,Hackett:2023nkr,Krutov:2020ewr, Xu:2023izo,Li:2023izn,Fujii:2024rqd,Broniowski:2024oyk,Choi:2025rto} (also see Ref.~\cite{Stegeman:2025tdl}), studies of the heavy mesonic sector remain relatively limited~\cite{Sultan:2024hep,Xu:2024hfx,Hu:2024edc}. In particular, pseudoscalar quarkonia—such as charmonium and bottomonium—offer a unique framework to probe the gravitational structure of hadrons in the heavy-quark sector. Due to the large masses of the constituent quarks, these systems are well-suited for investigating the interplay between nonperturbative QCD dynamics and heavy-quark effects. Therefore, exploring the GFFs of pseudoscalar charmonium and bottomonium states can provide valuable insights into confinement and the internal mechanical properties of heavy hadrons.

 Calculation of the GFFs, as well as other quark-quark correlations inside the hadron from the fundamental QCD lagrangian, is a formidable task because of the nonperturbative nature of the problem. Low-energy effective models are useful in this respect and have been widely used in studies of hadronic properties~\cite{Choi:1997iq,Broniowski:2008hx,Freese:2019bhb,Tanaka:2025pny}. One such model is the light-front constituent quark model (LFQM)~\cite{Choi:1997iq}. In LFQM, a meson state is expanded in the Fock basis containing the valence quark-antiquark pair as well as sea quarks and gluons. In the model scale (taken as of the order of the mass of the constituent quarks), it is customary to truncate these infinitely many Fock terms into the valence sector consisting of only a valence quark-antiquark pair. In this valence Fock state approximation, the meson states can be expressed as a two-body light-front wavefunction (LFWF) consisting of relative momenta of the quark-antiquark pair and their helicities. The hadronic properties are then readily obtained as the overlap of the LFWFs. In the LFQM, GFFs are expressed as the matrix element of the hadronic EMT tensor and subsequently expressed as the overlap integrals of LFWFs. However, it is worth noting that evaluating the $D$-term in this approach poses a considerable challenge, as it inevitably involves the so-called ``bad" current components, which are sensitive to LF zero modes~\cite{Choi:2013mda}. Over the years, a self-consistent LFQM has been developed that successfully addresses this zero-mode contamination in light-front (LF) dynamics in a variety of hadronic observables~\cite{Choi:2013mda,Choi:2014ifm,Choi:2017uos,Choi:2021mni,Choi:2021qza,Arifi:2022qnd,Arifi:2023uqc,Choi:2024ptc}. This approach gives a way of calculating physical observables in a current-component independent manner (including the ``bad" current components), which are otherwise hindered in standard LFQM. Within this approach, in particular, Ref.~\cite{Arifi:2022qnd} obtained the pseudoscalar and vector meson decay constants in a current-component- and polarization-independent manner. In Refs.~\cite{Choi:2014ifm,Choi:2017uos}, it has been discussed how this approach is consistent with the chiral symmetry of QCD in the context of pion and kaon distribution amplitudes and EMFFs.  More recently, the EMFFs~\cite{Choi:2024ptc} and GFFs~\cite{Choi:2025rto} of the pion were calculated within this approach, demonstrating the independence of the extracted form factors from the choice of current component. This approach, in particular, is applicable to symmetric-mass mesons, and it is followed here to extract the GFFs of $\eta_{c}$ and $\eta_{b}$. 
 
  In the present work, we calculate the GFFs, i.e., $A$ and $D$-terms of the $\eta_{c}$ and $\eta_{b}$ mesons by evaluating the matrix element of the EMT operator containing only the non-interacting quark fields. Various spatial mechanical distributions in the transverse plane are related to the Fourier transforms of these GFFs. We define these distributions in a completely relativistically invariant way, respecting the uncertainty principle, following the Refs.~\cite {Freese:2021czn,Freese:2021qtb,Freese:2021mzg,Freese:2022fat}. These distributions are subsequently evaluated in the transverse plane at a constant LF time. In particular, we determine the longitudinal momentum density, pressure distribution, shear stresses, force distribution, and internal energy distribution in the transverse plane. To check the sensitivity of the results on the quark-antiquark distribution inside the mesons, we use two sets of spatial wave functions in the present work, differing by the so-called ``jacobi factor"~\cite{Ma:1993ht,Choi:1997qh}.

The present work is organized as follows. In Secs.~\ref{Sec:LCQM}, \ref{sec:GFF}, and \ref{denstra}, we present the LFQM framework, the evaluation of the GFFs within the LFQM, and the spatial distributions inside the mesons, respectively. Section~\ref{results} contains the numerical results for the GFFs and the corresponding spatial distributions obtained using two different choices of wave functions. Finally, we summarize our findings and conclude in Sec.~\ref{Sec:conclusion}. Additional details and intermediate derivations are provided in Appendices~\ref{defspin}--\ref{apedensity}. Appendix~\ref{defspin} covers the LF conventions and spinors, Appendix~\ref{quarkcont} the EMT matrix elements, Appendix~\ref{BT_const} the BT construction and equivalent GFF expressions, and Appendix~\ref{apedensity} the associated spatial densities and useful formulae.

\section{Formalism}
\label{sec:Formalism}
\subsection{Description of pseudoscalar quarkonia in light-front quark model}
\label{Sec:LCQM}
We begin by describing the valence Fock sector of scalar quarkonia
($Q\bar{Q}$) in the LFQM. At fixed LF
time $z^{+} = t + z = 0$~\cite{Brodsky:1997de}, the valence Fock state
$\ket{Q\bar{Q}\left(x_{i}P^{+},\,\vec{k}_{\perp i} + x_{i}\vec{P}_{\perp},\,\lambda_{i}\right)}$ for a pseudoscalar quarkonium with total four-momentum
$P^\mu \equiv P= \left(P^{+},\,\vec{P}_{\perp},\, P^{-}\right)$ is expressed in terms of the internal variables $x_{i} = \frac{k_{i}^{+}}{P^{+}},~\vec{k}_{\perp i}$, and $\lambda_{i}$,  which represent the boost-invariant longitudinal momentum fraction, transverse internal momentum, and helicity carried by the $i$th constituent parton, respectively\footnote{The index $i$ in our case would belongs to either ($c,\bar{c}$) for charmonium or ($b, \bar{b}$)   for bottomonium. We generically call them ($q, \bar{q}$).}. These variables satisfy the momentum-conservation conditions
$\sum_{i} x_{i} = 1$ and $\sum_{i} \vec{k}_{\perp i} = 0$. The valence state of the quarkonium is expressed as,
\begin{widetext}
\be
|\mathcal{M} (P^+,\vec{P}_{\perp})\rangle &=&\sum_{\lambda_{q},\lambda_{\bar{q}}}\int \frac{dx_{q} dx_{\bar{q}} d^{2}{\vec{k}}_{q \perp}d^{2}{\vec {k}}_{\bar{q}\perp }}{2(2\pi)^{3}~2(2\pi)^{3}
\sqrt{x_{q}x_{\bar{q}}}}~2(2\pi)^{3}~\delta(1-x_{q}-x_{\bar{q}})~\delta^{2}(\vec{k}_{q \perp}+\vec{k}_{\bar{q} \perp})\Psi(x_{q},x_{\bar{q}},{\vec{k}}_{q \perp},{\vec{k}}_{\bar{q} \perp },\lambda_{q},\lambda_{\bar{q}})\nonumber\\
&&|x_{q}P^{+},x_{\bar{q}}P^{+},x_{q}\vec{P}_{\perp}+{\vec{k}}_{q \perp},x_{\bar{q}}\vec{P}_{\perp}+{\vec{k}}_{\bar{q} \perp }\lambda_{q},\lambda_{\bar{q}}\rangle\nonumber
\ee
where the subscripts $q$ and $\bar{q}$ are used for the quark and anti-quark, respectively. Simplifying the above integral by integrating over $x_{\bar{q}}$ and $\vec{k}_{\bar{q}\perp }$ we have,  
\be
|\mathcal{M} (P^+,\vec{P}_{\perp})\rangle &=& \sum_{\lambda_{q},\lambda_{\bar{q}}}\int \frac{dx \, d^2 \bfk}{  16 \pi^3 \sqrt{x(1-x)}} ~\Psi (x,\vec{k}_{\perp},\lambda_{q},\lambda_{\bar{q}})~ |x P^+, \vec{k}_{\perp}, \lambda_{q},\lambda_{\bar{q}} \rangle \label{eqnq} .
\ee 
where we redefine 
\be
&&x_{q}\equiv x, \vec{k}_{q\perp} \equiv \vec{k}_{\perp} \text{ and } |x_{q}P^{+},(1-x_{q})P^{+},x_{q}\vec{P}_{\perp}+{\vec{k}}_{q \perp},(1-x_{q})\vec{P}_{\perp}-{\vec{k}}_{q \perp }, \lambda_{q},\lambda_{\bar{q}}\rangle \equiv|x_{q}P^{+},x_{q}\vec{P}_{\perp}+{\vec{k}}_{q \perp},\lambda_{q},\lambda_{\bar{q}}\rangle.\nonumber
\ee
\end{widetext}
The quark and anti quark momentum can be identified as: $\vec{p}_{q}=(p_{q}^{+}\equiv x_{q} P^{+},\vec{p}_{q\perp}\equiv x_{q}\vec{P}_{\perp}+\vec{k}_{q\perp})$ and $\vec{p}_{\bar{q}}=(p_{\bar{q}}^{+}\equiv x_{\bar{q}}P^{+},\vec{p}_{\bar{q}\perp}\equiv x_{\bar{q}}\vec{P}_{\perp}+\vec{k}_{\bar{q}\perp})$.
The total wavefunction $\Psi$ is written as a product of the space wavefunction $\varphi$ and spin wavefunction $\mathcal{S}$, i.e., $\Psi(x,\vec{k}_{\perp},\lambda_{q},\lambda_{\bar{q}})=\varphi(x,\vec{k}_{\perp})~\mathcal{S}_{\lambda_{q}\lambda_{\bar{q}}}(x,\vec{k}_{\perp})$. The spin wavefunctions are obtained from the instant from spin states via Melosh-Wigner transformation~\cite{Melosh:1974cu,Xiao:2002iv} or equivalently by the choice of the appropriate quark-meson vertex~\cite{Jaus:1991cy} 
\be
&&\mathcal{S}_{\lambda_{q}\lambda_{\bar{q}}}=\frac{1}{\sqrt{2} M_{0}} \bar{u}_{\lambda_{q}}(p_{q})~\gamma_{5}~v_{\lambda_{\bar{q}}}(p_{\bar{q}}),\nonumber\\
&&\text{where } M_{0}^{2}=\frac{\vec{k}_{\perp}^{2}+m_{q}^{2}}{x(1-x)}\label{ex1}
\ee
is the squared invariant mass of the quarkonium.
The detailed form of the spin wavefunctions along with the definitions of the LF spinors are provided in appendix~\eqref{defspin}. The instant form momentum space wavefunction in the quarkonium rest frame or center-of-momentum (CM) frame can be assumed to be of the Harmonic oscillator form, i.e.,
\be
\varphi^{\rm CM}=\frac{1}{\pi^{3/4}\beta^{3/2}} e^{-(\vec{p}^{\rm CM})^{2}/2\beta^{2}}\label{ex2}, 
\ee
where $\beta$ is the harmonic scale parameter determining the width of the wavefunction. In order to obtain the momentum-space wave function in the front form,
$\varphi(x,\vec{k}_{\perp})$, from the CM frame wave
function in the instant form, $\varphi^{\rm CM}(\vec{p}^{\,\rm CM})$, one
employs the Brodsky--Huang--Lepage (BHL) prescription
\cite{Brodsky:1981jv,Brodsky:1980vj,Brodsky:1982nx}. According to this prescription, the correspondence between the CM and LF momentum variables is
given by $\vec{p}_{\perp}^{\,\rm CM} = \vec{k}_{\perp}$ and $p_{z}^{\,\rm CM} = \left(x - \frac{1}{2}\right) M_{0}$. Therefore, in the BHL prescription, we have 
\be
\varphi(x,\vec{k}_{\perp}) \propto \varphi_{\rm CM}((\vec{p}^{\rm CM})^{2}=\frac{\vec{k}_{\perp}^{2}+m_{q}^{2}}{4x(1-x)}-m_{q}^{2})~.\label{ex3}
\ee
In the current paper, two sets of momentum-space wavefunctions have been chosen for analysis, which mainly differ from $\varphi_{\,\rm CM}$ by a product of $(x,\vec{k}_{\perp})$ dependent factor (known as the ``jacobi factor"~\cite{Ma:1993ht,Choi:1997qh}).
The first set we use has been abundantly used in the LFQM studies of heavy mesons~\cite{CHUNG1988545,PhysRevC.71.028202,Jaus:1991cy,Choi:1997iq} 
\be
 \varphi^{\rm CJ} (x,\vec{k}_{\perp})&=&\frac{\sqrt{2(2\pi)^{3}}}{\pi^{3/4}\beta^{3/2}}\sqrt{\frac{\partial p_{z}^{\rm CM}}{\partial x}} \, exp \,\Biggl[\frac{m_{q}^{2}}{2\beta^{2}}\Biggr]\, \nonumber\\
 &&exp \,\Biggl[-\frac{m_{q}^{2}+\vec{k}_{\perp}^{2}}{8x(1-x)\beta^{2}}\Biggr]\nonumber\\
 &=&\frac{\sqrt{2(2\pi)^{3}}}{\pi^{3/4}\beta^{3/2}}\sqrt{\frac{M_{0}}{4x(1-x)}} \, exp \,\Biggl[\frac{m_{q}^{2}}{2\beta^{2}}\Biggr]\,\nonumber\\
 &&exp \,\Biggl[-\frac{m_{q}^{2}+\vec{k}_{\perp}^{2}}{8x(1-x)\beta^{2}}\Biggr], (\rm SET-I)\label{CJwfn}~.
\ee
This form was introduced by Chung-Coester-Polyzou~\cite{CHUNG1988545} and has been extensively used by C.R. Ji and collaborators~\cite{Choi:1998jd,Choi:2001hg,Choi:2007yu,Choi:2015ywa,Dhiman:2019ddr,Pandya:2024qoj,Acharyya:2024tql} to describe the properties of hadrons in LF models. The jacobian factor $M_{0}/4x(1-x)$ comes from relating the instant form differentials $d^{3}\vec{p}^{\rm CM}$ to $d^{2}\vec{k}_{\perp} dx$.
Another set which we use is introduced in Ref.~\cite{Terentev:1976jk,Carbonell:1998rj} and further used in the LF analysis~\cite{Ma:1993ht} (also see the review~\cite{KARMANOV1981331}),
\be
 \varphi^{\rm TK} (x,\vec{k}_{\perp})
 &=&A_{\rm TK}\sqrt{\frac{1}{2x(1-x)}} \, exp \,\Biggl[\frac{m_{q}^{2}}{2\beta^{2}}\Biggr]\,\nonumber\\
 &&exp \,\Biggl[-\frac{m_{q}^{2}+\vec{k}_{\perp}^{2}}{8x(1-x)\beta^{2}}\Biggr], (\rm SET-II)\label{TKwfn}~.
\ee
 The jacobian factor $1/2x(1-x)$ in Eq.~\eqref{TKwfn} comes from relating the instant form differentials $d^{3}\vec{p}^{\rm CM}/p_{0}^{\rm CM}$ to $d^{2}\vec{k}_{\perp} dx$.

One can observe that the spin wave functions in Eq.~\eqref{Ape8} satisfy the following normalization relation,
\be 
\sum_{\lambda_q \lambda_{\bar{q}}} \mathcal{S}^{\ast}_{\lambda_{q}\lambda_{\bar{q}}} (x,\vec{k}_{\perp}) \,\mathcal{S}_{\lambda_{q}\lambda_{\bar{q}}}(x,\vec{k}_{\perp}) = 1 \, \label{spin_nor}.
\ee
Since the total wavefunction (space and spin) has to be normalized, this gives the following normalization condition for the space part of the wave function,
\be
&&\sum_{\lambda_q \lambda_{\bar{q}}}\int \frac{d x d^2 \vec{k}_{\perp}}{2 (2 \pi)^3} \Psi^{\ast}(x,\vec{k}_{\perp},\lambda_q, \lambda_{\bar{q}}) \Psi(x,\vec{k}_{\perp},\lambda_q, \lambda_{\bar{q}})\nonumber\\
&=& \int \frac{{d x} d^2 \vec{k}_{\perp}}{2 (2 \pi)^3} \, |\varphi (x,\vec{k}_{\perp})|^2 =1~.
\ee

\subsection{Gravitational form factors in the self-consistent light-front quark model}\label{sec:GFF}
We consider a process in which a pseudoscalar quarkonia undergoes a transition from an initial state with momentum $P$ to a final state with momentum $P^{\prime}$. The
GFF of the quakonia can be obtained from the expression of the following matrix element of the EMT operator~\cite{Choi:2025rto}
\begin{equation}
   T^{\mu\nu}\equiv\langle \hat{T}^{\mu\nu}(0)\rangle =\langle \mathcal{M} (P^{\prime}) \vert \hat{T}^{\mu\nu}(0)\vert \mathcal{M}(P)\rangle~,\label{G1}
\end{equation}
where $\vert\mathcal{M}(P)\rangle$ and $\vert\mathcal{M}(P^{\prime})\rangle$ correspond to initial and final state of the meson. For our purpose, we choose the following form of the EMT operator, ignoring the gluon gauge fields~\cite{Choi:2025rto}
\begin{eqnarray}
&& \hat{T}^{\mu\nu}=\frac{1}{4}\bar{\psi}_{q}(z)\left(i\gamma^{\mu} \overleftrightarrow{\partial}^{\nu}+i\gamma^{\nu}\overleftrightarrow{\partial}^{\mu}\right)\psi_{q}(z)~,\label{G2}
\end{eqnarray}
where, $q$ stands for either charm ($c$) or bottom ($b$).
Lorentz symmetry gives the decomposition of the matrix element containing Lorentz structures and the corresponding GFF~\cite{Polyakov:2018zvc,Choi:2025rto},
\begin{eqnarray}
    &&T^{\mu\nu} =A~\Gamma_{A}^{\mu\nu}+D~\Gamma_{D}^{\mu\nu}~,\label{G3}
\end{eqnarray}
where $A$ describes the quark and anti-quark contribution to the energy and momentum densities, while $D$ characterizes the quark and anti-quark contribution to the internal pressure and shear stress distributions. Both the $A$ and $D$-terms can be written down as a sum of quark and anti-quark contributions as $A=A_{q}+A_{\bar{q}}$ and $D=D_{q}+D_{\bar{q}}$. One usually expresses the Lorentz structure with the help of the average momentum $\bar{P}=\frac{P+P^{\prime}}{2}$ and momentum transfer $\Delta=P^{\prime}-P$ as, $\Gamma_{A}^{\mu\nu}=2\bar{P}^{\mu}\bar{P}^{\nu}$ and $\Gamma_{D}^{\mu\nu}=\frac{1}{2}(\Delta^{\mu}\Delta^{\nu}-g^{\mu\nu} \Delta^{2})$. 

It is noteworthy that, while Eq.~\eqref{G3} provides the Lorentz decomposition of the matrix elements of the physical EMT and preserves complete Poincaré symmetry, a practical model calculation does not necessarily retain this property~\cite{Carbonell:1998rj}. In practice, the left-hand side of Eq.~\eqref{G3} is evaluated within LF dynamics using a suitable QCD-motivated model, typically with a truncated Fock space. In the LFQM considered in Sec.~\ref{Sec:LCQM}, for example, we restrict the Fock space to the valence sector. Such a truncation can give rise to several issues that require careful consideration, including zero-mode contributions~\cite{Jaus:1999zv}, violation of Lorentz covariance~\cite{Carbonell:1998rj}, and frame dependence~\cite{Li:2017uug}. These issues are not unique to the calculation of GFFs, but also arise in the evaluation of other partonic observables, such as distribution amplitudes~\cite{Choi:2013mda} and EMFF~\cite{Choi:2014ifm}. A common strategy for mitigating these issues in LF approaches is to identify the so-called ``good components'' of the relevant currents and use them to extract the physical observables. Accordingly, the identification of the good current components and the characterization of spurious contributions associated with the ``bad components" have been investigated for various partonic observables. In particular, to account for the loss of Lorentz covariance in LF calculations, a light-like four-vector $\omega$ can be introduced into the Lorentz decomposition of the hadronic matrix element~\cite{Karmanov:1976iv,Carbonell:1998rj}. Its presence allows the genuine Lorentz structures, and hence the physical form factors, to be distinguished from spurious contributions induced by the LF formulation. For the hadronic EMT, this issue has been investigated in Refs.~\cite{Cao:2023ohj,Cao:2024rul,Cao:2024fto}, where the appropriate current components for extracting the physical form factors and isolating the spurious contributions were identified within a strongly coupled scalar theory in the LF Hamiltonian formalism.

A complementary approach to the treatment of spurious contributions was developed by Jaus~\cite{Jaus:1999zv}, who established a connection between covariant Bethe--Salpeter calculations and LFQMs. This framework was subsequently modified and extended by Choi and Ji~\cite{Choi:2013mda,Choi:2014ifm,Choi:2017uos,Choi:2021mni,Choi:2021qza} and further developed in Refs.~\cite{Arifi:2022qnd,Arifi:2023uqc,Choi:2024ptc}. In this approach, the LF vertex function obtained in the covariant Bethe--Salpeter approach was mapped onto a typical Gaussian trial wavefunction, with a consistent replacement of the meson mass by its invariant mass.
A remarkable finding of this analysis was that both the zero-mode and instantaneous contributions, which are present in the covariant Bethe--Salpeter model, are absent in the LFQM when the constituent quark and antiquark are treated as LF on-mass-shell degrees of freedom.
Readers interested in the details of the method are referred to the original works~\cite{Choi:2013mda,Choi:2014ifm,Choi:2017uos,Choi:2021mni,Choi:2021qza,Arifi:2022qnd,Arifi:2023uqc,Choi:2024ptc}, which collectively establish the framework known as the self-consistent LFQM to obtain the physical observables in a manner independent of the current components. 
In what follows, we briefly describe this method and discuss the extraction of the GFFs of quarkonia within this framework.

In the standard LFQM described in Sec .~\ref {Sec:LCQM}, one describes the meson state in a non-interacting $Q\bar{Q}$ Fock state. As a result, the total four momentum $P=p_{q}+p_{\bar{q}}$ is required to be conserved at a meson-quark vertex. In particular, the LF energy conservation $P^{-}=p_{q}^{-}+p_{\bar{q}}^{-}, \implies \frac{M_{0}^{2}+\vec{P}_{\perp}^{2}}{P^{+}}=\frac{m_{q}^{2}+\vec{p}_{q\perp}^{2}}{p_{q}^{+}}+\frac{m_{\bar{q}}^{2}+\vec{p}_{\bar{q}\perp}^{2}}{p_{\bar{q}}^{+}}$ necessitate the use of invariant mass $M_{0}$ in place of the physical meson mass $M$ in a meson-quark vertex~\cite{Choi:2024ptc}. In accordance with this, the spin wavefunction is obtained by interaction-independent Melosh rotation of on-mass-shell quarks and anti-quarks. One can notice that the spin wavefunction as specified in Eq.~\eqref{ex1} contains the invariant mass of the system $M_{0}$ and is automatically normalized (see Eq.~\eqref{spin_nor}). The effect of interaction is introduced only in the mass operator $M$ through the phenomenological $V_{Q\bar{Q}}$ potential~\cite{Godfrey:1985xj}, i.e., $M=M_{0}+V_{Q\bar{Q}}$ to be consistent with the Bakamjian--Thomas (BT) construction~\cite{PhysRev.92.1300,Keister:1991sb} and preserve the group structure or the commutation relation of the operators. The interaction in turn fixes the radial wavefunction parameters, $m_{q}$ and $\beta$ through variational analysis~\cite{Choi:1997iq,Choi:2015ywa,Arifi:2022pal,Dwibedi:2026rtm}.

The traditional LFQMs~\cite{Jaus:1991cy,Jaus:1989au,Cheng:1996if} use BT construction (the replacement $M\rightarrow M_{0}$) only in evaluating the matrix element, e.g., on Eq.~\eqref{G1} or the left-hand side of Eq.~\eqref{G3},  while leaving the Lorentz structure (right-hand side of Eq.~\eqref{G3}) intact. The selective application of BT construction only to the matrix elements yields LF zero modes that affect physical observables~\cite{Choi:2013mda,Choi:2014ifm,Choi:2017uos,Choi:2021mni,Choi:2021qza,Arifi:2022qnd,Arifi:2023uqc,Choi:2024ptc,Choi:2025rto}. This results in different values for the same observables when extracted from different current components (particularly when ``bad current components" are used). We have demonstrated this fact, in connection with EMT, in appendix~\eqref{BT_const} (see Eqs.~\eqref{Apbt9}--\eqref{Apbt15}). 
Therefore, it is also important to apply the BT construction to both the Lorentz factors and matrix elements~\cite{Choi:2013mda,Choi:2014ifm,Choi:2017uos,Choi:2021mni,Choi:2021qza,Arifi:2022qnd,Arifi:2023uqc,Choi:2024ptc,Choi:2025rto} to realize a consistent application of the BT construction up to the level of computation. The consistent application of BT construction within LFQM yields a self-consistent LFQM that can extract physical observables in a current-component-independent manner, including the so-called ``bad components". As noted in Ref.~\cite{Choi:2025rto}, the Lorentz structure $\Gamma_{A}^{\mu\nu}$ must be modified to properly implement the BT construction in LFQM studies, and the physical meson mass $M$ should be replaced with the invariant mass $M_{0}$ within both the Lorentz structures $\Gamma_{A}^{\mu\nu}$ and $\Gamma_{D}^{\mu\nu}$. Following Ref.~\cite{Choi:2025rto}, we modify the form of Lorentz structure $\Gamma_{A}^{\mu\nu}$ as
$\Gamma_{A}^{\mu\nu}=2\left[\bar{P}^{\mu}\bar{P}^{\nu}-\frac{\bar{P}\cdot\Delta}{\Delta^{2}}
\left(\Delta^{\mu}\bar{P}^{\nu}+\Delta^{\nu}\bar{P}^{\mu}-\bar{P}\cdot\Delta\, g^{\mu\nu}\right)\right]$.  Within the BT framework, the invariant masses of the initial and final quarkonia entering the Lorentz structure are generally different, i.e., $M^{2}_{0} \neq M_{0}^{\prime 2}=\frac{\vec{k}_{\perp}^{\prime 2}+m_{q}^{2}}{x(1-x)}$. Keeping this in mind, the different components of the Lorentz structures in the BT framework can be worked out easily and are provided in Table~\ref{tab:BTlor} of Appendix~\ref{BT_const}. The extraction of the GFFs of quarkonia in the self-consistent LFQM from different current components is illustrated in detail in Appendix~\ref{BT_const}. Particularly, we numerically establish in Appendix~\ref{BT_const} that different ways (using $++,+i,-i,ij$-EMT components\footnote{We also see that $(--)$-EMT component leads to different results in the self-consistent LFQM as found in Ref.~\cite{Choi:2025rto} for pion and should be discarded in studying the GFFs within the present framework.} and two different radial wavefunctions, CJ and TK) of extracting the GFFs lead to the same result in self-consistent LFQM, while the traditional LFQM results predict different values when different current components are used.

For the sake of illustration, we write the following expression for $A_{q}$ and $D_{q}$ extracted from $(++)$ and ($+-$) components using Eq.~\eqref{G1} along with Eq.~\eqref{G3},
\begin{eqnarray}
&& A_{q}=\left\langle\frac{ T_{q}^{++}}{\Gamma_{A}^{++}}\right\rangle_{\rm BT}\label{G4}\\
&& D_{q}=\left\langle\frac{\Gamma_{A}^{++}T_{q}^{+-}-\Gamma^{+-}_{A}T_{q}^{++}}{\Gamma_{D}^{+-}\Gamma_{A}^{++}}\right\rangle_{\rm BT}~.\label{G5}
\end{eqnarray}
As emphasized in the Appendix~\ref{BT_const}, the Lorentz structures, which now depend on the internal quark momenta, have to be integrated along with the matrix element $T_{q}^{++}$ and $T_{q}^{++}$ in the self-consistent LFQM. We provide the details of the evaluation of the matrix element of $T_{q}^{\mu\nu}$ in the appendix~\eqref{quarkcont}. Here, we briefly outline the steps needed to get the final expression for GFFs. The individual meson momentum in terms of the average momentum and momentum transfer is given by $P=\bar{P}-\frac{\Delta}{2}$ and $P^{\prime}=\bar{P}+\frac{\Delta}{2}$. We would select an asymmetric frame with zero skewness (no longitudinal momentum transfer $\Delta^{+}=0$). The choice of the frame render us with  $P^{\prime +}=P^{+}=\bar{P}^{+}$, $\vec{P}=(P^{+},\vec{0})$, $\vec{P}^{\prime}=(P^{+},\vec{\Delta}_{\perp})$ and $\vec{\bar{P}}=(P^{+},\frac{\vec{\Delta}_{\perp}}{2})$. 

The GFFs of the quarkonia can be obtained by calculating the matrix element of the EMT operator and substituting in the expressions ~\eqref{G4} and ~\eqref{G5}. It is worth mentioning that being a symmetric mass system, both the quark and antiquark contribute equally to the GFFs, i.e., $A=2A_{q}$ and $D=2D_{q}$ (see appendix \eqref{antiquarkcont} for a formal derivation). The matrix element of the EMT operator can be expressed as,
\begin{eqnarray}
 T_{q}^{\mu\nu}&\equiv&\langle \hat{T}^{\mu\nu}_{q}(0)\rangle=\langle \mathcal{M} (P^{\prime}) \vert \hat{T}^{\mu\nu}_{q}(0)\vert \mathcal{M}(P)\rangle\nonumber\\
 &=& \int \frac{dx~ d^{2}\vec{k}_{\perp}}{2(2\pi)^{3}} ~\varphi^{\ast}(x,\vec{k}_{\perp}^{\prime})\varphi(x,\vec{k}_{\perp}) \Tr [\mathcal{S} \mathcal{S}^{\prime \dag} U^{\mu\nu}]~,\nonumber\\
 \label{G8}   
\end{eqnarray}
where $\mathcal{S} (x,\vec{k}_{\perp})$ and $\mathcal{S}^{\prime}\equiv\mathcal{S} (x,\vec{k}_{\perp}^{\prime})$ are the $2\times 2$ matrices (in the helicity space) representing the initial and final state spin wavefunctions of the quarkonia.  $U^{\mu\nu}\equiv U^{\mu\nu}_{\lambda_{q}^{\prime}\lambda_{q}}= \frac{1}{4x}\bar{u}_{\lambda_{q}^{\prime}}(p_{q}^{\prime})[\gamma^{\mu}p_{sq}^{\nu}+\gamma^{\nu}p_{sq}^{\mu}]u_{\lambda_{q}}(p_{q})$ is a specific Dirac structure which determines the spin trace of the tensor current $S_{q}^{\mu\nu}\equiv \Tr [\mathcal{S} \mathcal{S}^{\prime \dag} U^{\mu\nu}]$\footnote{While the spin indices ($\lambda$) are explicitly shown in the derivation provided in the appendix, here we are only keeping the Dirac indices and ignoring the helicity or the spin indices.}. The momentum variables of the LF Dirac spinors occurring in the definition of the $U^{\mu\nu}$ represent the momentum of the active quark. The momenta of the incoming quark and the outgoing quark are, respectively, $p_{q}=(p_{q}^{-},p_{q}^{+},\vec{p}_{q\perp})=(\frac{\vec{p}_{q\perp}^{2}+m_{q}^{2}}{p_{q}^{+}},xP^{+},\vec{k}_{\perp})$ and $
 p_{q}^{\prime}=(p_{q}^{\prime-},p_{q}^{\prime+},\vec{p}_{q\perp}^{\prime})=(\frac{\vec{p}_{q\perp}^{\prime 2}+m_{q}^{2}}{p_{q}^{\prime +}},xP^{+},x\vec{P}_{\perp}^{\prime}+\vec{k}_{\perp}^{\prime})$. As the momentum of the spectator anti-quark remains unchanged, we have $\vec{k}_{\perp}+\vec{\Delta}_{\perp}=x\vec{P}_{\perp}^{\prime}+\vec{k}_{\perp}^{\prime}$ which implies $\vec{k}_{\perp}^{\prime}=\vec{k}_{\perp}+(1-x)\vec{\Delta}_{\perp}$. The total quark momentum is denoted by $p_{sq}= p_{q}+ p_{q}^{\prime}$.

\subsection{Distributions in transverse plane}\label{denstra}
The distribution of the spatial densities inside the hadrons, as the Fourier transformation of the corresponding form factors, is tempting, and, in fact, has been defined in this way in the past. In this definition, one takes a 3D Fourier transformation of the form factors in the Breit frame $\bar{P}=(E,0,0,0)$ and obtains the full 3D instant form spatial densities of the hadron~\cite{Polyakov:2018zvc,PhysRev.126.2256}. However, as pointed out in several references~\cite{Burkardt:2000za,Miller:2018ybm,Jaffe:2020ebz}, these densities suffer from relativistic and quantum delocalization effects~\cite{Jaffe:2020ebz}. Properly defined 3D densities can be obtained in the full non-relativistic limits as mentioned in the Ref.~\cite{Freese:2021mzg}. However, this field is active, and readers can consult Ref.~\cite{Lorce:2025oot} for further discussion. 

In this section, we discuss the transverse plane 2D densities obtained in the LF formalism, which are fully consistent with relativity. These 2D distributions for the EMT were introduced in Ref.~\cite{Lorce:2018egm} in the Wigner function formalism and later confirmed in Refs.~\cite{Freese:2021czn,Freese:2021qtb,Freese:2021mzg,Freese:2022fat}. We follow the Refs.~\cite{Freese:2021czn,Freese:2021qtb,Freese:2021mzg,Freese:2022fat} where the LF 2D density have been defined as the expectation value of the relevant operator in a localized hadron state (spatially localized in the transverse plane). In Refs.~\cite{Freese:2021czn,Freese:2021qtb,Freese:2021mzg,Freese:2022fat}, starting from the expectation value of the EMT operator in a localized wave function, the authors have shown that the internal momentum, stress, and energy distribution can be obtained from the 2D Fourier transformation of the matrix element of the EMT operator. Appendix~\eqref{apedensity} contains a brief discussion along these lines, along with derivations of some useful formulas to which readers may refer. 
Once the $A$ and $D$-terms of the quarkonia are known one can express the matrix element of the EMT as
\begin{eqnarray}
  T^{\mu\nu}  =2\bar{P}^{\mu}\bar{P}^{\nu}~A+\frac{\Delta^{\mu}\Delta^{\nu}-g^{\mu\nu}\Delta^{2}}{2}~D~.\label{G9}
\end{eqnarray}
The pure stress tensor is associated with the following Fourier transformation,
\begin{eqnarray}
  {S}^{ij}
  &=& \frac{1}{4P^{+}}\int \frac{d^{2}\vec{\Delta}_{\perp}}{(2\pi)^{2}}~e^{-i\vec{\Delta}_{\perp}\cdot \vec{z}_{\perp}}~(\Delta_{\perp}^{i}\Delta_{\perp}^{j}-\vec{\Delta}_{\perp}^{2}~\delta^{ij})\nonumber\\
  &&D(-\vec{\Delta}_{\perp}^{2})~.\label{G10}
  \end{eqnarray}
  While the pure stress tensor is dependent on the state of the quarkonia via $P^{+}$, an internal stress tensor defined by ${\mathcal{S}}^{ij}\equiv P^{+}{S}^{ij}$ is not. 
Drawing an analogy from continuum mechanics, we can define the pure stress tensor as the decomposition,
\begin{eqnarray}
S^{ij}&=& \left(\frac{z_{\perp}^{i}z_{\perp}^{j}}{z_{\perp}^{2}}-\frac{1}{2}\delta^{ij}\right)~s(z_{\perp})+ \delta^{ij}~p(z_{\perp})~,\label{G11}
\end{eqnarray}
where $p$ and $s$ are the LF static pressure and shear function (give rise to shear stresses), respectively, with $|\vec{z}_{\perp}|=z_{\perp}$ is the distance in the transverse plane. The shear function and pressure can be obtained by comparing the fundamental definition \eqref{G10} with the tensor decomposition \eqref{G11} as (see Appendix~\eqref{apedensity}),
\begin{eqnarray}
&&s(z_{\perp})=-z_{\perp}\frac{d}{dz_{\perp}}\left(\frac{1}{z_{\perp}}\frac{d\tilde{D}}{dz_{\perp}}\right),\label{ex4}\\
&&p(z_{\perp})=\frac{1}{2z_{\perp}}\frac{d}{dz_{\perp}}\left(z_{\perp}\frac{d\tilde{D}}{dz_{\perp}}\right),\label{ex5}\\
&&\text{where } \tilde{D}\equiv \frac{1}{4P^{+}} \int \frac{d^{2}\vec{\Delta}_{\perp}}{(2\pi)^{2}} e^{-i\vec{\Delta}_{\perp}\cdot \vec{z}_{\perp}} ~D(-\vec{\Delta}_{\perp}^{2}).\label{G12}
\end{eqnarray}
By virtue of the relation obtained above and EMT conservation, the shear function and pressure obey certain conditions~\cite{Freese:2021czn},
\begin{eqnarray}
\frac{d p(z_{\perp})}{d z_{\perp}}+\frac{1}{z_{\perp}}s(z_{\perp})+\frac{1}{2}\frac{d s(z_{\perp})}{d z_{\perp}}=0\label{ex6}\\
\text{and }\int d^{2}\vec{z}_{\perp}~p(z_{\perp})=0.\label{G13}
\end{eqnarray}
Eq.~\eqref{ex6} is a direct result of $\partial_{i}S^{ij}=0$, whereas the second equality~\eqref{G13}, known as the von Laue condition of stability, is an implication of the expression of the pressure stated in Eq.~\eqref{ex5}. The effective normal force density on a 1D surface\footnote{This is basically a line in 2D. We choose it to be a circular section.} is obtained as $\mathcal{F}^{i}=S^{ij}n^{j}=\frac{z_{\perp}^{j}}{z_{\perp}}(p+\frac{1}{2}s)$, where $n^{j}$ is the normal to the surface. Apart from the von Laue global stability condition, the local stability of the matter implies an outward force at any point~\cite{Perevalova:2016dln,Freese:2021czn}, i.e., $p+\frac{1}{2}s>0$. One defines the LF momentum ($P^{+}$) and energy ($P^{-}$) densities in a similar manner as~\cite{Freese:2022fat},
\begin{eqnarray}
\mathcal{P}&=& P^{+}\int \frac{d^{2}\vec{\Delta}_{\perp}}{(2\pi)^{2}}e^{-i\vec{\Delta}_{\perp}\cdot \vec{z}_{\perp}}~A(-\vec{\Delta}_{\perp}^{2})\label{G14}\\
\mathcal{E}&=& \frac{1}{2P^{+}}\int \frac{d^{2}\vec{\Delta}_{\perp}}{(2\pi)^{2}}e^{-i\vec{\Delta}_{\perp}\cdot \vec{z}_{\perp}}~\bigg(\big(M^{2}+\frac{\vec{\Delta}_{\perp}^{2}}{4}\big)\nonumber\\
&\times&A(-\vec{\Delta}_{\perp}^{2})
+\frac{\vec{\Delta}_{\perp}^{2}}{2}~D(-\vec{\Delta}_{\perp}^{2})\bigg)~.\label{G15}
\end{eqnarray}
The mean squared radius associated with momentum and force is given by,
\begin{eqnarray}
\langle z_{\perp}^{2}\rangle_{\mathcal{P}}&=&\frac{1}{P^{+}}\int d^{2}\vec{z}_{\perp} ~z_{\perp}^{2} \mathcal{P}(z_{\perp})\label{G16}\\
\langle z_{\perp}^{2}\rangle_{\mathcal{F}}&=& \frac{\int d^{2}\vec{z}_{\perp}~ z_{\perp}^{2}~\left(p(z_{\perp})+\frac{1}{2}s(z_{\perp})\right)}{\int d^{2}\vec{z}_{\perp} ~\left(p(z_{\perp})+\frac{1}{2}s(z_{\perp})\right)} ~.\label{G17}
\end{eqnarray}

\section{Results and Discussions}\label{results}
In this section, we describe the results of the GFFs and the corresponding spatial densities in the transverse plane using the expressions provided in Secs.~\ref{sec:GFF} and \ref{denstra}. Specifically, for $A_{q}$ and $D_{q}$-terms, we use Eq.~\eqref{G4} and Eq.~\eqref{G5}, respectively, where the components of the EMT tensor given in Eq.~\eqref{G8} are evaluated using the spin trace of the tensor current provided in Table~\ref{tab:spintraces}.
Two different sets of wavefunctions, i.e., SET-I corresponding to $\varphi^{\rm CJ}$ (Eq.~\eqref{CJwfn}) and SET-II corresponding to $\varphi^{\rm TK}$ (Eq.~\eqref{TKwfn}), are used for each result. It should be noted that the wavefunctions are characterized by two parameters, the mass of the quark (or antiquark) $m_{q}$ and the harmonic parameter $\beta$. In the following results and discussion, we use a common set of parameters for both wavefunctions to isolate the differences arising from their shapes. The parameters are adopted from Ref.~\cite{Arifi:2022pal}: $m_{c}=1.68$ GeV, $m_{b}=5.10$ GeV, $\beta_{\eta_{c}}=0.699$ GeV and $\beta_{\eta_{b}}=1.376$ GeV.
\begin{figure}[H]
	\centering
	\includegraphics[width=\columnwidth]{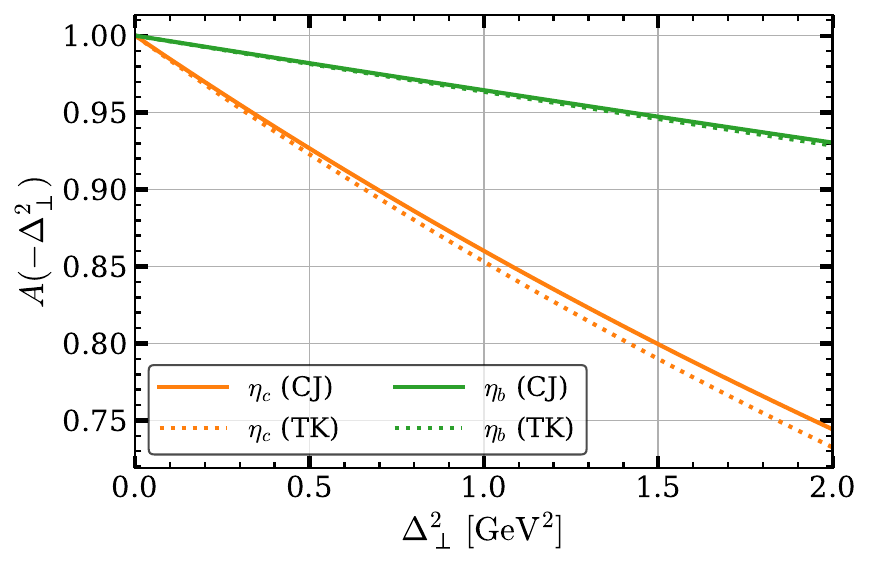}
	\caption{(Color online) Variation of the $A$-term with squared momentum transfer.}
	\label{fig:A(t)}
\end{figure}

\begin{figure}[H]
	\centering
	\includegraphics[width=\columnwidth]{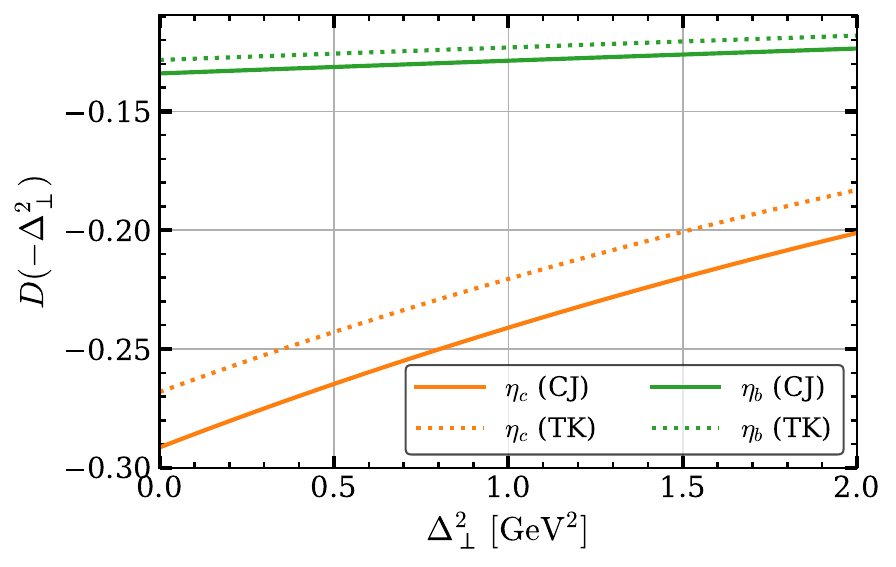}
	\caption{(Color online) Variation of the $D$-term with squared momentum transfer.}
	\label{fig:D(t)}
\end{figure}

\begin{figure}[H]
	\centering
	\includegraphics[width=\columnwidth]{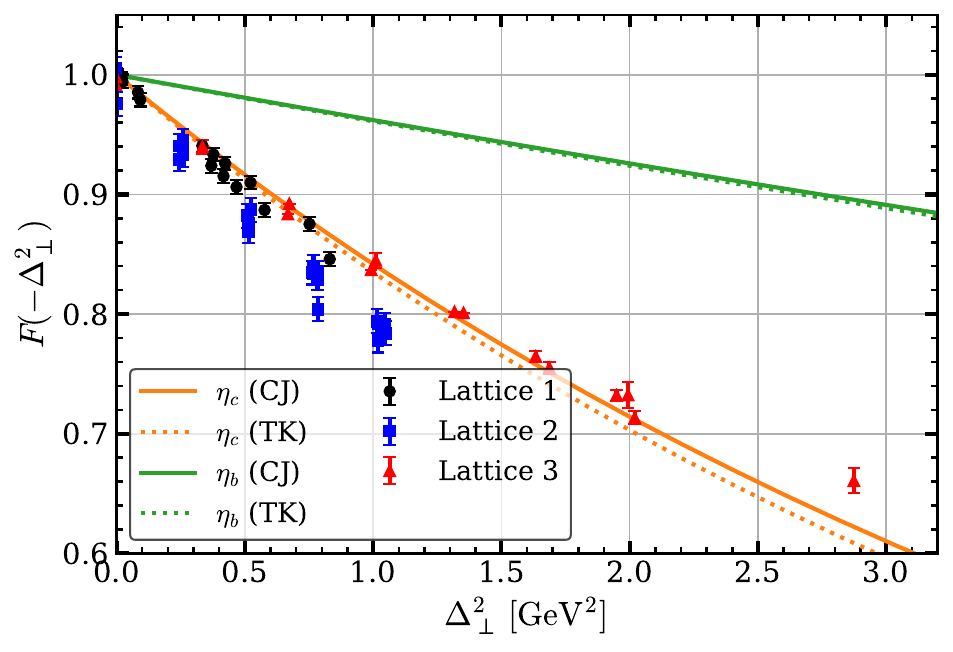}
	\caption{(Color online) Variation of the EMFFs with squared momentum transfer.  For the case of $\eta_{c}$ the data points of Lattice 1~\cite{Li:2020gau}, 2~\cite{Chen:2011kpa}, and 3~\cite{Delaney:2023fsc} are also put for comparison.}
	\label{fig:F(t)}
\end{figure}
To begin with, we display the $A$-term as a function of the momentum transfer. From Fig.~\ref{fig:A(t)}, we observe that $A(-\vec{\Delta}_{\perp}^{2})$ satisfies the momentum sum rule~\cite{Freese:2021czn}, namely $A(0)=1$, for both charmonium and bottomonium. We also find that the A-term decreases with increasing momentum transfer for both quarkonia. For charmonium, a clear difference between the results obtained from different wave functions appears at high momentum transfer. In contrast, this difference is only marginal for bottomonium. The slope of the momentum form factor at zero momentum transfer provides a measure of the longitudinal momentum distribution in the transverse plane~\cite{Freese:2021czn} (We discuss more regarding this at the end of this section, also see Eq.~\eqref{apd15} of Appendix~\eqref{apedensity}). From the slopes shown in Fig.~\ref{fig:A(t)}, it is evident that the longitudinal momentum in bottomonium is distributed over a smaller transverse area compared to charmonium.
\begin{figure}[H]
	\centering
	\begin{subfigure}{0.48\textwidth}
		\includegraphics[width=\linewidth]{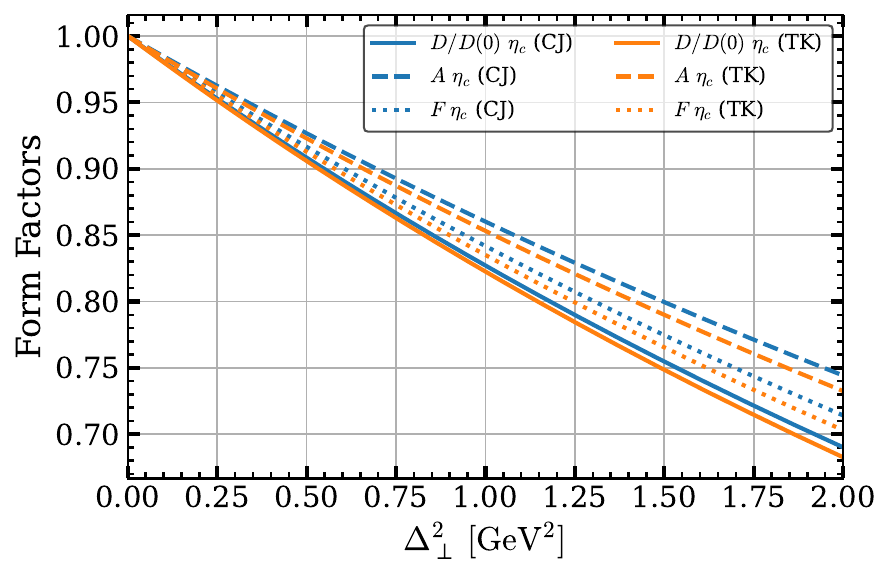}
	\end{subfigure}
	\hfill
	\begin{subfigure}{0.48\textwidth}
		\includegraphics[width=\linewidth]{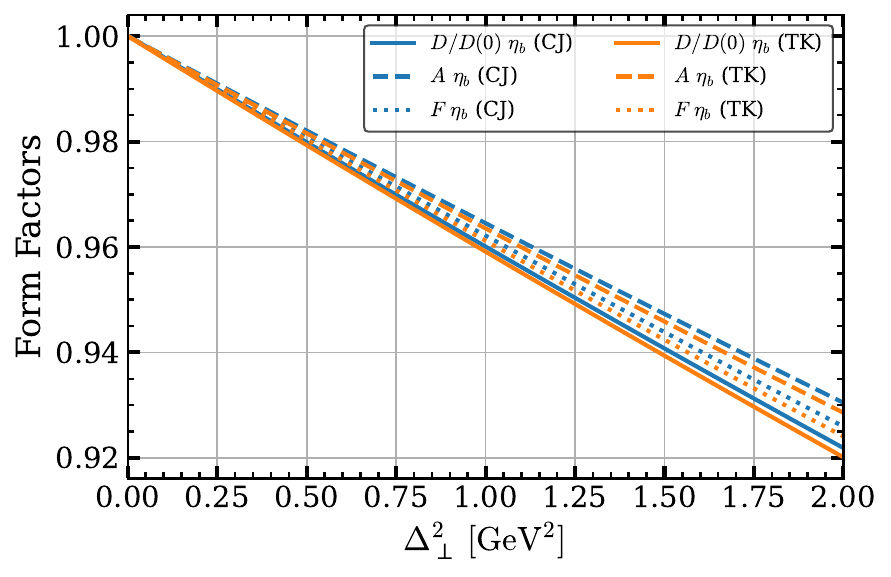}
	\end{subfigure}
	\caption{(Color online) Variation of all the form factors (GFFs and EMFFs) as a function of momentum transfer is shown for $\eta_{c}$ (upper panel) and $\eta_{b}$ (lower panel). Note that the normalized $D$-term, $D(-\vec{\Delta}_{\perp}^{2})/D(-\vec{\Delta}_{\perp}^{2}=0)$ is shown for a meaningful comparison.}
	\label{fig:allFFeta}
\end{figure}

In Fig.~\ref{fig:D(t)}, we display the so-called Druck term or $D$-term for quarkonia. 
Unlike the $A$-term, there are no established values or constraints for the magnitude of the $D$-term at zero momentum transfer except for the pion. For the pion, low-energy theorems can be used to dictate the value to be $-1$ in the chiral limit, i.e., $D_{\pi}(-\vec{\Delta}_{\perp}^{2}=0)=-1$ for $m_{\pi}\rightarrow 0$ with corrections terms at finite pion mass (See Ref.~\cite{Hudson:2017xug} and also articles cited there in). For heavy mesons made up of heavy quarks, the magnitude of $D(-\vec{\Delta}_{\perp}^{2}=0)$ can be different from one while still staying negative. We notice in Fig.~\ref{fig:D(t)} that the magnitude of the $D(-\vec{\Delta}_{\perp}^{2}=0)$ is far from the value $-1$ predicted for the Goldstone bosons. For bottomonium, which is heavier than for charmonium, the value of $D_{\pi}(-\vec{\Delta}_{\perp}^{2}=0)$ is even farther away from $-1$. Our results for the $D$-term are found to be consistent with the contact-interaction approach \cite{Sultan:2024hep}, while they deviate from the results reported in Refs.~\cite{Xu:2024hfx,Hu:2024edc}. 
At zero transverse momentum transfer, we observe the $D(-\vec{\Delta}_{\perp}^{2}=0)$ values are obtained as
\begin{equation}
\begin{aligned}
D^{\rm CJ}_{\eta_c}   &=-0.291, \qquad & D^{\rm CJ}_{\eta_b} &=-0.134 , \\
D^{\rm TK}_{\eta_c} &=-0.268 , \qquad & D^{\rm TK}_{\eta_b} &= -0.127 .
\end{aligned}
\end{equation}
We also see a clear dependence of the space wavefunction on the behavior of the $D$ as a function of momentum transfer. The magnitude of the $D$ decreases as the momentum transfer increases for both the quarkonia, and the wavefunction dependence is pronounced for $\eta_{c}$ than $\eta_{b}$.

\begin{figure*}[t]
	\centering
	
	\begin{subfigure}{0.95\linewidth}
		\centering
		\includegraphics[width=\linewidth]{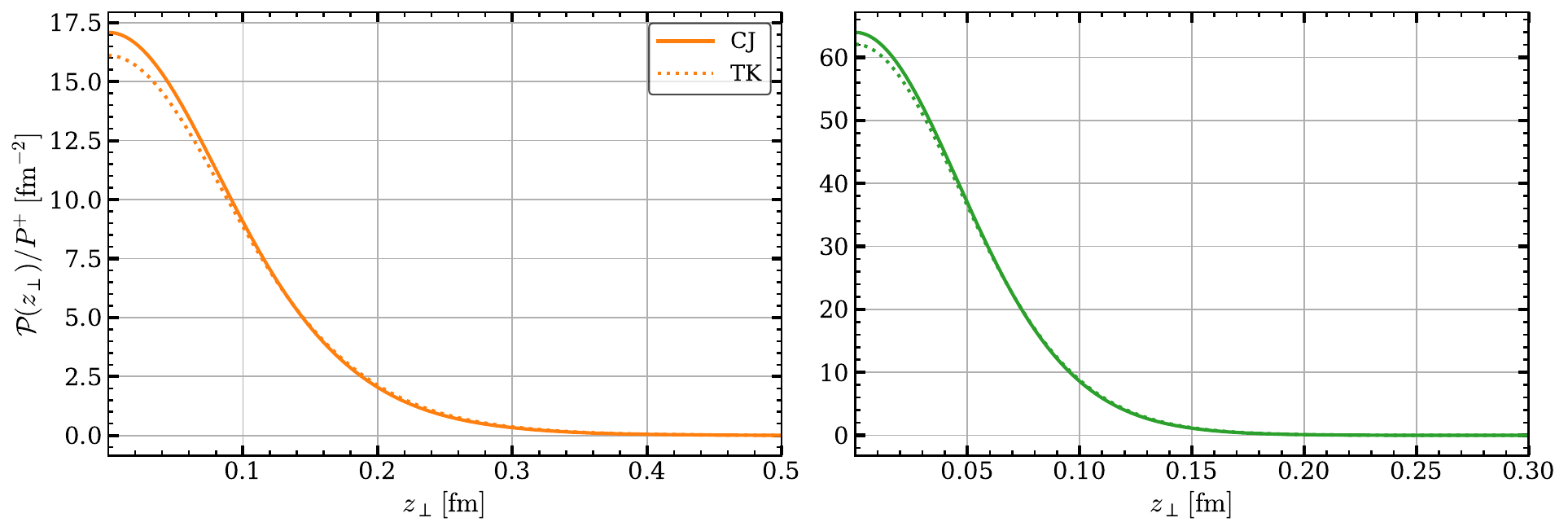}
		\caption{(Color online) Light-front longitudinal momentum density of the mesons $\eta_{c}$ (left) and $\eta_{b}$ (right).}
		\label{fig:mom}
	\end{subfigure}
	
	\vspace{0.5cm}
	
	\begin{subfigure}{0.95\linewidth}
		\centering
		\includegraphics[width=\linewidth]{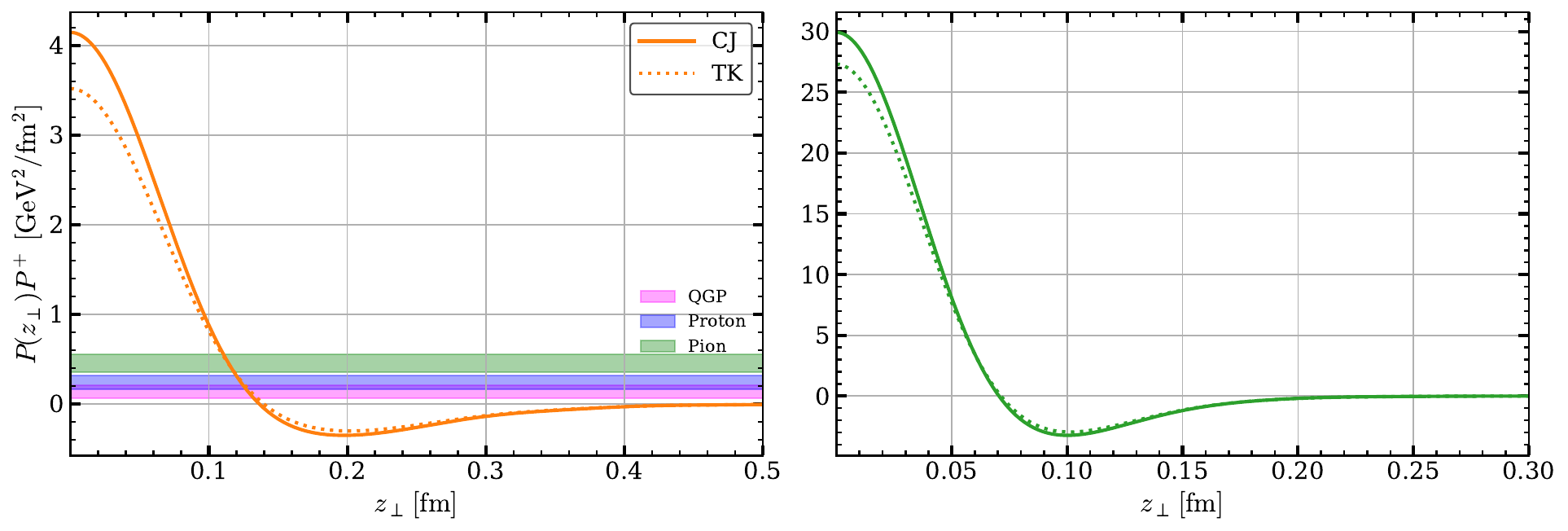}
		\caption{(Color online) The pressure distribution in the transverse plane for the mesons $\eta_{c}$ (left) and $\eta_{b}$ (right). Also shown are the pressure range of the quark--gluon plasma (QGP) from lattice QCD simulations~\cite{PhysRevLett.126.232001}, the peak pressure (with uncertainty) in the proton extracted from experimental measurements~\cite{Burkert:2018bqq}, and the range of peak pressure values in the pion obtained in various theoretical studies~\cite{Xu:2023izo,Liu:2024vkj,Choi:2025rto}.}
		\label{fig:press}
	\end{subfigure}
	
	\caption{Dependence of spatial distributions of mechanical properties (momentum and pressure) of mesons on their wavefunctions.}
	\label{mom_press}
\end{figure*}
\begin{figure*}[t]
	\centering
	
	\begin{subfigure}{0.95\linewidth}
		\centering
		\includegraphics[width=\linewidth]{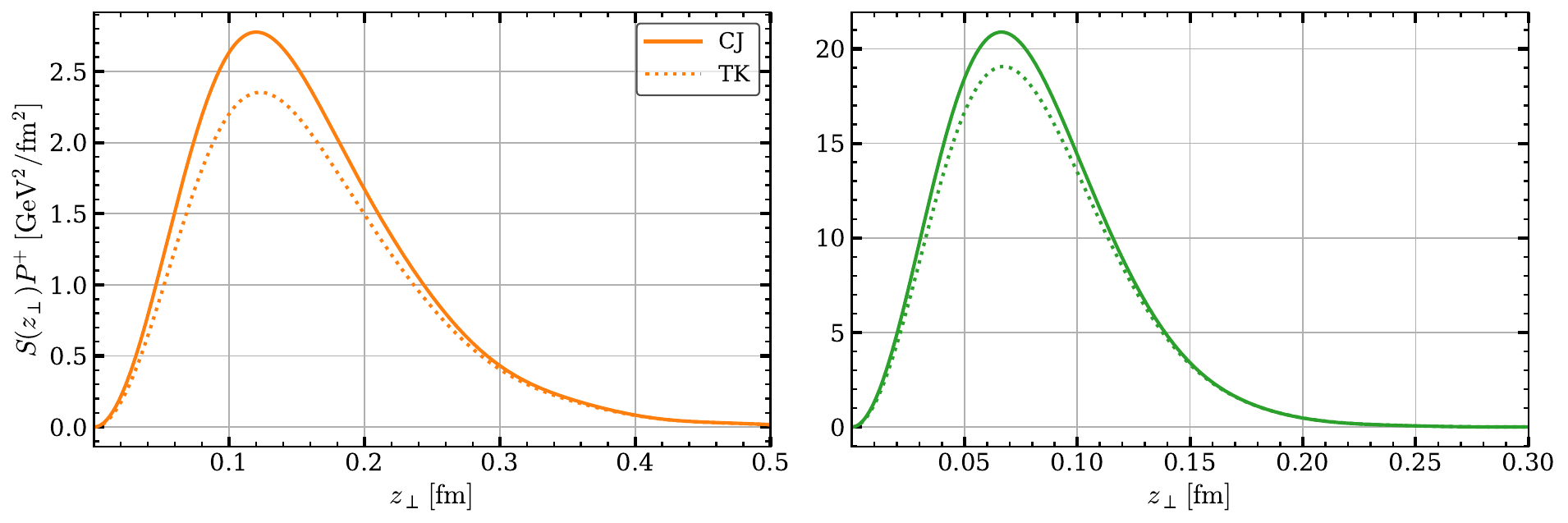}
		\caption{(Color online) Distribution of shear stresses in the transverse plane for the mesons $\eta_{c}$ (left) and $\eta_{b}$ (right).}
		\label{fig:shear}
	\end{subfigure}
	
	\vspace{0.5cm}
	
	\begin{subfigure}{0.95\linewidth}
		\centering
		\includegraphics[width=\linewidth]{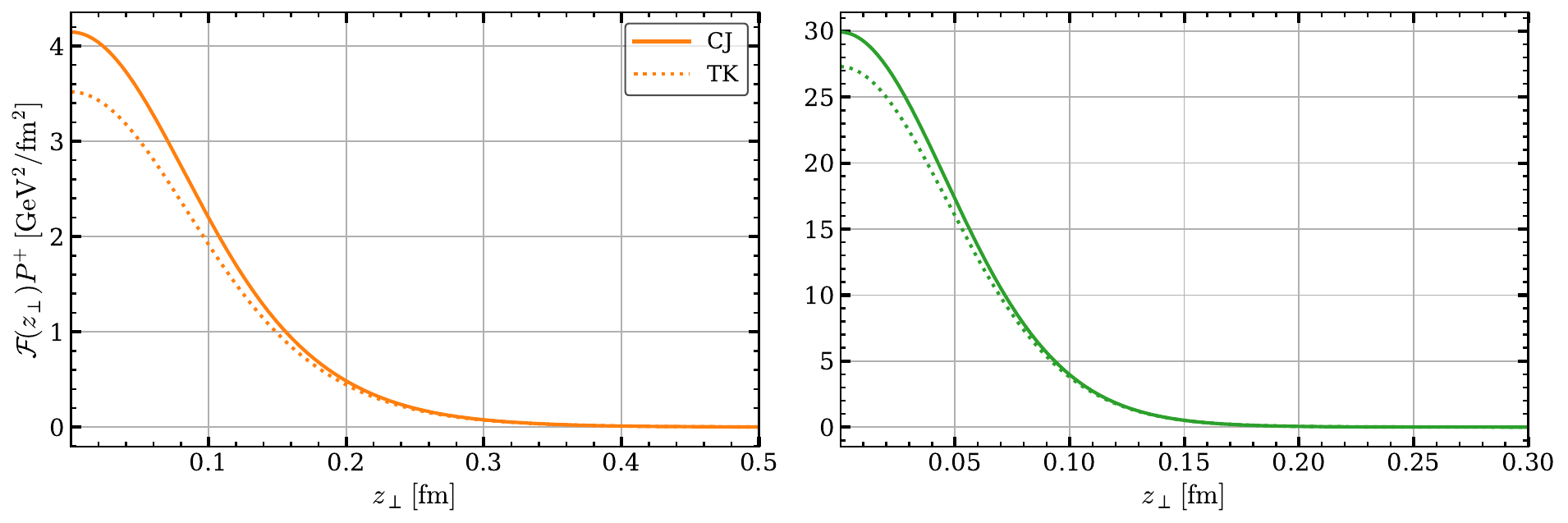}
		\caption{(Color online) Force density in the transverse plane for the mesons $\eta_{c}$ (left) and $\eta_{b}$ (right).}
		\label{fig:force}
	\end{subfigure}

    \begin{subfigure}{0.95\linewidth}
		\centering
		\includegraphics[width=\linewidth]{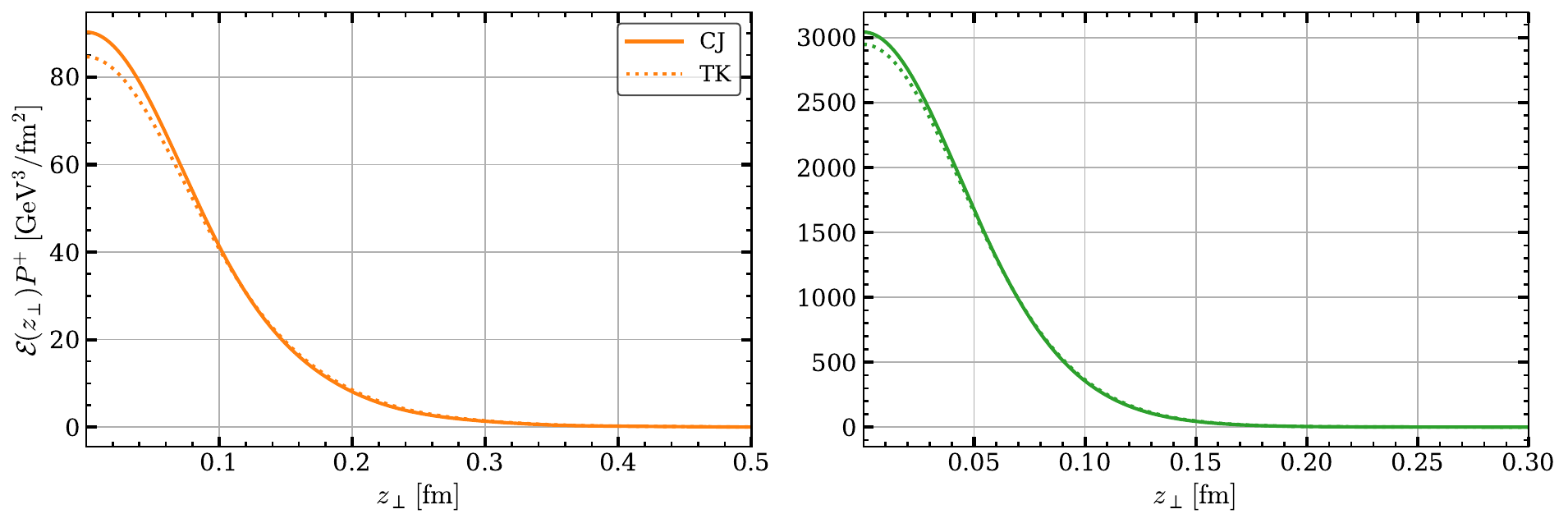}
		\caption{(Color online) Internal energy distribution in the transverse plane for the mesons $\eta_{c}$ (left) and $\eta_{b}$ (right).}
		\label{fig:energy}
	\end{subfigure}

	\caption{The wavefunction dependence of the mechanical distributions (shear, force, and energy) inside $\eta_{c}$ and $\eta_{b}$.}
    \label{fig:shaer_force_energy}
\end{figure*}

For completeness, we also show the EMFFs of the quarkonia as a function of momentum transfer in Fig.~\ref{fig:F(t)} as well as a unified comparison in Fig.~\ref{fig:allFFeta} for both the choices of wavefunctions. The expression of the EMFF in terms of an integral over the LFWFs is well established and can be found in Refs.~\cite{Puhan:2025kzz,Puhan:2025pfs}.
We notice a slight variation of the EMFF with the wave function for the case of $\eta_{c}$ and marginal variations in $\eta_{b}$. We compare our results for the EMFF with lattice simulations for the meson $\eta_{c}$. Our results are in good agreement with the lattice data of the Refs.~\cite{Li:2020gau,Delaney:2023fsc,Chen:2011kpa}. In Fig.~\ref{fig:allFFeta} we present all the form factors as a function of momentum transfer. This shows nicely all the form factors relative momentum dependence as well as the dependence on the spatial wavefunctions. The normalized value of the $D$-term, i.e., $D/D(0)$ as a function of momentum transfer, provides a measure of force distribution inside the meson. The mechanical radius is inversely proportional to the value of $D/D(0)$ (see Eq.~\eqref{apd14} of Appendix~\eqref{apedensity}). From the Fig.~\ref{fig:allFFeta}, one reckons that the transverse radius associated with $\eta_{b}$ is less than $\eta_{c}$, which inturn imply that the force distribution is more concentrated for $\eta_{b}$ than $\eta_{c}$. These facts would become clear as we display the mechanical distributions in the transverse plane in the next paragraphs.

In Fig.~\ref{mom_press}, we show the LF longitudinal momentum distribution (upper panel) and pressure in the transverse plane (lower panel), respectively. The distribution corresponding to the $\eta_{c}$ is shown on the left, and that of $\eta_{b}$ is shown on the right.
For the momentum distribution, we use the normalization factor $P^{+}$ so that the normalized distribution $\mathcal{P}/P^{+}$ integrated over the transverse plane would give one. Note that the momentum distribution is sensitive to the choice of wavefunction in the region near the center. This is a manifestation of the sensitivity of the $A$-term to the wavefunction for large momentum transfer (see Fig.~\ref{fig:A(t)}). Particularly, we notice that the CJ wavefunction gives higher momentum density for both the quarkonia. For $\eta_{c}$ and $\eta_{b}$, both the wavefunctions give approximately similar momentum density as the transverse distance increases further from $0.10$ fm and $0.05$ fm, respectively. We see for the heavier meson $\eta_{b}$, the momentum distribution is more concentrated on the central region than that of $\eta_{c}$. Turning our attention to the pressure inside the mesons, we note that it is obtained by taking appropriate spatial derivatives of the Fourier transform of the $D$-term (Eq.~\eqref{G12}). Therefore, the pressure inside the meson is determined by the variation of the $\tilde{D}$ with the transverse distance.
In the transverse plane, we observe a node at a finite distance for each meson, where the pressure turns from positive (i.e., attractive) to negative (i.e., repulsive), and finally vanishes as one moves towards the periphery. This effect is a result of the von Laue stability condition~\cite{Polyakov:2018zvc}, which states that $\int dz_{\perp} z_{\perp} p(z_{\perp})=0$ (see Eq.~\eqref{apd10} of Appendix~\eqref{apedensity} for derivation). We have numerically verified the validity of the von Laue stability condition for different transverse distances, thereby ensuring the consistency of the numerical analysis.
Although pressure in general is observed to depend on the chosen wavefunction type, at least in the central region, the position of the nodes is seen to be mostly independent. The nodes in the $\eta_{c}$ and $\eta_{b}$ mesons are located approximately at $0.14$ fm and $0.07$ fm, respectively.
To facilitate a direct comparison of numerical magnitudes, we also include pressure bands corresponding to the QGP, proton, and pion obtained in Refs.~\cite{Burkert:2018bqq,PhysRevLett.126.232001,Xu:2023izo,Liu:2024vkj,Choi:2025rto}, alongside the pressure distributions of the quarkonia. For a thermalized system pressure depends on both the temperature and chemical potential. For the QGP, the band is derived from the thermodynamic pressure calculated using lattice QCD simulations~\cite{PhysRevLett.126.232001} for temperatures in the range $T\approx0.200-0.240$ GeV at vanishing baryon chemical potential, $\mu_{B}=0$. For the proton, we display the pressure at $r=0.25$ fm together with its associated uncertainty, motivated by Ref.~\cite{Burkert:2018bqq}, where $r^{2}P(r)$ reaches its maximum near this radius. The pion band is constructed from the peak pressure values reported in Refs.~\cite{Xu:2023izo,Liu:2024vkj,Choi:2025rto}. We observe that the peak pressures of both the quarkonia are substantially larger than the corresponding values for the QGP, proton, and pion.

We analyze the shear function, force density, and internal energy density inside the quarkonia in Fig.~\ref{fig:shaer_force_energy} in the upper, middle, and lower panels, respectively. The distribution of the shear function shows a spike around the point where we observe the nodes in the pressure distribution. Interestingly, in the shear function, we see sensitivity to the wavefunction at intermediate distances, whereas at the center and periphery, the shear functions for the two sets of wavefunctions merge. Apart from the von Laue condition, which provides a global condition of stability, one can have the local stability condition, which states that force density should point radially outward, i.e., $\mathcal{F}\equiv p+\frac{1}{2}s>0$, as postulated in Ref.~\cite{Perevalova:2016dln}. As a local stability condition, it provides a stronger stability criterion. The force density displayed in the middle panel of Fig.~\ref{fig:shaer_force_energy} verifies this local stability criterion for both the mesons, irrespective of the choice of the wavefunctions. The internal energy density of the quarkonia, as defined in Eq.~\eqref{G15}, contains the Fourier transformation of both the GFFs in its expression. As pointed out in Ref.~\cite{Freese:2022fat}, this internal energy density includes kinetic energy of the constituents (including rest mass energies) relative to the center-of-momentum, and the potential energy. We observe that this internal energy is positive for both the quarkonia and decreases with distance from the center. For $\eta_{b}$, we notice that the energy density is very large at the center, which drops rapidly when one approaches the surface of the quarkonium. We also observe that the energy density is sensitive to the choice of the wavefunction in the central region.

We close the discussion by giving several useful length scales that define the spread of the mechanical properties of the quarkonia in the transverse plane. One can then compare it with the length associated with the EMFF, which decides the spread in the charge distribution. A useful quantity is the slope of the corresponding form factors at zero momentum transfer, from which one defines the length scale~\cite{Choi:2025rto}
\be
\langle r^{2}_{\mathscr{F}} \rangle=-\frac{6}{\mathscr{F}(0)}\frac{d\mathscr{F}}{d\vec{\Delta}_{\perp}^{2}}~,\label{res1}
\ee
where $\mathscr{F}$ can be the GFFs or the EMFFs. 
The generic radius associated with the $A$ and $F$-terms is the length associated with the mass and charge distributions. Similarly, we also provide the length associated with the $D$-term by finding the slope at zero momentum transfer as provided in Ref.~\cite{Choi:2025rto} for pions. 
\begin{table}[H]
\centering
\begin{tabular}{|c|c|ccc|cc|}
\hline
LFWFs & Meson
& $\sqrt{\langle r^{2}_{A} \rangle}$ 
& $\sqrt{\langle r^{2}_{D} \rangle}$ 
& $\sqrt{\langle r^{2}_{F} \rangle}$
& $\sqrt{\langle z^{2}_{\perp} \rangle}_{\mathcal{P}}$
& $\sqrt{\langle z^{2}_{\perp} \rangle}_{\mathcal{F}}$ \\ \hline

\multirow{2}{*}{{\rm CJ}}
& $\eta_c$
& 0.188 & 0.217 & 0.202
& 0.152 & 0.152 \\ \cline{2-7}

& $\eta_b$
& 0.092 & 0.076 & 0.095
& 0.075 & 0.075 \\ \hline

\multirow{2}{*}{{\rm TK}}
& $\eta_c$
& 0.193 & 0.237 & 0.207
& 0.156 & 0.155 \\ \cline{2-7}

& $\eta_b$
& 0.093 & 0.074 & 0.096
& 0.076 & 0.076 \\ \hline
\end{tabular}
\caption{Mean radii (measured in fm) corresponding to the three form factors and two transverse-plane distributions.}
\label{T1}
\end{table}
As described in Appendix~\eqref{apedensity}, the radii obtained with Eq.~\eqref{res1} correspond to the length scale associated with 3D distributions. Similarly, $2D$ mean radii associated with the momentum and force distribution in the transverse plane are defined in Sec.~\ref{denstra} in Eqs.~\eqref{G16} and \eqref{G17}. We provide a comparison of these radii for different choices of wavefunctions in the Table~\eqref{T1}.
We should mention that the $2D$ and $3D$ radius associated with the $A$-term are related by $\sqrt{\langle r^{2}_{A} \rangle}=\sqrt{\frac{3}{2}}\sqrt{\langle z^{2}_{\perp} \rangle}_{\mathcal{P}}$ (see Eq.~\eqref{apd15} of Appendix~\eqref{apedensity}). One can see that this equality holds to a good extent in our numerical evaluation of the radius parameters provided in Table~\eqref{T1}. Moreover, we see that similar to the pion case reported in Ref.~\cite{Choi:2025rto,Kumano:2017lhr} for charmoinium, the ranking $\sqrt{\langle r^{2}_{A_{\eta_{c}}}\rangle} < \sqrt{\langle r^{2}_{F_{\eta_{c}}}\rangle} < \sqrt{\langle r^{2}_{D_{\eta_{c}}}\rangle}$ hold. However, for the heavier meson, the bottomonium, it no longer holds. In the $2D$ plane, we notice that the momentum and force are distributed almost equally with $\sqrt{\langle z^{2}_{\perp} \rangle}_{\mathcal{P}}\approx \sqrt{\langle z^{2}_{\perp} \rangle}_{\mathcal{F}}$ for both the quarkonia. The mechanical properties for the bottomonium are spread in a region which is almost half of that of the charmonium, i.e., $\sqrt{\langle z^{2}_{\perp} \rangle}_{\eta_{b}} \approx \frac{1}{2} \sqrt{\langle z^{2}_{\perp} \rangle}_{\eta_{c}}$.

\section{Summary}\label{Sec:conclusion}
To summarize the work, we have investigated the mechanical properties and internal force distributions of the heavy pseudoscalar mesons by solving the energy-momentum tensors (EMTs) within the self-consistent light-front quark model (LFQM). This study provides insight into the spatial distributions of energy, pressure, and shear forces inside heavy quarkonia by extracting the gravitational form factors associated with the EMT.
In order to properly account for the zero-mode contributions that arise in the light-front dynamics, we adopted the Bakamjian–Thomas (BT) construction method following the Ref.~\cite{Choi:2025rto}. This implementation of the BT framework ensures the restoration of covariance and consistency between different current components for meson systems composed of equal-mass quark and anti-quark. 

By evaluating the matrix elements of the EMT ($\hat{T}^{\mu \nu}$) using the appropriate mesonic Fock-state expansion and quark-field operators, we derived the $A$- and $D$-term gravitational form factors (GFFs), which primarily encode the internal energy, momentum, pressure, and shear-force distributions inside the meson. The $A$-term, which is calculated from the good current components of the EMT, is associated with the momentum and mass-energy distribution carried by the constituents, while the $D$-term characterizes the mechanical properties of the hadron, including the internal pressure and stability conditions. With two distinct light-front wavefunctions, we have provided a comprehensive description of the mechanical distributions of heavy quarkonia. The \(A\)-term is found to satisfy the corresponding sum rule, and the calculated longitudinal momentum density exhibits positive distributions throughout the transverse position space. The pressure distribution, obtained from the \(D\)-term, shows positive values in the low-\(z_\perp\) region, indicating a strong repulsive force among the constituents, while it becomes negative at larger \(z_\perp\), suggesting the presence of an attractive force among the constituents. The stress distribution is found to exhibits the positive distribution all over the \(z_\perp\). Similarly, the force density and internal energy density are also found to exhibit positive distributions throughout the transverse plane. In the case of heavy bottomonium, the \(D\)-term is found to be of smaller magnitude than the charmonium case. The longitudinal momentum density, internal energy density, pressure, force density, and shear distributions for the bottomonium are found to be larger in magnitude in the central region than those of charmonium in both model wavefunctions. The mechanical and momentum radii of quarkonia are found to be the same, which are less than their charge radii. All the distributions in this paper are calculated using the 2D Fourier transformation rather than choosing the 3D transformations to avoid the quantum delocalization effects.

Overall, this study provides a comprehensive description of the internal mechanical distributions of heavy quarkonium systems within the framework of light-front dynamics with two distinct light-front wavefunctions. These findings may serve as useful theoretical inputs for future investigations of gravitational form factors, generalized parton distributions, and lattice QCD studies of heavy hadronic systems. The present work can also be extended to excited states, higher Fock sectors, and spin-dependent mechanical properties in future studies.

\acknowledgments
This work was partially supported by the Ministry of Education (MoE), Government of India (A.D.), and by the Board of Research in Nuclear Sciences (BRNS) and the Department of Atomic Energy (DAE), Government of India, under Grant No. 57/14/01/2024-BRNS/313 (S.G.). A.D. thanks the YITP International School 2026 (YITP-W-25-16), “International School on EIC Physics at Kyoto”, where discussions contributed to this work. He acknowledges financial support from the MoE through the International Conference Visit Scheme of IIT Bhilai for his visit to Japan. He also thanks the organizers of the YITP School for their hospitality in Kyoto, and Prof. Chiho Nonaka and Dr. Cho Win Aung for hosting his research visit at Hiroshima University. S.P. thanks Prof. Oleg V. Teryaev for helpful discussions during his visit to BLTP-JINR, Dubna, Russia, and acknowledges the hospitality of IIT Bhilai and BLTP-JINR, where part of this work was carried out.

\appendix
\section{Light-front spinors and Light-front spin wavefunctions of pseudoscalar mesons}\label{defspin}
We follow the Lepage-Brodsky (LB) convention~\cite{Lepage:1980fj,Brodsky:1997de} in our analysis. The LF position coordinates $z_{\rm LF}\equiv z=(z^{+},z^{1},z^{2},z^{-})$ and the instant from coordinates $x=(x^{0},x^{1},x^{2},x^{3})$ are related by $z^{+}=x^{0}+x^{3}$, $z^{1}=x^{1}$, $z^{2}=x^{2}$ and $z^{-}=x^{0}-x^{3}$. This transformation leads to a non-diagonal metric
\begin{eqnarray}
g_{\mu\nu}=
\begin{pmatrix}
0 & 0 & 0 & \frac{1}{2}\\
0 & -1 & 0 & 0\\
0 & 0 & -1 & 0\\
\frac{1}{2} & 0 & 0 & 0\\
\end{pmatrix},
~g^{\mu\nu}=
\begin{pmatrix}
0 & 0 & 0 & 2\\
0 & -1 & 0 & 0\\
0 & 0 & -1 & 0\\
2 & 0 & 0 & 0
\end{pmatrix}\label{Ape1}~.
\end{eqnarray}
The inner product between two vectors is defined with the help of the metric as $ A\cdot B\equiv g_{\mu\nu}A^{\mu}B^{\nu}=\frac{1}{2}(A^{+} B^{-}+A^{-} B^{+})-\vec{A}_{\perp}\cdot \vec{B}_{\perp}$. We use the Dirac matrices in the Dirac representation, i.e.,
\begin{eqnarray}
&&  \gamma^{0}\equiv\boldsymbol{\beta}=
  \begin{pmatrix}
I & 0\\
0 & -I\\
\end{pmatrix},~
 \gamma^{k}=
  \begin{pmatrix}
0 & \sigma^{k}\\
-\sigma^{k} & 0\\
\end{pmatrix},\nonumber\\
&&\gamma^{5}=
\begin{pmatrix}
0 & I\\
I & 0\\
\end{pmatrix},~\alpha^{k}\equiv \boldsymbol{\beta} \gamma^{k}=
\begin{pmatrix}
0 & \sigma^{k}\\
\sigma^{k} & 0\\
\end{pmatrix}\label{Ape2}~,
\end{eqnarray}
where $I$ and $\sigma^{k}$ are the two dimensional identity and Pauli matrices. The LF Dirac matrices $\gamma^{\pm}$ are defined as $\gamma^{0}\pm \gamma^{3}$. The LF Dirac spinors are defined as
\begin{eqnarray}
  && u_{\lambda_{q}=\pm 1}(p_{q})=\frac{1}{\sqrt{p_{q}^{+}}}(p_{q}^{+}+\boldsymbol{\beta} m_{q} + \vec{\alpha}_{\perp}\cdot \vec{p}_{q\perp})~ \chi_{\pm},\nonumber\\
  \label{Ape3}\\
  && v_{\lambda_{\bar{q}}=\pm 1}(p_{\bar{q}})=\frac{1}{\sqrt{p_{\bar{q}}^{+}}}(p_{\bar{q}}^{+}-\boldsymbol{\beta} m_{\bar{q}} + \vec{\alpha}_{\perp}\cdot \vec{p}_{\bar{q}\perp}) ~\chi_{\mp}~,\nonumber\\
  \label{Ape4}
\end{eqnarray}
where $\chi_{+}^{T}=1/\sqrt{2}~(1~~0~~1~~0)$ and $\chi_{-}^{T}=1/\sqrt{2}~(0~~1~~0~~-1)$. Upon simplification, we can get the following four Dirac spinors:
\begin{eqnarray}
 && u_{\lambda_{q}=+1}(p_{q})=\frac{1}{\sqrt{2p_{q}^{+}}}
 \begin{pmatrix}
p_{q}^{+}+m_{q} \\
p_{q}^{R}\\
p_{q}^{+}-m_{q}\\
p_{q}^{R}\\
\end{pmatrix},\label{Ape05}\\
&&u_{\lambda_{q}=-1}(p_{q})=\frac{1}{\sqrt{2p_{q}^{+}}}
 \begin{pmatrix}
-p_{q}^{L} \\
p_{q}^{+}+m_{q}\\
p_{q}^{L}\\
-(p_{q}^{+}-m_{q})\\
\end{pmatrix}\label{Ape5}\\
 && v_{\lambda_{\bar{q}}=+1}(p_{\bar{q}})=\frac{1}{\sqrt{2p_{\bar{q}}^{+}}}
 \begin{pmatrix}
-p_{\bar{q}}^{L} \\
p_{\bar{q}}^{+}-m_{\bar{q}}\\
p_{\bar{q}}^{L}\\
-(p_{\bar{q}}^{+}+m_{{\bar{q}}})\\
\end{pmatrix},\label{Ape06}\\
&&v_{\lambda_{\bar{q}}=-1}(p_{\bar{q}})=\frac{1}{\sqrt{2p_{\bar{q}}^{+}}}
 \begin{pmatrix}
p_{\bar{q}}^{+}-m_{\bar{q}} \\
p_{\bar{q}}^{R}\\
p_{\bar{q}}^{+}+m_{\bar{q}}\\
p_{\bar{q}}^{R}\\
\end{pmatrix}~,\label{Ape6}
\end{eqnarray}
where we use the notation $p_{q(\bar{q})}^{R}\equiv p_{q(\bar{q})}^{x}+i p_{q(\bar{q})}^{y}$ and $p_{q(\bar{q})}^{L}\equiv p_{q(\bar{q})}^{x}-i p_{q(\bar{q})}^{y}$.
For a meson traveling with momenta $\vec{P}=(P^{+},\vec{P}_{\perp})$, the momentum of the quark and anti-quark comprising the valence state of the meson in the LF is expressed as, $\vec{p}_{q}=(p_{q}^{+}\equiv x_{q} P^{+},\vec{p}_{q\perp}\equiv x_{q}\vec{P}_{\perp}+\vec{k}_{q\perp})$ and $\vec{p}_{\bar{q}}=(p_{\bar{q}}^{+}\equiv x_{\bar{q}}P^{+},\vec{p}_{\bar{q}\perp}\equiv x_{\bar{q}}\vec{P}_{\perp}+\vec{k}_{\bar{q}\perp})$ such that $\vec{P}=\vec{p}_{q}+\vec{p}_{\bar{q}}$. The last condition renders the possibility of only two independent variables characterizing the longitudinal and transverse motion, i.e., $x\equiv x_{q}$ and $\vec{k}_{\perp}\equiv \vec{k}_{q\perp}$. The anti-quark momentum is constrained to be $\vec{p}_{\bar{q}}=(p_{\bar{q}}^{+}= (1-x)P^{+},\vec{p}_{\bar{q}\perp}=(1-x)\vec{P}_{\perp}-\vec{k}_{\perp})$. One writes the total valence state wavefunction $\Psi$ as a product of space $\varphi$ and spin wavefunctions $\mathcal{S}$, i.e., $\Psi_{\lambda_{q}\lambda_{\bar{q}}}(x, \vec{k}_{\perp})=\varphi(x, \vec{k}_{\perp}) ~\mathcal{S}_{\lambda_{q}\lambda_{\bar{q}}}(x, \vec{k}_{\perp})$. The spin wavefunction of a pseudoscalar meson is obtained by the choice of an appropriate vertex in terms of the quark and antiquark spinors as 
\begin{eqnarray}
&&\mathcal{S}_{\lambda_{q}\lambda_{\bar{q}}}=\frac{1}{\sqrt{2}} \frac{\bar{u}_{\lambda_{q}}(p_{q})~\gamma_{5}~v_{\lambda_{\bar{q}}}(p_{\bar{q}})}{\sqrt{M_{0}^{2}-(m_{q}-m_{\bar{q}})^{2}}},\nonumber\\
&&\text{ where } M_{0}^{2}=\frac{\vec{k}_{\perp}^{2}+m_{q}^{2}}{x}+\frac{\vec{k}_{\perp}^{2}+m_{\bar{q}}^{2}}{1-x}~.  \label{Ape7}  
\end{eqnarray} 
Upon explicitly performing the spinor multiplications in Eq.~\eqref{Ape7} using Eqs.~\eqref{Ape05} to \eqref{Ape6} we get 
\begin{eqnarray}
   &&\mathcal{S}_{\lambda_{q}\lambda_{\bar{q}}}=\frac{1}{\sqrt{2}\sqrt{\vec{k}_{\perp}^{2}+((1-x)m_{q}+x m_{\bar{q}})^{2}}}\nonumber\\
  && \times \begin{pmatrix}
-k^{R} & - ((1-x)m_{q}+x m_{\bar{q}})  \\
((1-x)m_{q}+x m_{\bar{q}}) & -k^{L}~
\end{pmatrix}~,\nonumber\\
\label{Ape8}
\end{eqnarray}
where we label the row (column) starting from $\lambda_{q(\bar{q})}=-1$ to ending at  $\lambda_{q(\bar{q})}=1$. It is worth pointing out that in the spin wavefunctions given in Eq.~\eqref{Ape8} the dependence on the total meson momenta $\vec{P}=(P^{+},\vec{P}_{\perp})$ cancels out. 

\section{Evaluation of matrix elements of EMT operator}\label{quarkcont} 
To get the EMT operator, we first write down the free quark fields and their derivatives. The field operators and their derivatives are given by,
\begin{eqnarray}
\psi_{q}(z)&=&\int \frac{dp_{q}^{+} d^{2}\vec{p}_{q}}{2p_{q}^{+}(2\pi)^{3}}~\sum_{\lambda_{q}}u_{\lambda_{q}}(p_{q})~ b_{q}(p_{q},\lambda_{q})~ e^{-ip_{q}\cdot z}\nonumber\\
&+&\int \frac{dp_{\bar{q}}^{\prime +} d^{2}\vec{p}_{\bar{q}}^{\prime}}{2p_{\bar{q}}^{\prime +}(2\pi)^{3}}\sum_{\lambda_{\bar{q}}^{\prime}} v_{\lambda_{\bar{q}}^{\prime}}(p_{\bar{q}}^{\prime}) ~d^{\dag}_{\bar{q}}(p_{\bar{q}^{\prime}},\lambda_{\bar{q}}^{\prime})~ e^{ip_{\bar{q}}^{\prime}\cdot z}~,\nonumber\\
\label{Ap1}\\
\bar{\psi}_{q}(z)&=&\int \frac{dp_{q}^{\prime+} d^{2}\vec{p}^{\prime}_{q}}{2p_{q}^{\prime+}(2\pi)^{3}}~\sum_{\lambda_{q}^{\prime}} \bar{u}_{\lambda_{q}^{\prime}}(p_{q}^{\prime})~ b^{\dag}_{q}(p_{q}^{\prime},\lambda_{q}^{\prime})~ e^{ip_{q}^{\prime}\cdot z}\nonumber\\
&+& \int \frac{dp_{\bar{q}}^{+} d^{2}\vec{p}_{\bar{q}}}{2p_{\bar{q}}^{+}(2\pi)^{3}}\sum_{\lambda_{\bar{q}}} \bar{v}_{\lambda_{\bar{q}}}(p_{\bar{q}}) ~d_{\bar{q}}(p_{\bar{q}},\lambda_{\bar{q}})~ e^{-ip_{\bar{q}}\cdot z}~,\nonumber\\
\label{Ap2}
\end{eqnarray}
\begin{eqnarray}
\partial^{\nu}\psi_{q}(z)&=&-i\int \frac{dp_{q}^{+} d^{2}\vec{p}_{q}}{2p_{q}^{+}(2\pi)^{3}}~p_{q}^{\nu}~\sum_{\lambda_{q}} u_{\lambda_{q}}(p_{q})~ b_{q}(p_{q},\lambda_{q})\nonumber\\
&&e^{-ip_{q}\cdot z}\nonumber\\
&+& i \int \frac{dp_{\bar{q}}^{\prime +} d^{2}\vec{p}_{\bar{q}}^{\prime}}{2p_{\bar{q}}^{\prime +}(2\pi)^{3}} p_{\bar{q}}^{\prime \nu} \sum_{\lambda_{\bar{q}}^{\prime}} v_{\lambda_{\bar{q}}^{\prime}}(p_{\bar{q}}^{\prime}) ~d^{\dag}_{\bar{q}}(p_{\bar{q}^{\prime}},\lambda_{\bar{q}}^{\prime})\nonumber\\ &&e^{ip_{\bar{q}}^{\prime}\cdot z},\label{Ap3}\\
\partial^{\nu}\bar{\psi}_{q}(z)&=&i\int \frac{dp_{q}^{\prime+} d^{2}\vec{p}^{\prime}_{q}}{2p_{q}^{\prime+}(2\pi)^{3}} p^{\prime\nu}_{q}~\sum_{\lambda_{q}^{\prime}} \bar{u}_{\lambda_{q}^{\prime}}(p_{q}^{\prime})
b^{\dag}_{q}(p_{q}^{\prime},\lambda_{q}^{\prime})\nonumber\\
&&e^{ip_{q}^{\prime}\cdot z} \nonumber\\
&-& i \int \frac{dp_{\bar{q}}^{+} d^{2}\vec{p}_{\bar{q}}}{2p_{\bar{q}}^{+}(2\pi)^{3}}p^{\nu}_{\bar{q}}\sum_{\lambda_{\bar{q}}} \bar{v}_{\lambda_{\bar{q}}}(p_{\bar{q}})
d_{\bar{q}}(p_{\bar{q}},\lambda_{\bar{q}})\nonumber\\
&&e^{-ip_{\bar{q}}\cdot z}~.\label{Ap4}
\end{eqnarray}

The EMT operator, in terms of the quark field operators, is given by,
\begin{eqnarray}
\hat{T}^{\mu\nu}&=&\frac{1}{4}\bar{\psi}_{q}(z)\left(i\gamma^{\mu} \overleftrightarrow{\partial}^{\nu}+i\gamma^{\nu}\overleftrightarrow{\partial}^{\mu}\right)\psi_{q}(z)\nonumber\\
\implies \hat{T}^{\mu\nu}&=&\frac{1}{4}\bar{\psi}_{q}(z)i\gamma^{\mu} \partial^{\nu}\psi_{q}(z)+\frac{1}{4}\bar{\psi}_{q}(z)i\gamma^{\nu} \partial^{\mu}\psi_{q}(z)\nonumber\\
&-&\frac{1}{4}\partial^{\nu}\bar{\psi}_{q}(z) ~i\gamma^{\mu}\psi_{q}(z)
-\frac{1}{4}\partial^{\mu}\bar{\psi}_{q}(z) ~i\gamma^{\nu}\psi_{q}(z),\nonumber
\end{eqnarray}
\begin{widetext}
Substituting the quark field operators and their derivative from Eqs.~\eqref{Ap1} to \eqref{Ap4} we have,
\begin{eqnarray}
\implies \hat{T}^{\mu\nu}&=&\frac{1}{4} \bigg[\int \frac{dp_{q}^{\prime+} d^{2}\vec{p}^{\prime}_{q}}{2p_{q}^{\prime+}(2\pi)^{3}}~\sum_{\lambda_{q}^{\prime}} \bar{u}_{\lambda_{q}^{\prime}}(p_{q}^{\prime})~ b^{\dag}_{q}(p_{q}^{\prime},\lambda_{q}^{\prime})~ e^{ip_{q}^{\prime}\cdot z} + \int \frac{dp_{\bar{q}}^{+} d^{2}\vec{p}_{\bar{q}}}{2p_{\bar{q}}^{+}(2\pi)^{3}}\sum_{\lambda_{\bar{q}}} \bar{v}_{\lambda_{\bar{q}}}(p_{\bar{q}}) ~d_{\bar{q}}(p_{\bar{q}},\lambda_{\bar{q}})~ e^{-ip_{\bar{q}}\cdot z}\bigg]\nonumber\\
&&\bigg[\int \frac{dp_{q}^{+} d^{2}\vec{p}_{q}}{2p_{q}^{+}(2\pi)^{3}}~\gamma^{\mu}p_{q}^{\nu}~\sum_{\lambda_{q}} u_{\lambda_{q}}(p_{q})~ b_{q}(p_{q},\lambda_{q})~ e^{-ip_{q}\cdot z} -  \int \frac{dp_{\bar{q}}^{\prime +} d^{2}\vec{p}_{\bar{q}}^{\prime}}{2p_{\bar{q}}^{\prime +}(2\pi)^{3}} \gamma^{\mu}p_{\bar{q}}^{\prime \nu} \sum_{\lambda_{\bar{q}}^{\prime}} v_{\lambda_{\bar{q}}^{\prime}}(p_{\bar{q}}^{\prime}) ~d^{\dag}_{\bar{q}}(p_{\bar{q}^{\prime}},\lambda_{\bar{q}}^{\prime})~ e^{ip_{\bar{q}}^{\prime}\cdot z}\bigg]\nonumber\\
&+& (\mu\leftrightarrow \nu)\nonumber\\
&+&\frac{1}{4} \bigg[\int \frac{dp_{q}^{\prime+} d^{2}\vec{p}^{\prime}_{q}}{2p_{q}^{\prime+}(2\pi)^{3}}~\sum_{\lambda_{q}^{\prime}} \bar{u}_{\lambda_{q}^{\prime}}(p_{q}^{\prime})~ b^{\dag}_{q}(p_{q}^{\prime},\lambda_{q}^{\prime})e^{ip_{q}^{\prime}\cdot z}~p^{\prime\nu}_{q}\gamma^{\mu}- \int \frac{dp_{\bar{q}}^{+} d^{2}\vec{p}_{\bar{q}}}{2p_{\bar{q}}^{+}(2\pi)^{3}}~\sum_{\lambda_{\bar{q}}} \bar{v}_{\lambda_{\bar{q}}}(p_{\bar{q}}) ~d_{\bar{q}}(p_{\bar{q}},\lambda_{\bar{q}})e^{-ip_{\bar{q}}\cdot z}~p^{\nu}_{\bar{q}}\gamma^{\mu}\bigg]\nonumber\\
&& \bigg[\int \frac{dp_{q}^{+} d^{2}\vec{p}_{q}}{2p_{q}^{+}(2\pi)^{3}}~\sum_{\lambda_{q}}u_{\lambda_{q}}(p_{q})~ b_{q}(p_{q},\lambda_{q})~ e^{-ip_{q}\cdot z} +\int \frac{dp_{\bar{q}}^{\prime +} d^{2}\vec{p}_{\bar{q}}^{\prime}}{2p_{\bar{q}}^{\prime +}(2\pi)^{3}}\sum_{\lambda_{\bar{q}}^{\prime}} v_{\lambda_{\bar{q}}^{\prime}}(p_{\bar{q}}^{\prime}) ~d^{\dag}_{\bar{q}}(p_{\bar{q}^{\prime}},\lambda_{\bar{q}}^{\prime})~ e^{ip_{\bar{q}}^{\prime}\cdot z}\bigg] \nonumber\\
&+&(\mu\leftrightarrow \nu)\nonumber\\
&=&\frac{1}{4}\sum_{\lambda_{q},\lambda_{q}^{\prime}}\int \frac{dp_{q}^{+} d^{2}\vec{p}_{q}}{2p_{q}^{+}(2\pi)^{3}} \frac{dp_{q}^{\prime+} d^{2}\vec{p}^{\prime}_{q}}{2p_{q}^{\prime+}(2\pi)^{3}} \bar{u}_{\lambda_{q}^{\prime}}(p_{q}^{\prime})[\gamma^{\mu}(p_{q}^{\nu}+p_{q}^{\prime \nu})+\gamma^{\nu}(p_{q}^{\mu}+p_{q}^{\prime \mu})]u_{\lambda_{q}}(p_{q})\nonumber\\
&&b^{\dag}_{q}(p_{q}^{\prime},\lambda_{q}^{\prime})b_{q}(p_{q},\lambda_{q})~e^{-i(p_{q}-p_{q}^{\prime})\cdot z}\nonumber\\
&-&\frac{1}{4}\sum_{\lambda_{\bar{q}},\lambda_{\bar{q}}^{\prime}}\int \frac{dp_{\bar{q}}^{+} d^{2}\vec{p}_{\bar{q}}}{2p_{\bar{q}}^{+}(2\pi)^{3}} \frac{dp_{\bar{q}}^{\prime+} d^{2}\vec{p}^{\prime}_{\bar{q}}}{2p_{\bar{q}}^{\prime+}(2\pi)^{3}}\bar{v}_{\lambda_{\bar{q}}}(p_{\bar{q}})[\gamma^{\mu}(p_{\bar{q}}^{\nu}+p_{\bar{q}}^{\prime \nu})+\gamma^{\nu}(p_{\bar{q}}^{\mu}+p_{\bar{q}}^{\prime \mu})]v_{\lambda_{\bar{q}}^{\prime}}(p_{\bar{q}}^{\prime})\nonumber\\
&&d_{\bar{q}}(p_{\bar{q}},\lambda_{\bar{q}}) d^{\dag}_{\bar{q}}(p_{\bar{q}}^{\prime},\lambda_{\bar{q}}^{\prime}) ~e^{-i(p_{\bar{q}}-p_{\bar{q}}^{\prime})\cdot z}~.\label{Ap5}
\end{eqnarray}
In the above expression, we ignore terms that do not contribute to the matrix elements. We re-express Eq.~\eqref{Ap5} by normal ordering the 2nd term,
\begin{eqnarray}
\implies \hat{T}^{\mu\nu}&=&\frac{1}{4}\sum_{\lambda_{q},\lambda_{q}^{\prime}}\int \frac{dp_{q}^{+} d^{2}\vec{p}_{q}}{2p_{q}^{+}(2\pi)^{3}} \frac{dp_{q}^{\prime+} d^{2}\vec{p}^{\prime}_{q}}{2p_{q}^{\prime+}(2\pi)^{3}}\bar{u}_{\lambda_{q}^{\prime}}(p_{q}^{\prime})[\gamma^{\mu}(p_{q}^{\nu}+p_{q}^{\prime \nu})+\gamma^{\nu}(p^{\mu}_{q}+p^{\prime \mu}_{q})]u_{\lambda_{q}}(p_{q})b^{\dag}_{q}(p_{q}^{\prime},\lambda_{q}^{\prime})b_{q}(p_{q},\lambda_{q})~ e^{-i(p_{q}-p_{q}^{\prime})\cdot z}\nonumber\\
&&+\frac{1}{4}\sum_{\lambda_{\bar{q}},\lambda_{\bar{q}}^{\prime}}\int \frac{dp_{\bar{q}}^{+} d^{2}\vec{p}_{\bar{q}}}{2p_{\bar{q}}^{+}(2\pi)^{3}} \frac{dp_{\bar{q}}^{\prime+} d^{2}\vec{p}^{\prime}_{\bar{q}}}{2p_{\bar{q}}^{\prime+}(2\pi)^{3}}\bar{v}_{\lambda_{\bar{q}}}(p_{\bar{q}})[\gamma^{\mu}(p_{\bar{q}}^{\nu}+p_{\bar{q}}^{\prime \nu})+\gamma^{\nu}(p_{\bar{q}}^{\mu}+p_{\bar{q}}^{\prime \mu})]v_{\lambda_{\bar{q}}^{\prime}}(p_{\bar{q}}^{\prime})d^{\dag}_{\bar{q}}(p_{\bar{q}}^{\prime},\lambda_{\bar{q}}^{\prime})d_{\bar{q}}(p_{\bar{q}},\lambda_{\bar{q}})~ e^{-i(p_{\bar{q}}-p_{\bar{q}}^{\prime})\cdot z}~.\nonumber\\
\label{Ap6}
\end{eqnarray}
From Eq.~\eqref{Ap6} it is clear that in the matrix element the first term would correspond to the quark contribution ($\hat{T}^{\mu\nu}_{q}$) and the second would correspond to the anti quark contribution ($\hat{T}^{\mu\nu}_{\bar{q}}$). The pseudoscalar quarkonia states in the LFQM are expressed as,
\begin{eqnarray}
    \langle \mathcal{M}(P^{\prime})\vert&=&\langle 0\vert\int \frac{dx^{\prime}d^{2}\vec{k}_{\perp}^{\prime}}{2(2\pi)^{3}\sqrt{x^{\prime}(1-x^{\prime})}}\sum_{\lambda_{1}^{\prime},\lambda_{2}^{\prime}}\Psi^{\ast}_{\lambda_{1}^{\prime}\lambda_{2}^{\prime}}d_{\bar{q}}((1-x^{\prime})P^{\prime +}, -\vec{k}_{\perp}^{\prime}+(1-x^{\prime})\vec{P}_{\perp}^{\prime}, \lambda_{2}^{\prime})b_{q}(x^{\prime}P^{\prime +}, \vec{k}_{\perp}^{\prime}+x^{\prime}\vec{P}_{\perp}^{\prime}, \lambda_{1}^{\prime})\nonumber\\
    \label{Ap7}\\
     \vert \mathcal{M}(P)\rangle&=&\int \frac{dxd^{2}\vec{k}_{\perp}}{2(2\pi)^{3}\sqrt{x(1-x)}}\sum_{\lambda_{1},\lambda_{2}}\Psi_{\lambda_{1}\lambda_{2}}b_{q}^{\dag}(xP^{+}, \vec{k}_{\perp}+x\vec{P}_{\perp}, \lambda_{1})d_{\bar{q}}^{\dag}((1-x)P^{+}, -\vec{k}_{\perp}+(1-x)\vec{P}_{\perp}, \lambda_{2})~\vert 0\rangle\label{Ap8}  
\end{eqnarray}
Using Eqs.~\eqref{Ap6}, \eqref{Ap7} and \eqref{Ap8} we can write down the matrix element as sum of quark and anti-quark contributions,
\begin{eqnarray}
    \langle \hat{T}^{\mu\nu}_{q}\rangle&=&\langle 0\vert\int \frac{dx^{\prime}d^{2}\vec{k}_{\perp}^{\prime}}{2(2\pi)^{3}\sqrt{x^{\prime}(1-x^{\prime})}}\sum_{\lambda_{1}^{\prime},\lambda_{2}^{\prime}}\Psi^{\ast}_{\lambda_{1}^{\prime}\lambda_{2}^{\prime}}d_{\bar{q}}((1-x^{\prime})P^{\prime +}, -\vec{k}_{\perp}^{\prime}+(1-x^{\prime})\vec{P}_{\perp}^{\prime}, \lambda_{2}^{\prime})~b_{q}(x^{\prime}P^{\prime +}, \vec{k}_{\perp}^{\prime}+x^{\prime}\vec{P}_{\perp}^{\prime}, \lambda_{1}^{\prime})\nonumber\\
    &&\frac{1}{4}\sum_{\lambda_{q},\lambda_{q}^{\prime}}\int \frac{dp_{q}^{+} d^{2}\vec{p}_{q}}{2p_{q}^{+}(2\pi)^{3}} \frac{dp_{q}^{\prime+} d^{2}\vec{p}^{\prime}_{q}}{2p_{q}^{\prime+}(2\pi)^{3}} ~\bar{u}_{\lambda_{q}^{\prime}}(p_{q}^{\prime})[\gamma^{\mu}(p_{q}^{\nu}+p_{q}^{\prime \nu})+\gamma^{\nu}(p^{\mu}_{q}+p^{\prime \mu}_{q})]u_{\lambda_{q}}(p_{q})~b^{\dag}_{q}(p_{q}^{\prime},\lambda_{q}^{\prime})b_{q}(p_{q},\lambda_{q})\nonumber \\
    &&\int \frac{dxd^{2}\vec{k}_{\perp}}{2(2\pi)^{3}\sqrt{x(1-x)}}\sum_{\lambda_{1},\lambda_{2}}\Psi_{\lambda_{1}\lambda_{2}}b_{q}^{\dag}(xP^{+}, \vec{k}_{\perp}+x\vec{P}_{\perp}, \lambda_{1})~d_{\bar{q}}^{\dag}((1-x)P^{+}, -\vec{k}_{\perp}+(1-x)\vec{P}_{\perp}, \lambda_{2})~\vert 0\rangle\nonumber \\
    &=&\frac{1}{4}\int \frac{dx^{\prime}d^{2}\vec{k}_{\perp}^{\prime}~dp_{q}^{+} d^{2}\vec{p}_{q}~dp_{q}^{\prime+} d^{2}\vec{p}^{\prime}_{q}dxd^{2}\vec{k}_{\perp}}{(2(2\pi)^{3})^{4}\sqrt{x(1-x)}\sqrt{x^{\prime}(1-x^{\prime})}p_{q}^{+}p_{q}^{\prime +}}\sum_{\{\lambda\}}\Psi^{\ast}_{\lambda_{1}^{\prime}\lambda_{2}^{\prime}}\Psi_{\lambda_{1}\lambda_{2}} \bar{u}_{\lambda_{q}^{\prime}}(p_{q}^{\prime})[\gamma^{\mu}(p_{q}^{\nu}+p_{q}^{\prime \nu})+\gamma^{\nu}(p^{\mu}_{q}+p^{\prime \mu}_{q})]u_{\lambda_{q}}(p_{q})\nonumber\\
    &&\langle 0\vert d_{\bar{q}}((1-x^{\prime})P^{\prime +}, -\vec{k}_{\perp}^{\prime}+(1-x^{\prime})\vec{P}_{\perp}^{\prime}, \lambda_{2}^{\prime})~b_{q}(x^{\prime}P^{\prime +}, \vec{k}_{\perp}^{\prime}+x^{\prime}\vec{P}_{\perp}^{\prime}, \lambda_{1}^{\prime})~b^{\dag}_{q}(p_{q}^{\prime},\lambda_{q}^{\prime})b_{q}(p_{q},\lambda_{q})\nonumber\\
    &&b_{q}^{\dag}(xP^{+}, \vec{k}_{\perp}+x\vec{P}_{\perp}, \lambda_{1})~d_{\bar{q}}^{\dag}((1-x)P^{+}, -\vec{k}_{\perp}+(1-x)\vec{P}_{\perp}, \lambda_{2})\vert 0\rangle~,\label{Ap9}\\
     \langle \hat{T}^{\mu\nu}_{\bar{q}}\rangle&=&\langle 0\vert\int \frac{dx^{\prime}d^{2}\vec{k}_{\perp}^{\prime}}{2(2\pi)^{3}\sqrt{x^{\prime}(1-x^{\prime})}}\sum_{\lambda_{1}^{\prime},\lambda_{2}^{\prime}}\Psi^{\ast}_{\lambda_{1}^{\prime}\lambda_{2}^{\prime}}d_{\bar{q}}((1-x^{\prime})P^{\prime +}, -\vec{k}_{\perp}^{\prime}+(1-x^{\prime})\vec{P}_{\perp}^{\prime}, \lambda_{2}^{\prime})~b_{q}(x^{\prime}P^{\prime +}, \vec{k}_{\perp}^{\prime}+x^{\prime}\vec{P}_{\perp}^{\prime}, \lambda_{1}^{\prime})\nonumber\\
    &&\frac{1}{4}\sum_{\lambda_{\bar{q}},\lambda_{\bar{q}}^{\prime}}\int \frac{dp_{\bar{q}}^{+} d^{2}\vec{p}_{\bar{q}}}{2p_{\bar{q}}^{+}(2\pi)^{3}} \frac{dp_{\bar{q}}^{\prime+} d^{2}\vec{p}^{\prime}_{\bar{q}}}{2p_{\bar{q}}^{\prime+}(2\pi)^{3}}~ \bar{v}_{\lambda_{\bar{q}}}(p_{\bar{q}})[\gamma^{\mu}(p_{\bar{q}}^{\nu}+p_{\bar{q}}^{\prime \nu})+\gamma^{\nu}(p_{\bar{q}}^{\mu}+p_{\bar{q}}^{\prime \mu})]v_{\lambda_{\bar{q}}^{\prime}}(p_{\bar{q}}^{\prime}) ~d^{\dag}_{\bar{q}}(p_{\bar{q}}^{\prime},\lambda_{\bar{q}}^{\prime})d_{\bar{q}}(p_{\bar{q}},\lambda_{\bar{q}})\nonumber \\
    &&\int \frac{dxd^{2}\vec{k}_{\perp}}{2(2\pi)^{3}\sqrt{x(1-x)}}\sum_{\lambda_{1},\lambda_{2}}\Psi_{\lambda_{1}\lambda_{2}}b_{q}^{\dag}(xP^{+}, \vec{k}_{\perp}+x\vec{P}_{\perp}, \lambda_{1})~d_{\bar{q}}^{\dag}((1-x)P^{+}, -\vec{k}_{\perp}+(1-x)\vec{P}_{\perp}, \lambda_{2})~\vert 0\rangle\nonumber \\
    &=&\frac{1}{4}\int \frac{dx^{\prime}d^{2}\vec{k}_{\perp}^{\prime}~dp_{q}^{+} d^{2}\vec{p}_{q}~dp_{q}^{\prime+} d^{2}\vec{p}^{\prime}_{q}dxd^{2}\vec{k}_{\perp}}{(2(2\pi)^{3})^{4}\sqrt{x(1-x)}\sqrt{x^{\prime}(1-x^{\prime})}p_{\bar{q}}^{+}p_{\bar{q}}^{\prime +}}\sum_{\{\bar{\lambda}\}}\Psi^{\ast}_{\lambda_{1}^{\prime}\lambda_{2}^{\prime}}\Psi_{\lambda_{1}\lambda_{2}} \bar{v}_{\lambda_{\bar{q}}}(p_{\bar{q}})[\gamma^{\mu}(p_{\bar{q}}^{\nu}+p_{\bar{q}}^{\prime \nu})+\gamma^{\nu}(p_{\bar{q}}^{\mu}+p_{\bar{q}}^{\prime \mu})]v_{\lambda_{\bar{q}}^{\prime}}(p_{\bar{q}}^{\prime})\nonumber\\
    &&\langle 0\vert d_{\bar{q}}((1-x^{\prime})P^{\prime +}, -\vec{k}_{\perp}^{\prime}+(1-x^{\prime})\vec{P}_{\perp}^{\prime}, \lambda_{2}^{\prime})~b_{q}(x^{\prime}P^{\prime +}, \vec{k}_{\perp}^{\prime}+x^{\prime}\vec{P}_{\perp}^{\prime}, \lambda_{1}^{\prime})~d^{\dag}_{\bar{q}}(p_{\bar{q}}^{\prime},\lambda_{\bar{q}}^{\prime})d_{\bar{q}}(p_{\bar{q}},\lambda_{\bar{q}})\nonumber\\
    &&b_{q}^{\dag}(xP^{+}, \vec{k}_{\perp}+x\vec{P}_{\perp}, \lambda_{1})~d_{\bar{q}}^{\dag}((1-x)P^{+}, -\vec{k}_{\perp}+(1-x)\vec{P}_{\perp}, \lambda_{2})\vert 0\rangle~.\label{Ap10}
\end{eqnarray}
For the sum over helicity indices we choose the short hand $\{\lambda\}\equiv(\lambda_{1},\lambda_{2},\lambda_{1}^{\prime},\lambda_{2}^{\prime}, \lambda_{q}, \lambda_{q}^{\prime})$  for quark contribution and $\{\bar{\lambda}\}=(\lambda_{1},\lambda_{2},\lambda_{1}^{\prime},\lambda_{2}^{\prime}, \lambda_{\bar{q}}$, $\lambda_{\bar{q}}^{\prime}$) for anti quark contribution.

To evaluate the quark contribution to the EMT matrix element, we have to simplify the following vacuum expectation value, 
\begin{eqnarray}
&&\langle 0\vert d_{\bar{q}}((1-x^{\prime})P^{\prime +}, -\vec{k}_{\perp}^{\prime}+(1-x^{\prime})\vec{P}_{\perp}^{\prime}, \lambda_{2}^{\prime})~b_{q}(x^{\prime}P^{\prime +}, \vec{k}_{\perp}^{\prime}+x^{\prime}\vec{P}_{\perp}^{\prime}, \lambda_{1}^{\prime})~b^{\dag}_{q}(p_{q}^{\prime},\lambda_{q}^{\prime})b_{q}(p_{q},\lambda_{q})\nonumber\\
    &&b_{q}^{\dag}(xP^{+}, \vec{k}_{\perp}+x\vec{P}_{\perp}, \lambda_{1})~d_{\bar{q}}^{\dag}((1-x)P^{+}, -\vec{k}_{\perp}+(1-x)\vec{P}_{\perp}, \lambda_{2})\vert 0\rangle~.\label{Ap11}
\end{eqnarray}
This expectation value can be simplified by adhering to the LF commutation relations involving the creation and annihilation operators: $\{b_{q}(p, \lambda),b_{q}^{\dag}(p^{\prime}, \lambda^{\prime})\}=2(2\pi)^{3}~p^{+}\delta(p^{+}-p^{\prime +})\delta^{2}(\vec{p}_{\perp}-\vec{p}^{\prime}_{\perp}) \delta_{\lambda \lambda^{\prime}}$ and $\{d_{\bar{q}}(p, \lambda),d_{\bar{q}}^{\dag}(p^{\prime}, \lambda^{\prime})\}=2(2\pi)^{3}~p^{+}\delta(p^{+}-p^{\prime +})\delta^{2}(\vec{p}_{\perp}-\vec{p}^{\prime}_{\perp}) \delta_{\lambda \lambda^{\prime}}$ with all other anticommutators being zero. One can now calculate the expectation value using Wick's theorem; nevertheless, we work it out explicitly.
Let us evaluate the vacuum expectation value by collectively defining the arguments in the operators as, $d_{\bar{q}}((1-x^{\prime})P^{\prime +}, -\vec{k}_{\perp}^{\prime}+(1-x^{\prime})\vec{P}_{\perp}^{\prime}, \lambda_{2}^{\prime})~b_{q}(x^{\prime}P^{\prime +}, \vec{k}_{\perp}^{\prime}+x^{\prime}\vec{P}_{\perp}^{\prime}, \lambda_{1}^{\prime})~b^{\dag}_{q}(p_{q}^{\prime},\lambda_{q}^{\prime})b_{q}(p_{q},\lambda_{q})
    b_{q}^{\dag}(xP^{+}, \vec{k}_{\perp}+x\vec{P}_{\perp}, \lambda_{1})~d_{\bar{q}}^{\dag}((1-x)P^{+}, -\vec{k}_{\perp}+(1-x)\vec{P}_{\perp}, \lambda_{2})\equiv d_{\bar{q}}(1)b_{q}(2)b_{q}^{\dag}(3)b_{q}(4)b_{q}^{\dag}(5)d_{\bar{q}}^{\dag}(6)$. The expectation value becomes
    \begin{eqnarray}
   \langle 0 \vert d_{\bar{q}}(1)b_{q}(2)b_{q}^{\dag}(3)b_{q}(4)b_{q}^{\dag}(5)d_{\bar{q}}^{\dag}(6)\vert 0 \rangle&=& \langle 0 \vert d_{\bar{q}}(1)b_{q}(2)b_{q}^{\dag}(3)b_{q}(4)b_{q}^{\dag}(5)d_{\bar{q}}^{\dag}(6)\vert 0 \rangle\nonumber\\
   &=& \langle 0 \vert d_{\bar{q}}(1)d_{\bar{q}}^{\dag}(6)b_{q}(2)b_{q}^{\dag}(3)b_{q}(4)b_{q}^{\dag}(5)\vert 0 \rangle\nonumber\\
   &=&  \langle 0 \vert (\delta(16)-d_{\bar{q}}^{\dag}(6)d_{\bar{q}}(1))(\delta(23)-b^{\dag}_{q}(3)b_{q}(2))(\delta(45)-b^{\dag}_{q}(5)b_{q}(4))\vert 0 \rangle\nonumber\\
   &=&  \langle 0 \vert (\delta(16)-d_{\bar{q}}^{\dag}(6)d_{\bar{q}}(1))\big(\delta(23)\delta(45)-\delta(23)b^{\dag}_{q}(5)b_{q}(4)-\delta(45)b^{\dag}_{q}(3)b_{q}(2)\nonumber\\
   &+&b^{\dag}_{q}(3)b_{q}(2)b^{\dag}_{q}(5)b_{q}(4)\big)\vert 0 \rangle\nonumber\\
    &=&  \langle 0 \vert (\delta(16)-d_{\bar{q}}^{\dag}(6)d_{\bar{q}}(1))\big(\delta(23)\delta(45)-\delta(23)b^{\dag}_{q}(5)b_{q}(4)-\delta(45)b^{\dag}_{q}(3)b_{q}(2)\nonumber\\
   &+&b^{\dag}_{q}(3)(\delta(25)-b^{\dag}_{q}(5)b_{q}(2))b_{q}(4)\big)\vert 0 \rangle\nonumber\\
   &=&  \langle 0 \vert (\delta(16)-d_{\bar{q}}^{\dag}(6)d_{\bar{q}}(1))\big(\delta(23)\delta(45)-\delta(23)b^{\dag}_{q}(5)b_{q}(4)-\delta(45)b^{\dag}_{q}(3)b_{q}(2)\nonumber\\
   &+&b^{\dag}_{q}(3)b_{q}(4)\delta(25)-b^{\dag}_{q}(3)b^{\dag}_{q}(5)b_{q}(2)b_{q}(4)\big)\vert 0 \rangle\nonumber\\
    &=&  \langle 0 \vert \big(\delta(16)\delta(23)\delta(45)-\delta(16)\delta(23)b^{\dag}_{q}(5)b_{q}(4)-\delta(16)\delta(45)b^{\dag}_{q}(3)b_{q}(2)\nonumber\\
   &+&\delta(16)\delta(25)b^{\dag}_{q}(3)b_{q}(4)-\delta(16)b^{\dag}_{q}(3)b^{\dag}_{q}(5)b_{q}(2)b_{q}(4)\big)\nonumber\\
   &+&\big(-\delta(23)\delta(45)d_{\bar{q}}^{\dag}(6)d_{\bar{q}}(1)
   -\delta(23)d_{\bar{q}}^{\dag}(6)b^{\dag}_{q}(5)d_{\bar{q}}(1)b_{q}(4)\nonumber\\
   &-&\delta(45)d_{\bar{q}}^{\dag}(6)b^{\dag}_{q}(3)d_{\bar{q}}(1)b_{q}(2)+\delta(25)d_{\bar{q}}^{\dag}(6)b^{\dag}_{q}(3)d_{\bar{q}}(1)b_{q}(4)\nonumber\\
   &+&d_{\bar{q}}^{\dag}(6)b^{\dag}_{q}(3)b^{\dag}_{q}(5)d_{\bar{q}}(1)b_{q}(2)b_{q}(4)\big)\vert 0 \rangle\nonumber\\
   &=&\langle 0 \vert \delta(16)\delta(23)\delta(45) \vert 0 \rangle\nonumber\\
   &=& \delta(16)\delta(23)\delta(45)\label{Ap12}
\end{eqnarray}
Substituting it in the Eq.~\eqref{Ap11} we have,
\begin{eqnarray}
&&\langle 0\vert d_{\bar{q}}((1-x^{\prime})P^{\prime +}, -\vec{k}_{\perp}^{\prime}+(1-x^{\prime})\vec{P}_{\perp}^{\prime}, \lambda_{2}^{\prime})~b_{q}(x^{\prime}P^{\prime +}, \vec{k}_{\perp}^{\prime}+x^{\prime}\vec{P}_{\perp}^{\prime}, \lambda_{1}^{\prime})~b^{\dag}_{q}(p_{q}^{\prime},\lambda_{q}^{\prime})b_{q}(p_{q},\lambda_{q})\nonumber\\
    &&b_{q}^{\dag}(xP^{+}, \vec{k}_{\perp}+x\vec{P}_{\perp}, \lambda_{1})~d_{\bar{q}}^{\dag}((1-x)P^{+}, -\vec{k}_{\perp}+(1-x)\vec{P}_{\perp}, \lambda_{2})\vert 0\rangle\nonumber\\
    &&=\big(2(2\pi)^{3}\big)^{3} (1-x)P^{+}\delta((1-x^{\prime})P^{\prime +}-(1-x)P^{+})~ \delta^{2}(-\vec{k}^{\prime}_{\perp}+(1-x^{\prime})\vec{P}_{\perp}^{\prime} +\vec{k}_{\perp}-(1-x)\vec{P}_{\perp}) ~\delta_{\lambda_{2}\lambda_{2}^{\prime}}\nonumber\\
    && p_{q}^{\prime +} \delta(x^{\prime}P^{\prime +}-p_{q}^{\prime +})~\delta^{2}(\vec{k}^{\prime}_{\perp}+x^{\prime}\vec{P}_{\perp}^{\prime}-\vec{p}_{q\perp}^{\prime}) \delta_{\lambda^{\prime}_{1}\lambda^{\prime}_{q}}~p_{q}^{ +} \delta(xP^{+}-p_{q}^{ +})~\delta^{2}(\vec{k}_{\perp}+x \vec{P}_{\perp}-\vec{p}_{q\perp}) \delta_{\lambda_{q}\lambda_{1}}\label{Ap13}~.
\end{eqnarray}
Integrating over the variables $x^{\prime}$, $\vec{k}_{\perp}^{\prime}$, $p_{q}^{\prime +}$, $\vec{p}_{q \perp }^{\prime}$, $p_{q}^{+}$, and $\vec{p}_{q \perp}$ and summing over $\lambda_{1}$, $\lambda_{1}^{\prime}$, and $\lambda_{2}^{\prime}$ in Eq.~\eqref{Ap9}  with the use of Eq.~\eqref{Ap13} we have,
\begin{eqnarray}
 \langle \hat{T}^{\mu\nu}_{q}\rangle &=&  \int \frac{dx~ d^{2}\vec{k}_{\perp}}{2(2\pi)^{3}} ~\varphi^{\ast}(x,\vec{k}_{\perp}^{\prime})\varphi(x,\vec{k}_{\perp})\sum_{\lambda_{q},\lambda_{q}^{\prime},\lambda_{2}}\mathcal{S}^{\ast}_{\lambda_{q}^{\prime}\lambda_{2}}(x,\vec{k}_{\perp}^{\prime}) U^{\mu\nu}_{\lambda_{q}^{\prime}\lambda_{q}} \mathcal{S}_{\lambda_{q}\lambda_{2}}(x,\vec{k}_{\perp})\nonumber\\
 &=&  \int \frac{dx~ d^{2}\vec{k}_{\perp}}{2(2\pi)^{3}} ~\varphi^{\ast}(x,\vec{k}_{\perp}^{\prime})\varphi(x,\vec{k}_{\perp}) ~S^{\mu\nu}_{q}\label{Ap14}~,
\end{eqnarray}
where the following definitions are used
\begin{eqnarray}
&&U^{\mu\nu}_{\lambda_{q}^{\prime}\lambda_{q}}\equiv \frac{1}{4x}\bar{u}_{\lambda_{q}^{\prime}}(p_{q}^{\prime})[\gamma^{\mu}p_{sq}^{\nu}+\gamma^{\nu}p_{sq}^{\mu}]u_{\lambda_{q}}(p_{q})~,\nonumber\\
&&S^{\mu\nu}_{q}\equiv\sum_{\lambda_{q},\lambda_{q}^{\prime},\lambda_{2}}\mathcal{S}^{\ast}_{\lambda_{q}^{\prime}\lambda_{2}}(x,\vec{k}_{\perp}^{\prime}) U^{\mu\nu}_{\lambda_{q}^{\prime}\lambda_{q}} \mathcal{S}_{\lambda_{q}\lambda_{2}}(x,\vec{k}_{\perp})~,\nonumber\\
&& p_{q}=(p_{q}^{-},p_{q}^{+},\vec{p}_{q\perp})=(\frac{\vec{p}_{q\perp}^{2}+m_{q}^{2}}{p_{q}^{+}},xP^{+},\vec{k}_{\perp})~,\nonumber\\
&& p_{q}^{\prime}=(p_{q}^{\prime-},p_{q}^{\prime+},\vec{p}_{q\perp}^{\prime})=(\frac{\vec{p}_{q\perp}^{\prime 2}+m_{q}^{2}}{p_{q}^{\prime +}},xP^{+},\vec{k}_{\perp}+\vec{\Delta}_{\perp})~,\nonumber\\
&& p_{sq}= p_{q}+ p_{q}^{\prime}\nonumber\\
&&\vec{k}_{\perp}^{\prime}=\vec{k}_{\perp}+(1-x)\vec{\Delta}_{\perp}~.\label{Ap15}
\end{eqnarray}

For the evaluation of the antiparticle contribution to the matrix element provided in Eq.~\eqref{Ap10}, we need to evaluate the following vacuum expectation value,
\begin{eqnarray}
   &&\langle 0\vert d_{\bar{q}}((1-x^{\prime})P^{\prime +}, -\vec{k}_{\perp}^{\prime}+(1-x^{\prime})\vec{P}_{\perp}^{\prime}, \lambda_{2}^{\prime})~b_{q}(x^{\prime}P^{\prime +}, \vec{k}_{\perp}^{\prime}+x^{\prime}\vec{P}_{\perp}^{\prime}, \lambda_{1}^{\prime})~d^{\dag}_{\bar{q}}(p_{\bar{q}}^{\prime},\lambda_{\bar{q}}^{\prime})d_{\bar{q}}(p_{\bar{q}},\lambda_{\bar{q}})\nonumber\\
    &&b_{q}^{\dag}(xP^{+}, \vec{k}_{\perp}+x\vec{P}_{\perp}, \lambda_{1})~d_{\bar{q}}^{\dag}((1-x)P^{+}, -\vec{k}_{\perp}+(1-x)\vec{P}_{\perp}, \lambda_{2})\vert 0\rangle \nonumber\\
  \equiv && \langle 0 \vert  d_{\bar{q}}(1)b_{q}(2)d^{\dag}_{\bar{q}}(3)d_{\bar{q}}(4)b_{q}^{\dag}(5)d_{\bar{q}}^{\dag}(6) \vert 0\rangle \nonumber\\
  =&& \langle 0\vert b_{q}(2) b_{q}^{\dag}(5)d_{\bar{q}}(1) d^{\dag}_{\bar{q}}(3) d_{\bar{q}}(4) d_{\bar{q}}^{\dag}(6)\vert 0\rangle \nonumber\\
   =&& \langle 0\vert :b_{q}(2) b_{q}^{\dag}(5)d_{\bar{q}}(1) d^{\dag}_{\bar{q}}(3) d_{\bar{q}}(4) d_{\bar{q}}^{\dag}(6): + ~\delta(25) \delta(13)\delta(46)\vert 0\rangle\nonumber\\
   =&& \delta(25) \delta(13)\delta(46)\nonumber\\
   =&& (2(2\pi)^{3})^{3}~xP^{+} \delta(x^{\prime}P^{\prime +}-xP^{+})~\delta^{2}(\vec{k}_{\perp}^{\prime}+x^{\prime}\vec{P}_{\perp}^{\prime}-\vec{k}_{\perp}-x\vec{P}_{\perp})~\delta_{\lambda_{1}^{\prime}\lambda_{1}}\nonumber\\
   &&2p_{\bar{q}}^{\prime +} \delta(p_{\bar{q}}^{\prime +}-(1-x^{\prime})P^{\prime +}) ~\delta^{2}(-\vec{k}_{\perp}^{\prime}+(1-x^{\prime})\vec{P}_{\perp}^{\prime}-\vec{p}_{\bar{q}\perp}^{\prime})~ \delta_{\lambda_{2}^{\prime}\lambda_{\bar{q}}^{\prime}}\nonumber\\
   && 2p_{\bar{q}}^{+} \delta(p_{\bar{q}}^{+}-(1-x)P^{ +})~ \delta^{2}(-\vec{k}_{\perp}+(1-x)\vec{P}_{\perp}-\vec{p}_{\bar{q}\perp})~ \delta_{\lambda_{\bar{q}}\lambda_{2}}~.\label{Ap16}
\end{eqnarray}
Integrating over the variables $x^{\prime}$, $\vec{k}_{\perp}^{\prime}$, $p_{\bar{q}}^{\prime +}$, $\vec{p}_{\bar{q} \perp }^{\prime}$, $p_{\bar{q}}^{+}$, and $\vec{p}_{\bar{q} \perp}$ and summing over $\lambda_{2}$, $\lambda_{1}^{\prime}$, and $\lambda_{2}^{\prime}$ in Eq.~\eqref{Ap10}  with the use of Eq.~\eqref{Ap16} we have,
\begin{eqnarray}
 \langle \hat{T}^{\mu\nu}_{\bar{q}}\rangle &=& \int \frac{dx~ d^{2}\vec{k}_{\perp}}{2(2\pi)^{3}} ~\varphi^{\ast}(x,\vec{\Tilde{k}}_{\perp})\varphi(x,\vec{k}_{\perp})\sum_{\lambda_{\bar{q}},\lambda_{\bar{q}}^{\prime},\lambda_{1}}\mathcal{S}_{\lambda_{1}\lambda_{\bar{q}}}(x,\vec{k}_{\perp})V^{\mu\nu}_{\lambda_{\bar{q}}\lambda_{\bar{q}}^{\prime}} \mathcal{S}^{\ast}_{\lambda_{1}\lambda_{\bar{q}}^{\prime}}(x,\vec{\Tilde{k}}_{\perp})  \nonumber\\
 &=& \int \frac{dx~ d^{2}\vec{k}_{\perp}}{2(2\pi)^{3}} ~\varphi^{\ast}(x,\vec{\Tilde{k}}_{\perp})\varphi(x,\vec{k}_{\perp}) ~S^{\mu\nu}_{\bar{q}}\label{Ap17}~,
\end{eqnarray}
where we use the following definitions
\begin{eqnarray}
&&V^{\mu\nu}_{\lambda_{\bar{q}}\lambda_{\bar{q}}^{\prime}}\equiv \frac{1}{4(1-x)}\bar{v}_{\lambda_{\bar{q}}}(p_{\bar{q}})[\gamma^{\mu}p_{s\bar{q}}^{\nu}+\gamma^{\nu}p_{s\bar{q}}^{\mu}]v_{\lambda_{\bar{q}}^{\prime}}(p_{\bar{q}}^{\prime})~,\nonumber\\
&&S^{\mu\nu}_{\bar{q}}\equiv\sum_{\lambda_{\bar{q}},\lambda_{\bar{q}}^{\prime},\lambda_{1}}\mathcal{S}_{\lambda_{1}\lambda_{\bar{q}}}(x,\vec{k}_{\perp})V^{\mu\nu}_{\lambda_{\bar{q}}\lambda_{\bar{q}}^{\prime}} \mathcal{S}^{\ast}_{\lambda_{1}\lambda_{\bar{q}}^{\prime}}(x,\vec{\Tilde{k}}_{\perp}),\nonumber\\
&& p_{\bar{q}}=(p_{\bar{q}}^{-},p_{\bar{q}}^{+},\vec{p}_{\bar{q}\perp})=(\frac{\vec{p}_{\bar{q}\perp}^{2}+m_{\bar{q}}^{2}}{p_{\bar{q}}^{+}},(1-x)P^{+},-\vec{k}_{\perp})~,\nonumber\\
&& p_{q}^{\prime}=(p_{\bar{q}}^{\prime-},p_{\bar{q}}^{\prime+},\vec{p}_{\bar{q}\perp}^{\prime})=(\frac{\vec{p}_{\bar{q}\perp}^{\prime 2}+m_{\bar{q}}^{2}}{p_{\bar{q}}^{\prime +}},(1-x)P^{+},-\vec{k}_{\perp}+\vec{\Delta}_{\perp})~,\nonumber\\
&& p_{s\bar{q}}= p_{\bar{q}}+ p_{\bar{q}}^{\prime}\nonumber\\
&&\vec{\Tilde{k}}_{\perp}=\vec{k}_{\perp}-x\vec{\Delta}_{\perp}~.\label{Ap18}
\end{eqnarray}
\subsection{Evaluation of the quark contribution}\label{quarkcontex} 
Using Eqs.~\eqref{Ap14} and \eqref{Ap15} we write the quark contribution as 
\begin{eqnarray}
 \langle \hat{T}^{\mu\nu}_{q}\rangle= \int \frac{dx~ d^{2}\vec{k}_{\perp}}{2(2\pi)^{3}} ~\varphi^{\ast}(x,\vec{k}_{\perp}^{\prime})\varphi(x,\vec{k}_{\perp}) ~S^{\mu\nu}_{q},~S^{\mu\nu}_{q}=\sum_{\lambda_{q},\lambda_{q}^{\prime},\lambda_{2}}\mathcal{S}^{\ast}_{\lambda_{q}^{\prime}\lambda_{2}}(x,\vec{k}_{\perp}^{\prime}) U^{\mu\nu}_{\lambda_{q}^{\prime}\lambda_{q}} \mathcal{S}_{\lambda_{q}\lambda_{2}}(x,\vec{k}_{\perp})~.\label{Ap19}   
\end{eqnarray}
Apart from space wave functions (which are model-dependent), this contribution is directly proportional to the spin trace for the tensor current $S_{q}^{\mu\nu}$.  The spin trace of the tensor current can be performed as follows:
\begin{eqnarray}
S^{\mu\nu}_{q}= \Tr [\mathcal{S} \mathcal{S}^{\prime \dag} U^{\mu\nu}], \text{ where } \mathcal{S} \text{ are matrices corresponding to spin wavefunctions.}\label{Ap20}
\end{eqnarray}
The spin wavefunction $\mathcal{S}$ is the same as Eq.~\eqref{Ape8} whereas the spin wavefunction $\mathcal{S}^{\prime}=\mathcal{S}(x, \vec{k}^{\prime}_{\perp}=\vec{k}_{\perp} + (1-x)\vec{\Delta}_{\perp})$. The multiplication of the spin wavefunctions can be obtained as,
\begin{eqnarray}
    \mathcal{S}\mathcal{S}^{\prime\dag}&=&\mathcal{N}
    \begin{pmatrix}
-k^{R} & - ((1-x)m_{q}+x m_{\bar{q}})  \\
((1-x)m_{q}+x m_{\bar{q}}) & -k^{L}\\
\end{pmatrix}
\begin{pmatrix}
-k^{\prime L} &  ((1-x)m_{q}+x m_{\bar{q}}) \\
- ((1-x)m_{q}+x m_{\bar{q}}) & -k^{\prime R}\\
\end{pmatrix}\nonumber\\
&=& \mathcal{N}
\begin{pmatrix}
k^{R}k^{\prime L} + ((1-x)m_{q}+x m_{\bar{q}})^{2} &  ((1-x)m_{q}+x m_{\bar{q}})(k^{\prime R}-k^{R}) \\
((1-x)m_{q}+x m_{\bar{q}})(k^{ L}-k^{\prime L}) & k^{L}k^{\prime R}+((1-x)m_{q}+x m_{\bar{q}})^{2}\\
\end{pmatrix}\label{Ap21}~,
\end{eqnarray}
where the the shorthand $\mathcal{N}\equiv 2\sqrt{(\vec{k}_{\perp}^{2}+((1-x)m_{q}+x m_{\bar{q}})^{2})(\vec{k}^{\prime 2}_{\perp}+((1-x)m_{q}+x m_{\bar{q}})^{2})}$. Substituting the results $k^{R}k^{\prime L}=\vec{k}_{\perp}\cdot\vec{k}_{\perp}^{\prime}+i (\vec{k}_{\perp}^{\prime}\times\vec{k}_{\perp})_{z}=k_{\perp}^{2}+(1-x)\vec{\Delta}\cdot \vec{k}_{\perp}+i(1-x)(\vec{\Delta}_{\perp}\times \vec{k}_{\perp})_{z}$, $k^{\prime R}-k^{R}=(1-x)\Delta_{\perp}^{R}$, and $k^{L}-k^{\prime L}=-(1-x)\Delta_{\perp}^{L}$, the product of spin matrices in Eq.~\eqref{Ap21} becomes,
\begin{eqnarray}
  &&  \mathcal{S}\mathcal{S}^{\prime\dag}=\mathcal{N}
  \begin{pmatrix}
 \vec{k}_{\perp}^{2}+(1-x)\vec{\Delta}\cdot \vec{k}_{\perp}+i(1-x)(\vec{\Delta}_{\perp}\times \vec{k}_{\perp})_{z}+\tilde{m}^{2} & \tilde{m}(1-x)\Delta_{\perp}^{R} \\
-\tilde{m}(1-x)\Delta_{\perp}^{L} &  \vec{k}_{\perp}^{2}+(1-x)\vec{\Delta}\cdot \vec{k}_{\perp}-i(1-x)(\vec{\Delta}_{\perp}\times \vec{k}_{\perp})_{z}+\tilde{m}^{2}\\
\end{pmatrix}~,\nonumber\\
\label{Ap22} 
\end{eqnarray}
where $\tilde{m}=(1-x)m_{q}+x m_{\bar{q}}$ which becomes equal to the quark masses in a symmetric mass system such as quarkonia, i.e., $\tilde{m}=m_{\bar{q}}=m_{q}$. For our purpose, we need the explicit evaluation of the components of $U^{\mu\nu}$. Multiplying the appropriate spinors (with the help of the spinor multiplication table provided in Ref.~\cite{Lepage:1980fj}), these components are given by
\begin{eqnarray}
   U^{++}_{\lambda_{q}^{\prime}\lambda_{q}}&=& \frac{1}{4x}\bar{u}_{\lambda_{q}^{\prime}}(p_{q}^{\prime})[\gamma^{+}p_{sq}^{+}+\gamma^{+}p_{sq}^{+}]u_{\lambda_{q}}(p_{q}) \nonumber\\
   &=& P^{+} \bar{u}_{\lambda_{q}^{\prime}}(p_{q}^{\prime})\gamma^{+}u_{\lambda_{q}}(p_{q})\nonumber\\
   &=& 2xP^{+2} I~\label{Ap23},\\
   U^{+-}_{\lambda_{q}^{\prime}\lambda_{q}}&=& \frac{1}{4x}\bar{u}_{\lambda_{q}^{\prime}}(p_{q}^{\prime})[\gamma^{+}p_{sq}^{-}+\gamma^{-}p_{sq}^{+}]u_{\lambda_{q}}(p_{q}) \nonumber\\
   &=& \frac{2\vec{k}_{\perp}^{2}+\vec{\Delta}_{\perp}^{2}+2\vec{k}_{\perp}\cdot \vec{\Delta}_{\perp}+2m_{q}^{2}}{2x} I+ \frac{1}{x}
   \begin{pmatrix}
     \vec{k}_{\perp}^{2}+\vec{k}_{\perp}\cdot \vec{\Delta}_{\perp}+m_{q}^{2} -i (\vec{\Delta}_{\perp}\times \vec{k}_{\perp})_{z} & -m_{q}\Delta_{\perp}^{R} \\
     m_{q}\Delta_{\perp}^{L} & \vec{k}_{\perp}^{2}+\vec{k}_{\perp}\cdot \vec{\Delta}_{\perp}+m_{q}^{2} +i (\vec{\Delta}_{\perp}\times \vec{k}_{\perp})_{z}\\    
   \end{pmatrix}\nonumber\\
   &=& \frac{4\vec{k}_{\perp}^{2}+\vec{\Delta}_{\perp}^{2}+4\vec{k}_{\perp}\cdot \vec{\Delta}_{\perp}+4m_{q}^{2}}{2x} I+ \frac{1}{x}
   \begin{pmatrix}
      -i (\vec{\Delta}_{\perp}\times \vec{k}_{\perp})_{z} & -m_{q}\Delta_{\perp}^{R} \\
     m_{q}\Delta_{\perp}^{L} & +i (\vec{\Delta}_{\perp}\times \vec{k}_{\perp})_{z}\\  \end{pmatrix}\label{Ap24}
\end{eqnarray}
\begin{eqnarray}
   U^{+i}_{\lambda_{q}^{\prime}\lambda_{q}}&=& \frac{1}{4x}\bar{u}_{\lambda_{q}^{\prime}}(p_{q}^{\prime})[\gamma^{+}p_{sq}^{i}+\gamma^{i}p_{sq}^{+}]u_{\lambda_{q}}(p_{q}) \nonumber\\
    &=&\frac{P^{+}}{2}(2k_{\perp}^{i}+\Delta_{\perp}^{i})I+\frac{P^{+}}{2}
    \begin{pmatrix}
     2k_{\perp}^{i}+\Delta_{\perp}^{i}+ i \epsilon^{ij} \Delta_{\perp}^{j} & 0 \\
     0 & 2k_{\perp}^{i}+\Delta_{\perp}^{i}- i \epsilon^{ij} \Delta_{\perp}^{j}\\    
   \end{pmatrix} \nonumber\\
   &=& P^{+}(2k_{\perp}^{i}+\Delta_{\perp}^{i}) I+\frac{P^{+}}{2}
    \begin{pmatrix}
     + i \epsilon^{ij} \Delta_{\perp}^{j} & 0 \\
     0 & - i \epsilon^{ij} \Delta_{\perp}^{j}\\    
   \end{pmatrix}, \label{ApR25}\\
    U^{ij}_{\lambda_{q}^{\prime}\lambda_{q}}&=& \frac{1}{4x}\bar{u}_{\lambda_{q}^{\prime}}(p_{q}^{\prime})[\gamma^{i}p_{sq}^{j}+\gamma^{j}p_{sq}^{i}]u_{\lambda_{q}}(p_{q}) \nonumber\\
    &=&\frac{1}{4x}\left[p_{sq}^{j}\begin{pmatrix}
     2k_{\perp}^{i}+\Delta_{\perp}^{i}+ i \epsilon^{ik} \Delta_{\perp}^{k} & 0 \\
     0 & 2k_{\perp}^{i}+\Delta_{\perp}^{i}- i \epsilon^{ik} \Delta_{\perp}^{k}\\    
   \end{pmatrix}+ p_{sq}^{i}
    \begin{pmatrix}
     2k_{\perp}^{j}+\Delta_{\perp}^{j}+ i \epsilon^{jk} \Delta_{\perp}^{k} & 0 \\
     0 & 2k_{\perp}^{j}+\Delta_{\perp}^{j}- i \epsilon^{jk} \Delta_{\perp}^{k}\\    
   \end{pmatrix}\right] \nonumber\\ 
  &=& \frac{1}{4x}\left[2(2k_{\perp}^{i}+\Delta_{\perp}^{i})(2k_{\perp}^{j}+\Delta_{\perp}^{j})I+p_{sq}^{j}\begin{pmatrix}
      i \epsilon^{ik} \Delta_{\perp}^{k} & 0 \\
     0 & - i \epsilon^{ik} \Delta_{\perp}^{k}\\    
   \end{pmatrix}+ p_{sq}^{i}
    \begin{pmatrix}
      i \epsilon^{jk} \Delta_{\perp}^{k} & 0 \\
     0 & - i \epsilon^{jk} \Delta_{\perp}^{k}\\    
   \end{pmatrix}\right],\label{ApR26}\\
   U^{-i}_{\lambda_{q}^{\prime}\lambda_{q}}&=& \frac{1}{4x}\bar{u}_{\lambda_{q}^{\prime}}(p_{q}^{\prime})[\gamma^{-}p_{sq}^{i}+\gamma^{i}p_{sq}^{-}]u_{\lambda_{q}}(p_{q}) \nonumber\\
    &=&\frac{1}{4x}\bigg[\frac{2p_{sq}^{i}}{xP^{+}}\begin{pmatrix}
     \vec{k}_{\perp}^{2}+\vec{k}_{\perp}\cdot \vec{\Delta}_{\perp}+m_{q}^{2} -i (\vec{\Delta}_{\perp}\times \vec{k}_{\perp})_{z} & -m_{q}\Delta_{\perp}^{R} \\
     m_{q}\Delta_{\perp}^{L} & \vec{k}_{\perp}^{2}+\vec{k}_{\perp}\cdot \vec{\Delta}_{\perp}+m_{q}^{2} +i (\vec{\Delta}_{\perp}\times \vec{k}_{\perp})_{z}\\    
   \end{pmatrix}\nonumber\\
   &+&p_{sq}^{-} 
    \begin{pmatrix}
     2k_{\perp}^{i}+\Delta_{\perp}^{i}+ i \epsilon^{ik} \Delta_{\perp}^{k} & 0 \\
     0 & 2k_{\perp}^{i}+\Delta_{\perp}^{i}- i \epsilon^{ik} \Delta_{\perp}^{k}\\    
   \end{pmatrix}\bigg]\nonumber\\
   &=&\frac{1}{4x}\bigg[\frac{2k_{\perp}^{i}+\Delta_{\perp}^{i}}{xP^{+}}(4\vec{k}_{\perp}^{2}+\vec{\Delta}_{\perp}^{2}+4\vec{k}_{\perp}\cdot \vec{\Delta}_{\perp}+4m_{q}^{2})I+\frac{2(2k_{\perp}^{i}+\Delta_{\perp}^{i})}{xP^{+}}\begin{pmatrix}
      -i (\vec{\Delta}_{\perp}\times \vec{k}_{\perp})_{z} & -m_{q}\Delta_{\perp}^{R} \\
     m_{q}\Delta_{\perp}^{L} &  +i (\vec{\Delta}_{\perp}\times \vec{k}_{\perp})_{z}\\    
   \end{pmatrix}\nonumber\\
   &+&p_{sq}^{-} 
    \begin{pmatrix}
     + i \epsilon^{ik} \Delta_{\perp}^{k} & 0 \\
     0 & - i \epsilon^{ik} \Delta_{\perp}^{k}\\    
   \end{pmatrix}\bigg],\label{ApR27}\\
    U^{--}_{\lambda_{q}^{\prime}\lambda_{q}}&=& \frac{1}{4x}\bar{u}_{\lambda_{q}^{\prime}}(p_{q}^{\prime})[\gamma^{-}p_{sq}^{-}+\gamma^{-}p_{sq}^{-}]u_{\lambda_{q}}(p_{q}) \nonumber\\
   &=&  \frac{p^{-}_{sq}}{x^{2}P^{+}}
   \begin{pmatrix}
     \vec{k}_{\perp}^{2}+\vec{k}_{\perp}\cdot \vec{\Delta}_{\perp}+m_{q}^{2} -i (\vec{\Delta}_{\perp}\times \vec{k}_{\perp})_{z} & -m_{q}\Delta_{\perp}^{R} \\
     m_{q}\Delta_{\perp}^{L} & \vec{k}_{\perp}^{2}+\vec{k}_{\perp}\cdot \vec{\Delta}_{\perp}+m_{q}^{2} +i (\vec{\Delta}_{\perp}\times \vec{k}_{\perp})_{z}\\    
   \end{pmatrix}\nonumber\\
   &=& \frac{(2\vec{k}_{\perp}^{2}+\vec{\Delta}_{\perp}^{2}+2\vec{k}_{\perp}\cdot \vec{\Delta}_{\perp}+2m_{q}^{2})(\vec{k}_{\perp}^{2}+\vec{k}_{\perp}\cdot \vec{\Delta}_{\perp}+m_{q}^{2})}{x^{3}(P^{+})^{2}} I+ \frac{2\vec{k}_{\perp}^{2}+\vec{\Delta}_{\perp}^{2}+2\vec{k}_{\perp}\cdot \vec{\Delta}_{\perp}+2m_{q}^{2}}{x^{3}(P^{+})^{2}}\nonumber\\
   &\times&\begin{pmatrix}
      -i (\vec{\Delta}_{\perp}\times \vec{k}_{\perp})_{z} & -m_{q}\Delta_{\perp}^{R} \\
     m_{q}\Delta_{\perp}^{L} & +i (\vec{\Delta}_{\perp}\times \vec{k}_{\perp})_{z}\\  \end{pmatrix}~.\label{ApR28}
\end{eqnarray}

The $``I"$ in the above result is the identity matrix and $\epsilon^{ij}$ is the antisymmetric tensor. These forms of $U^{\mu\nu}$ are particularly useful in calculating the spin trace of the tensor current $S^{\mu\nu}_{q}$.
Substituting the results of \eqref{Ap23}--\eqref{ApR28} in Eq.~\eqref{Ap20}, different components of $S_{q}^{\mu\nu}$ are obtained and provided in the Table~\ref{tab:spintraces}.

\begin{table*}[t]
\caption{The spin traces for the tensor current $S^{\mu\nu}_{q}= \Tr [\mathcal{S} \mathcal{S}^{\prime \dag} U^{\mu\nu}]$ is given below for easy reference. We use the following definition $\underline{k}_{\perp}^{2}\equiv \vec{k}_{\perp}^{2}+ \vec{k}_{\perp}\cdot \vec{\Delta}_{\perp}+m_{q}^{2}$~.}
\label{tab:spintraces}
\begin{ruledtabular}
\begin{tabular}{c c }
$(\mu\nu)$
&
$S_{q}^{\mu\nu}$
\\
\hline
$(++)$
&
$\displaystyle \frac{2x(P^{+})^{2}(\vec{k}_{\perp}\cdot \vec{k}_{\perp}^{\prime} + m_{q}^{2})}{\sqrt{\vec{k}_{\perp}^{2}+m_{q}^{2}}\sqrt{\vec{k}_{\perp}^{\prime 2}+m_{q}^{2}}}$
\\[7ex]
$(+-)$
&
$\displaystyle
\frac{2(\vec{k}_{\perp}\cdot \vec{k}_{\perp}^{\prime} + m_{q}^{2})\left(\underline{k}_{\perp}^{2}+\frac{\vec{\Delta}_{\perp}^{2}}{4}\right)+m_{q}^{2}(1-x)\vec{\Delta}_{\perp}^{2}+(1-x)(\vec{\Delta}_{\perp}\times \vec{k}_{\perp})^{2}_{z}}{x  \sqrt{\vec{k}_{\perp}^{2}+m_{q}^{2}}\sqrt{\vec{k}_{\perp}^{\prime 2}+m_{q}^{2}}}
$
\\[7ex]
$(+i)$
&
$\displaystyle
\frac{P^{+}(2k_{\perp}^{i}+\Delta_{\perp}^{i})(\vec{k}_{\perp}\cdot \vec{k}_{\perp}^{\prime} + m_{q}^{2})}{\sqrt{\vec{k}_{\perp}^{2}+m_{q}^{2}}\sqrt{\vec{k}_{\perp}^{\prime 2}+m_{q}^{2}}}-\frac{P^{+}\epsilon^{ij}\Delta_{\perp}^{j}(1-x)(\vec{\Delta}_{\perp}\times \vec{k}_{\perp})_{z}}{2\sqrt{\vec{k}_{\perp}^{2}+m_{q}^{2}}\sqrt{\vec{k}_{\perp}^{\prime 2}+m_{q}^{2}}}
$
\\[7ex]
$(ij)$
&
$\displaystyle
\frac{(\vec{k}_{\perp}\cdot \vec{k}_{\perp}^{\prime} + m_{q}^{2})(2k_{\perp}^{i}+\Delta_{\perp}^{i})(2k_{\perp}^{j}+\Delta_{\perp}^{j})}{2x\sqrt{\vec{k}_{\perp}^{2}+m_{q}^{2}}\sqrt{\vec{k}_{\perp}^{\prime 2}+m_{q}^{2}}}-\frac{(1-x)(\vec{\Delta}_{\perp}\times \vec{k}_{\perp})_{z}\Delta_{\perp}^{k}(\epsilon^{ik}(2k_{\perp}^{j}+\Delta_{\perp}^{j})+\epsilon^{jk}(2k_{\perp}^{i}+\Delta_{\perp}^{i}))}{4x\sqrt{\vec{k}_{\perp}^{2}+m_{q}^{2}}\sqrt{\vec{k}_{\perp}^{\prime 2}+m_{q}^{2}}}
$
\\[7ex]
$(-i)$
&
\makecell{$\displaystyle
\frac{(2k_{\perp}^{i}+\Delta_{\perp}^{i})\left(m_{q}^{2}(1-x)\vec{\Delta}_{\perp}^{2}+(\vec{k}_{\perp}\cdot \vec{k}_{\perp}^{\prime} + m_{q}^{2})(2\underline{k}_{\perp}^{2}+\frac{\vec{\Delta}_{\perp}^{2}}{2})+(1-x)(\vec{\Delta}_{\perp}\times \vec{k}_{\perp})^{2}_{z}\right)}{2x^{2}P^{+}\sqrt{\vec{k}_{\perp}^{2}+m_{q}^{2}}\sqrt{\vec{k}_{\perp}^{\prime 2}+m_{q}^{2}}}$\\
$\displaystyle-\frac{(2\underline{k}_{\perp}^{2}+\vec{\Delta}_{\perp}^{2})(1-x)(\vec{\Delta}_{\perp}\times \vec{k}_{\perp})_{z}\epsilon^{ik}\Delta_{\perp}^{k}}{4x^{2}P^{+}\sqrt{\vec{k}_{\perp}^{2}+m_{q}^{2}}\sqrt{\vec{k}_{\perp}^{\prime 2}+m_{q}^{2}}}$
}

\\[9ex]
$(--)$
&
$\displaystyle
\frac{(2\underline{k}_{\perp}^{2}+\vec{\Delta}_{\perp}^{2})\bigg(2m_{q}^{2}(1-x)\vec{\Delta}_{\perp}^{2}+2(1-x)(\vec{\Delta}_{\perp}\times \vec{k}_{\perp})^{2}_{z}+2\underline{k}_{\perp}^{2}(\vec{k}_{\perp}\cdot \vec{k}_{\perp}^{\prime}+m_{q}^{2})\bigg)}{2x^{3} (P^{+})^{2} \sqrt{\vec{k}_{\perp}^{2}+m_{q}^{2}}\sqrt{\vec{k}_{\perp}^{\prime 2}+m_{q}^{2}}}$
\end{tabular}

\end{ruledtabular}
\end{table*}

\end{widetext}
\subsection{Connecting the anti-quark contribution to the quark contribution}\label{antiquarkcont}
In this section, we connect the anti-quark contribution to the matrix element of the EMT  operator provided in Eqs.~\eqref{Ap16} and \eqref{Ap17} to the quark contribution provided in Eqs.~\eqref{Ap14} and \eqref{Ap15}. Even though we are only dealing with a symmetric mass system (quarkonia), we keep the discussion more general. To see the connection, we write the integral in Eq.~\eqref{Ap14} in terms of the transformed variables, $x\rightarrow (1-x)$ and $\vec{k}_{\perp} \rightarrow -\vec{k}_{\perp}$ and with a real replacement $m_{q}\leftrightarrow m_{\bar{q}}$. The replacement of the mass only matters when $m_{q}\neq m_{\bar{q}}$. Under these transformations, the space wavefunction transforms as, $\varphi(-\vec{k}_{\perp}, 1-x,m_{\bar{q}},m_{q})=\varphi(\vec{k}_{\perp}, x, m_{q}, m_{\bar{q}})$. Similarly, the spin trace part $S^{\mu\nu}_{q}$, will undergo a change. The transformed spin matrices $\mathcal{S}\rightarrow \tilde{\mathcal{S}}$ and $\mathcal{S}^{\prime \dag}\rightarrow \tilde{\mathcal{S}}^{\prime \dag}$ are,
\begin{eqnarray}
  &&\mathcal{S}=\frac{1}{\sqrt{2}\sqrt{\vec{k}_{\perp}^{2}+\tilde{m}^{2}}}
   \begin{pmatrix}
-k^{R} & - \tilde{m}  \\
\tilde{m} & -k^{L}
\end{pmatrix}\rightarrow\tilde{\mathcal{S}}=\frac{1}{\sqrt{2}}\nonumber\\
&&\times \frac{1}{\sqrt{\vec{k}_{\perp}^{2}+\tilde{m}^{2}}}
  \begin{pmatrix}
  k^{R} & -\tilde{m}\\
  \tilde{m} & k^{L}
  \end{pmatrix},~
   \tilde{\mathcal{S}}^{\prime\dag}=\frac{1}{\sqrt{2}}\frac{1}{\sqrt{\vec{\tilde{k}}_{\perp}^{2}+\tilde{m}^{2}}}\nonumber\\
   &&
 \times \begin{pmatrix}
  \tilde{k}^{L} & \tilde{m}\\
  -\tilde{m} & \tilde{k}^{R}
  \end{pmatrix},~\text{such that their product}\nonumber
  \end{eqnarray}
\begin{eqnarray}
&&\tilde{\mathcal{S}}\tilde{\mathcal{S}}^{\prime\dag}=\frac{1}{2\sqrt{\vec{k}_{\perp}^{2}+\tilde{m}^{2}}\sqrt{\vec{\tilde{k}}_{\perp}^{2}+\tilde{m}^{2}}} \begin{pmatrix}
   C & \tilde{m} \Delta_{\perp}^{R} x\\
   -\tilde{m} \Delta_{\perp}^{L} x &  C^{\ast} \nonumber
  \end{pmatrix},\\
  \label{Ap27}
\end{eqnarray}
where we use the short hand $C(x,\vec{k}_{\perp})\equiv \vec{k}_{\perp}\cdot \vec{\tilde{k}}_{\perp} -ix(\vec{\Delta}\times \vec{k}_{\perp})_{z}+\tilde{m}^{2} $ and 
$\vec{\tilde{k}}=\vec{k}_{\perp}-x\vec{\Delta}_{\perp}$ is defined in the same way as in Eq.~\eqref{Ap18}. The matrix $U^{\mu\nu}$ transforms as 
\begin{eqnarray}
U^{\mu\nu}_{\lambda_{q}^{\prime}\lambda_{q}}\rightarrow \frac{1}{4(1-x)}\bar{u}_{\lambda_{q}^{\prime}}(p_{\bar{q}}^{\prime})[\gamma^{\mu}p_{s\bar{q}}^{\nu}+\gamma^{\nu}p_{s\bar{q}}^{\mu}]u_{\lambda_{q}}(p_{\bar{q}})\nonumber\\
=  \frac{1}{4(1-x)}\bar{v}_{\lambda_{q}}(p_{\bar{q}})[\gamma^{\mu}p_{s\bar{q}}^{\nu}+\gamma^{\nu}p_{s\bar{q}}^{\mu}]v_{\lambda_{q}^{\prime}}(p_{\bar{q}}^{\prime})=V^{\mu\nu}_{\lambda_{q}\lambda_{q}^{\prime}}~,\nonumber\\
\label{Ap28}
\end{eqnarray}
where to get the last line we use the spinor identity~\cite{Brodsky:1997de} $\bar{u}_{\lambda_{q}^{\prime}}(p_{\bar{q}}^{\prime})\gamma^{\mu}u_{\lambda_{q}}(p_{\bar{q}})=\bar{v}_{\lambda_{q}}(p_{\bar{q}})\gamma^{\mu}v_{\lambda_{q}^{\prime}}(p_{\bar{q}}^{\prime})$. The transformed quark contribution to the matrix element with the renaming of the dummy summation variables $\lambda_{q}\rightarrow \lambda_{\bar{q}}$, $\lambda_{q}^{\prime}\rightarrow \lambda_{\bar{q}}^{\prime}$, and $\lambda_{2}\rightarrow \lambda_{1}$ becomes,
\begin{eqnarray}
 \langle \hat{T}^{\mu\nu}_{q}(m_{q}\leftrightarrow m_{\bar{q}})\rangle =&& \int \frac{dx~ d^{2}\vec{k}_{\perp}}{2(2\pi)^{3}} ~\varphi^{\ast}(x,\vec{\tilde{k}}_{\perp})\varphi(x,\vec{k}_{\perp})\nonumber\\
 \sum_{\lambda_{\bar{q}},\lambda_{\bar{q}}^{\prime},\lambda_{1}}&&\tilde{\mathcal{S}}^{\ast}_{\lambda_{\bar{q}}^{\prime}\lambda_{1}}(x,\vec{\tilde{k}}_{\perp}) V^{\mu\nu}_{\lambda_{\bar{q}}\lambda_{\bar{q}}^{\prime}} \tilde{\mathcal{S}}_{\lambda_{\bar{q}}\lambda_{1}}(x,\vec{k}_{\perp})  \nonumber\\
 =&&\int \frac{dx~ d^{2}\vec{k}_{\perp}}{2(2\pi)^{3}} ~\varphi^{\ast}(x,\vec{\tilde{k}}_{\perp})\varphi(x,\vec{k}_{\perp})\nonumber\\
 && \times \Tr[\tilde{\mathcal{S}}^{\ast\prime}\mathcal{S}^{T}V^{\mu\nu}].\label{Ap29}
\end{eqnarray}
The anti-quark contribution to the matrix element from Eqs.~\eqref{Ap17} and \eqref{Ap18} is expressed as,
\begin{eqnarray}
 \langle \hat{T}^{\mu\nu}_{\bar{q}}\rangle &=& \int \frac{dx~ d^{2}\vec{k}_{\perp}}{2(2\pi)^{3}} ~\varphi^{\ast}(x,\vec{\Tilde{k}}_{\perp})\varphi(x,\vec{k}_{\perp})\nonumber\\
 &&\times \Tr[\bar{\mathcal{S}}^{\prime \dag}\mathcal{S}V^{\mu\nu}]\label{Ap30}~,
\end{eqnarray}
where we have $\bar{\mathcal{S}}^{\prime}=\mathcal{S}(x,\vec{\Tilde{k}}_{\perp}=\vec{k}_{\perp}-x\vec{\Delta}_{\perp})$. Now, writing explicitly the product of the spin wavefunctions occurring in the right-hand side of the Eq.~\eqref{Ap30},
\begin{eqnarray}
 \bar{\mathcal{S}}^{\prime \dag}\mathcal{S}&=&\frac{1}{2\sqrt{\vec{k}_{\perp}^{2}+\tilde{m}^{2}}\sqrt{\vec{\tilde{k}}_{\perp}^{2}+\tilde{m}^{2}}}\nonumber\\
  &\times& \begin{pmatrix}
   C& -\tilde{m} \Delta_{\perp}^{L} x\\
   \tilde{m} \Delta_{\perp}^{R} x & C^{\ast}
  \end{pmatrix},\nonumber\\
  \text{where, }\bar{\mathcal{S}}^{\prime \dag}&=& \frac{1}{\sqrt{2}}\frac{1}{\sqrt{\vec{\tilde{k}}_{\perp}^{2}+\tilde{m}^{2}}}
  \begin{pmatrix}
 - k^{\prime L} & \tilde{m}\\
  -\tilde{m} & -k^{\prime R}
  \end{pmatrix}.\nonumber\\
  \label{Ap31}
\end{eqnarray}
Using Eqs.~\eqref{Ap27} and \eqref{Ap31} we get $ [\tilde{\mathcal{S}}\tilde{\mathcal{S}}^{\prime\dag}]^{T}=\tilde{\mathcal{S}}^{\ast \prime} \tilde{\mathcal{S}}^{T}=\bar{\mathcal{S}}^{\prime \dag}\mathcal{S}$, which alongwith Eqs.~\eqref{Ap29} and \eqref{Ap30} proves $ \langle \hat{T}^{\mu\nu}_{\bar{q}}\rangle= \langle \hat{T}^{\mu\nu}_{q}(m_{q}\leftrightarrow m_{\bar{q}})\rangle$. In the case of the symmetric mass system we have $ \langle \hat{T}^{\mu\nu}_{\bar{q}}\rangle= \langle \hat{T}^{\mu\nu}_{q}\rangle$.

\section{BT construction and different expression of \(A\) and \(D\) term}\label{BT_const}
\begin{figure*}[!t]
		\centering
		\includegraphics[width=0.95\linewidth]{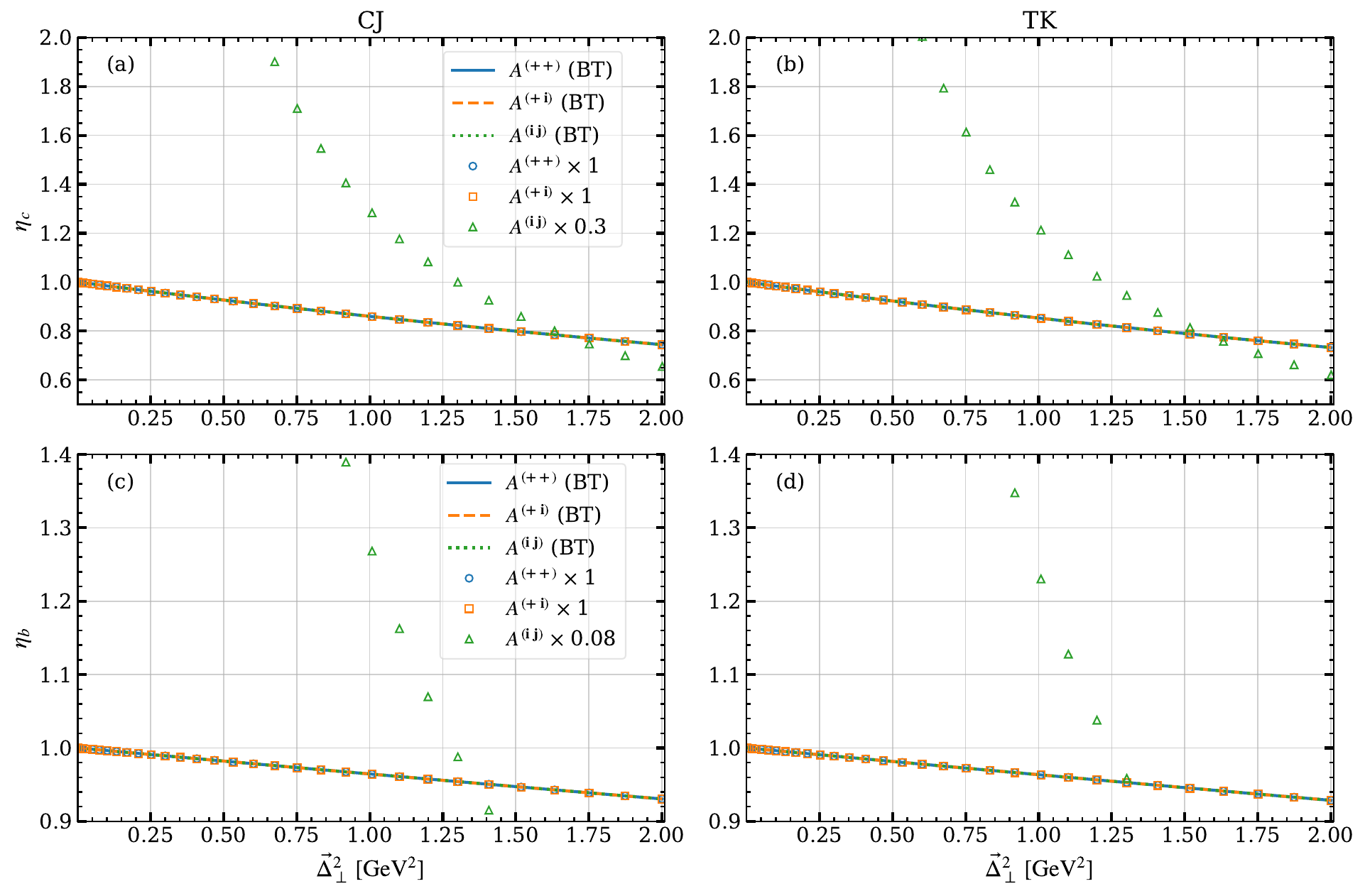}
		\caption{(Color online) Different extraction of $A$-term for $\eta_{c}$ (upper panel) and $\eta_{b}$ (lower panel) with different radial wavefunctions CJ (left) and TK (right). Figures in the upper panel (a and b) share the same legend (the same is true for Figures c and d).}
		\label{fig:AtermBT}
\end{figure*}
\begin{figure*}[!t]
		\centering
		\includegraphics[width=0.95\linewidth]{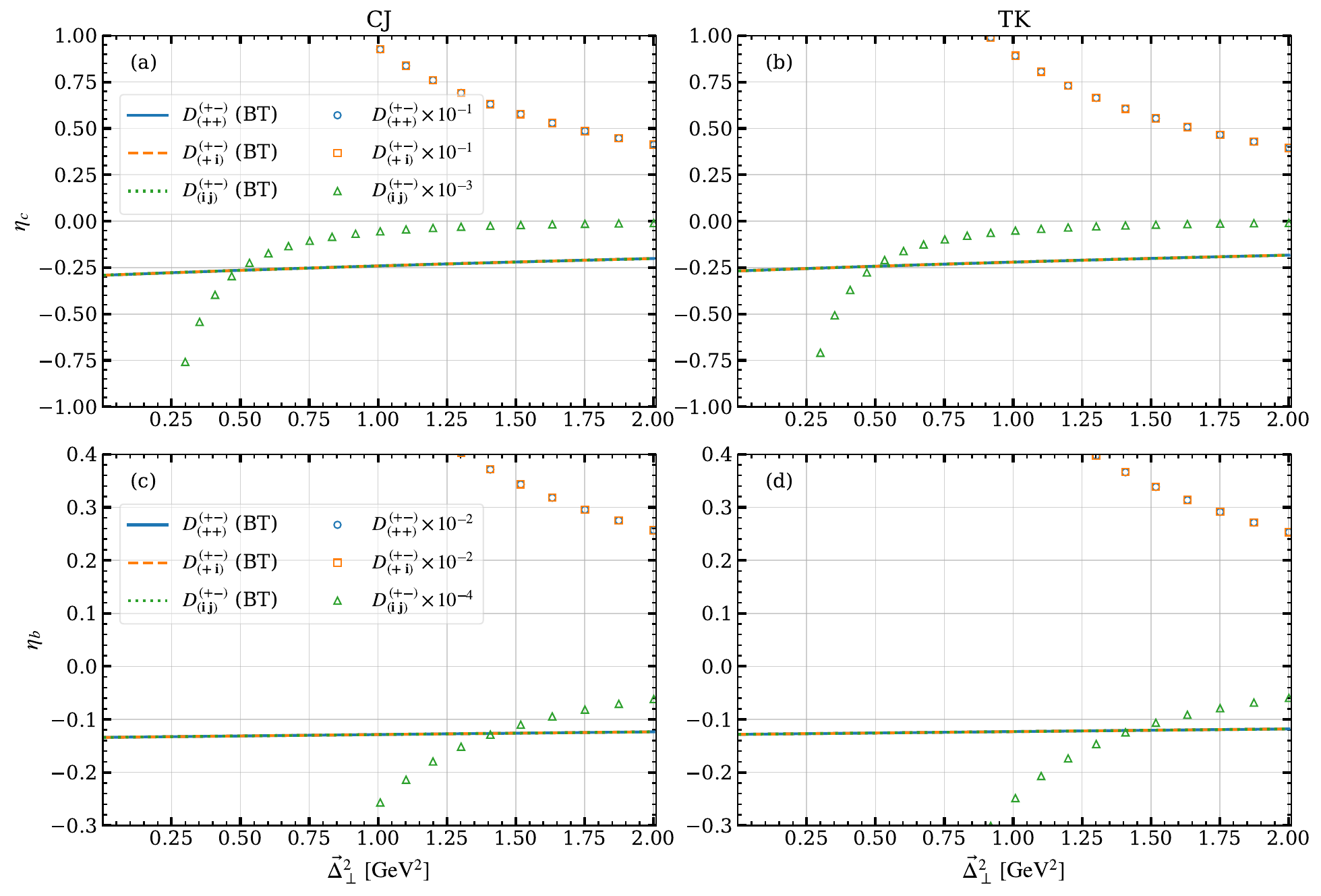}
		\caption{(Color online) Different extraction of $D$-term in the ($+-$) sector for $\eta_{c}$ (upper panel) and $\eta_{b}$ (lower panel) with different radial wavefunctions CJ (left) and TK (right). Figures in the upper panel (a and b) share the same legend (the same is true for Figures c and d).}
		\label{fig:DpmBT}
\end{figure*}  

\begin{figure*}[!t]
		\centering
		\includegraphics[width=0.95\linewidth]{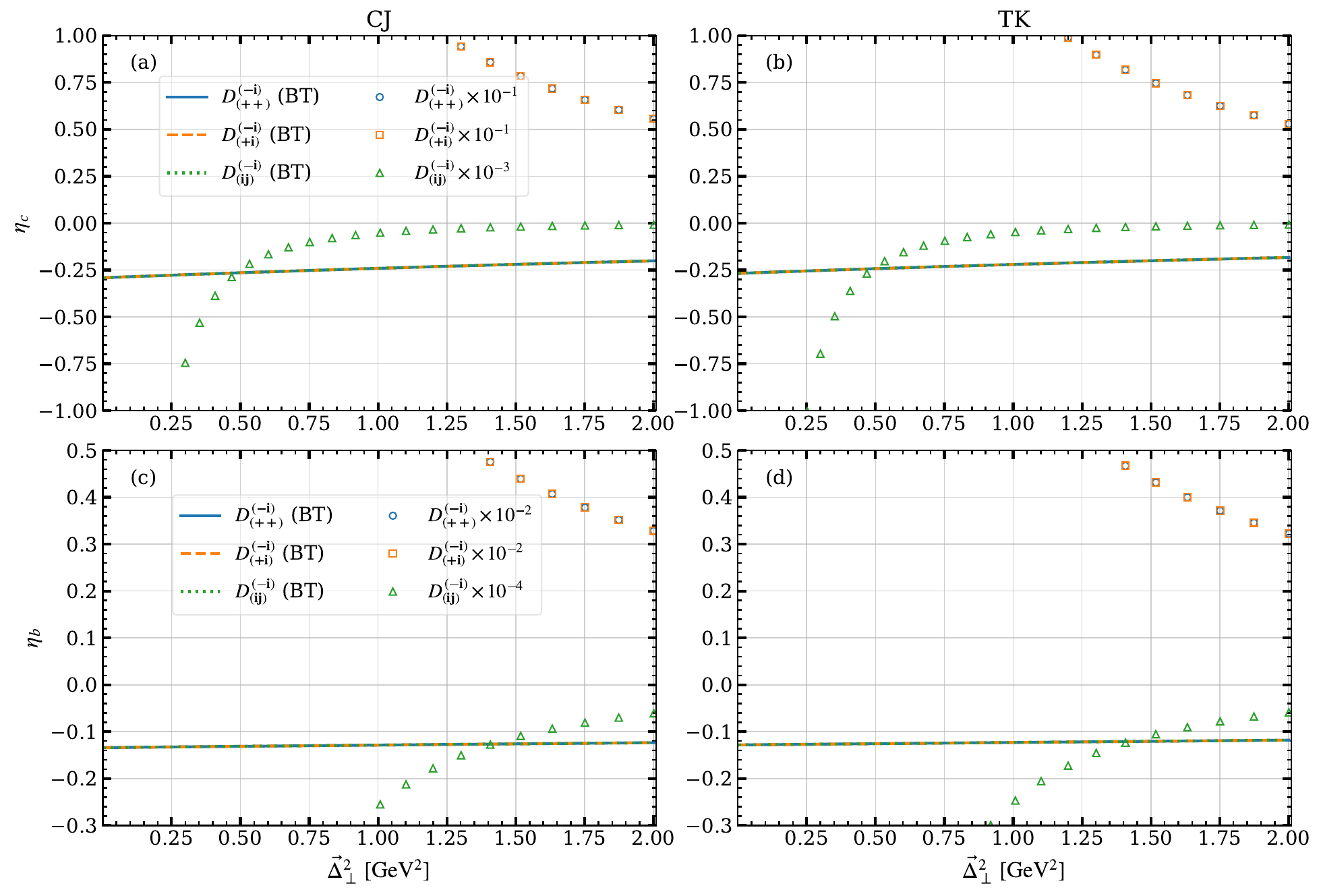}
		\caption{(Color online) Different extraction of $D$-term in the ($-i$) sector for $\eta_{c}$ (upper panel) and $\eta_{b}$ (lower panel) with different radial wavefunctions CJ (left) and TK (right). Figures in the upper panel (a and b) share the same legend (the same is true for Figures c and d).}
		\label{fig:DmiBT}
\end{figure*} 

\begin{figure*}[!t]
		\centering
		\includegraphics[width=0.95\linewidth]{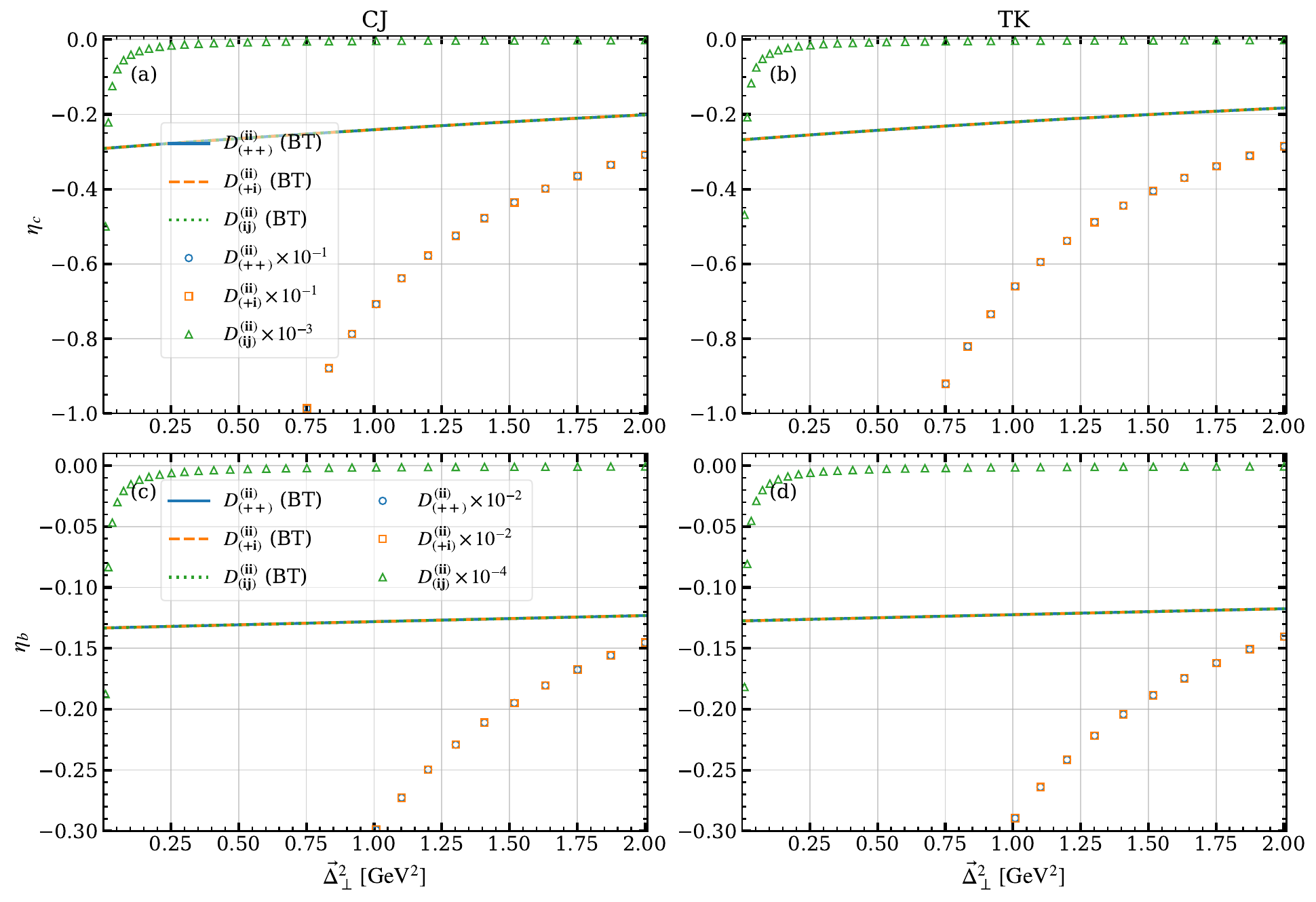}
		\caption{(Color online) Different extraction of $D$-term in the ($ii$) sector for $\eta_{c}$ (upper panel) and $\eta_{b}$ (lower panel) with different radial wavefunctions CJ (left) and TK (right). Figures in the upper panel (a and b) share the same legend (the same is true for Figures c and d).}
		\label{fig:DiiBT}
\end{figure*}
The Lorentz decomposition of the matrix element of EMT for the spin-0 mesons is given by,
\begin{equation}
 T^{\mu\nu}  =\Gamma_{A}^{\mu\nu}~A+\Gamma_{D}^{\mu\nu}~D~,\label{Apbt1}
\end{equation}
where we have $\Gamma_{A}^{\mu\nu}=2\bar{P}^{\mu}\bar{P}^{\nu}$ and $\Gamma_{D}^{\mu\nu}=\frac{\Delta^{\mu}\Delta^{\nu}-g^{\mu\nu}\Delta^{2}}{2}$.
While Eq.~\eqref{Apbt1} as it stands is generally valid, as we have discussed in Sec.~\ref {sec:GFF}, the manifest Lorentz covariance is lost in the phenomenological LFQM used in the present work. This, in turn, means that one can extract the GFFs from different components of $T^{\mu\nu}$, and their expression may differ depending on the component of EMT used to extract them. The Eq.~\eqref{Apbt1} gives us the following seven independent equations in a frame with $\Delta^{+}=0$~\cite{Choi:2025rto},
\begin{eqnarray}
    T^{++}&=&\Gamma_{A}^{++} A,\label{Apbt2}\\
    T^{+-}&=& \Gamma_{A}^{+-} A + \Gamma_{D}^{+-} D~,\label{Apbt3}\\
   \Delta_{\perp}^{i} T^{+i}&=& \Delta_{\perp}^{i}\Gamma_{A}^{+i} A~,\label{Apbt4}\\
   \Delta_{\perp}^{i}  \Delta_{\perp}^{j}  T^{ij}&=& \Delta_{\perp}^{i}\Delta_{\perp}^{i}\Gamma_{A}^{ij} A~,\label{Apbt5}\\
   T^{ii}&=& \Gamma_{A}^{ii} A+ \Gamma_{D}^{ii} D~,\label{Apbt6}  \\
   \Delta_{\perp}^{i} T^{-i}&=& \Delta_{\perp}^{i}\Gamma_{A}^{-i} A+ \Delta_{\perp}^{i}\Gamma_{D}^{-i} D~,\label{Apbt7}\\
  T^{--}&=& \Gamma_{A}^{--} A+ \Gamma_{D}^{--} D~,\label{Apbt8}
\end{eqnarray}
where we used $\Gamma_{D}^{++}=\Gamma_{D}^{+i}=\Delta_{\perp}^{i}\Delta_{\perp}^{j}\Gamma_{D}^{ij}=0$. One sees that $A$-term can be obtained in three independent ways using Eq.~\eqref{Apbt2}, Eq.~\eqref{Apbt4}, or Eq.~\eqref{Apbt5} irrespective of $D$. Moreover, once one determines $A$, then $D$ can be evaluated independently using either Eq.~\eqref{Apbt3}, Eq.~\eqref{Apbt6}, Eq.~\eqref{Apbt7} or Eq.~\eqref{Apbt8}. Let us write down the expressions of the $A$-term evaluated in this way with the BT construction applied only to the matrix element,
\begin{eqnarray}
    A^{(++)}&=&\frac{T^{++}}{\Gamma_{A}^{++}}=\frac{1}{2(P^{+})^{2}} \int_{\phi} S^{++}~, \label{Apbt9}\\
    A^{(+i)}&=&\frac{ \Delta_{\perp}^{i} T^{+i}}{\Delta_{\perp}^{i}\Gamma_{A}^{+i}}=\frac{1}{P^{+}\vec{\Delta}_{\perp}^{2}}\int_{\phi} \Delta_{\perp}^{i} S^{+i}~,\label{Apbt10}\\
    A^{(ij)}&=&\frac{ \Delta_{\perp}^{i} \Delta_{\perp}^{i}T^{ij}}{\Delta_{\perp}^{i}\Delta_{\perp}^{j}\Gamma_{A}^{ij}}=\frac{2}{\vec{\Delta}_{\perp}^{4}}\int_{\phi} \Delta_{\perp}^{i} \Delta_{\perp}^{j} S^{ij}~,\label{Apbt11}
\end{eqnarray}
where the integral $\int_{\phi} \equiv \int \frac{dx~ d^{2}\vec{k}_{\perp}}{2(2\pi)^{3}} ~\varphi^{\ast}(x,\vec{k}_{\perp}^{\prime})\varphi(x,\vec{k}_{\perp})$. One can also define the integrands that lead to the corresponding expression of $A$-term in Eqs.~\eqref{Apbt9}--\eqref{Apbt11} as, $A^{(\ast \ast)}=\int_{\phi} \mathcal{A}^{(\ast \ast)}$, where $(\ast \ast)\in \{++, +i, ij\}$. 
Now, since $D$ can be evaluated in four independent ways, and each of the Eqs.~\eqref{Apbt3}, \eqref{Apbt6}, \eqref{Apbt7} and \eqref{Apbt8} contains $A$, we obtain twelve independent expressions for $D$,
\begin{eqnarray}
    D^{(+-)}_{(\ast \ast)}&=&\frac{T^{+-}}{\Gamma_{D}^{+-}}-\frac{\Gamma_{A}^{+-}}{\Gamma_{D}^{+-}}A^{(\ast \ast)}\nonumber\\
    &=&\frac{1}{\vec{\Delta}_{\perp}^{2}} \int_{\phi} S^{+-}-\frac{2M^{2}+\vec{\Delta}_{\perp}^{2}}{\vec{\Delta}_{\perp}^{2}} \int_{\phi} \mathcal{A}^{(\ast \ast)}~, \label{Apbt12}\\
    D^{(ii)}_{(\ast \ast)}&=&\frac{T^{ii}}{\Gamma_{D}^{ii}}-\frac{\Gamma_{A}^{ii}}{\Gamma_{D}^{ii}} A^{(\ast \ast)}\nonumber\\
    &=&-\frac{2}{\vec{\Delta}_{\perp}^{2}}\int_{\phi} S^{ii}+\int_{\phi} \mathcal{A}^{(\ast \ast)}~,\label{Apbt13}\\
    D^{(-i)}_{(\ast \ast)}&=&\frac{\Delta_{\perp}^{i}T^{-i}}{\Delta_{\perp}^{i}\Gamma_{D}^{-i}}-\frac{\Delta_{\perp}^{i}\Gamma_{A}^{-i}}{\Delta_{\perp}^{i}\Gamma_{D}^{-i}}A^{(\ast \ast)}\nonumber\\
    &=&\frac{2P^{+}}{\vec{\Delta}_{\perp}^{4}}\int_{\phi} \Delta_{\perp}^{i} S^{-i}-\frac{2M^{2}+\vec{\Delta}_{\perp}^{2}}{\vec{\Delta}_{\perp}^{2}}\int_{\phi} \mathcal{A}^{(\ast \ast)}~,\nonumber\\
    \label{Apbt14}\\
    D^{(--)}_{(\ast \ast)}&=&\frac{T^{--}}{\Gamma_{D}^{--}}-\frac{\Gamma_{A}^{--}}{\Gamma_{D}^{--}}A^{(\ast \ast)}\nonumber\\
    &=&\frac{2(P^{+})^{2}}{\vec{\Delta}_{\perp}^{4}} \int_{\phi} S^{--}-\frac{(2M^{2}+\vec{\Delta}_{\perp}^{2})^{2}}{\vec{\Delta}_{\perp}^{4}} \int_{\phi} \mathcal{A}^{(\ast \ast)}~,\nonumber\\ \label{Apbt15}
\end{eqnarray}
where $M$ is the physical mass of the meson. In arriving  at Eqs.~\eqref{Apbt12}--\eqref{Apbt15} we used the explicit expression of the Lorentz structures $\Gamma_{A}^{\mu\nu}$ and $\Gamma_{D}^{\mu\nu}$.

Within the ambit of the BT construction, as discussed in Sec.~\ref {sec:GFF}, the form of the $\Gamma_{A}^{\mu\nu}$ changes and the physical meson mass is replaced with the invariant mass. Therefore, the meson mass associated with in the initial $P$ and final momentum $P^{\prime}$ differs, i.e., $P^{2}=M_{0}^{2}$ and $P^{\prime 2}=M_{0}^{\prime 2}$. The modification of the Lorentz structures is provided in Table~\ref{tab:BTlor}.
We write the different expressions of $A$ and $D$-terms within the BT framework using Eqs.~\eqref{Apbt2}--\eqref{Apbt8} and Table~\ref{tab:BTlor} as,
\begin{eqnarray}
    A^{(++)}\vert_{\rm BT}&=&\frac{T^{++}}{\Gamma_{A}^{++}}\bigg|_{\rm BT}=\int_{\phi} \frac{S^{++}}{2(P^{+})^{2}}~, \label{Apbt16}\\
    A^{(+i)}\vert_{\rm BT}&=&\frac{ \Delta_{\perp}^{i} T^{+i}}{\Delta_{\perp}^{i}\Gamma_{A}^{+i}}\bigg|_{\rm BT}\nonumber\\
    &=&\int_{\phi} \frac{\Delta_{\perp}^{i} S^{+i}}{P^{+}(M_{0}^{\prime 2}-M_{0}^{2}+\vec{\Delta}_{\perp}^{2})}~,\label{Apbt17}\\
    A^{(ij)}\vert_{\rm BT}&=&\frac{ \Delta_{\perp}^{i} \Delta_{\perp}^{i}T^{ij}}{\Delta_{\perp}^{i}\Delta_{\perp}^{j}\Gamma_{A}^{ij}}\bigg|_{\rm BT}\nonumber\\
    &=&\int_{\phi} \frac{2\Delta_{\perp}^{i} \Delta_{\perp}^{j} S^{ij}}{\vec{\Delta}_{\perp}^{4}+(M_{0}^{\prime 2}-M_{0}^{2})(M_{0}^{\prime 2}-M_{0}^{2}+2\vec{\Delta}_{\perp}^{2})}~,\nonumber\\
    \label{Apbt18}\\
    D^{(+-)}_{(\ast \ast)}\vert_{\rm BT}&=&\frac{T^{+-}}{\Gamma_{D}^{+-}}-\frac{\Gamma_{A}^{+-}}{\Gamma_{D}^{+-}}A^{(\ast \ast)}\bigg|_{\rm BT}\nonumber\\
    &=& \int_{\phi} \frac{S^{+-}}{\vec{\Delta}_{\perp}^{2}}- \int_{\phi} \frac{2M_{0}^{\prime 2}+\vec{\Delta}_{\perp}^{2}}{\vec{\Delta}_{\perp}^{2}}\mathcal{A}^{(\ast \ast)}\vert_{\rm BT}~, \label{Apbt19}\\
    D^{(ii)}_{(\ast \ast)}\vert_{\rm BT}&=&\frac{T^{ii}}{\Gamma_{D}^{ii}}-\frac{\Gamma_{A}^{ii}}{\Gamma_{D}^{ii}} A^{(\ast \ast)}\bigg|_{\rm BT}\nonumber\\
    &=&-\int_{\phi} \frac{2S^{ii}}{\vec{\Delta}_{\perp}^{2}} + \int_{\phi} \mathcal{A}^{(\ast \ast)}\vert_{\rm BT}  \nonumber\\
    &\times&\frac{\vec{\Delta}_{\perp}^{4}+2(M_{0}^{\prime 2}-M_{0}^{2})(\vec{\Delta}_{\perp}^{2}+M_{0}^{\prime 2}-M_{0}^{2})}{\vec{\Delta}_{\perp}^{4}},\nonumber\\
    \label{Apbt20}
    \end{eqnarray}
    \begin{eqnarray}
    D^{(-i)}_{(\ast \ast)}\vert_{\rm BT}&=&\frac{\Delta_{\perp}^{i}T^{-i}}{\Delta_{\perp}^{i}\Gamma_{D}^{-i}}-\frac{\Delta_{\perp}^{i}\Gamma_{A}^{-i}}{\Delta_{\perp}^{i}\Gamma_{D}^{-i}}A^{(\ast \ast)}\bigg|_{\rm BT}\nonumber\\
    &=&\int_{\phi} \frac{2P^{+}\Delta_{\perp}^{i} S^{-i}}{(M_{0}^{\prime 2}-M_{0}^{2}+\vec{\Delta}_{\perp}^{2})\vec{\Delta}_{\perp}^{2}}\nonumber\\
    &-&\int_{\phi} \frac{2M_{0}^{\prime 2}+\vec{\Delta}_{\perp}^{2}}{\vec{\Delta}_{\perp}^{2}} \mathcal{A}^{(\ast \ast)}\vert_{\rm BT}~,\label{Apbt21}\\
    D^{(--)}_{(\ast \ast)}\vert_{\rm BT}&=&\frac{T^{--}}{\Gamma_{D}^{--}}-\frac{\Gamma_{A}^{--}}{\Gamma_{D}^{--}}A^{(\ast \ast)}\bigg|_{\rm BT}\nonumber\\
    &=& \int_{\phi} \frac{2(P^{+})^{2} S^{--}}{(M_{0}^{\prime 2}-M_{0}^{2}+\vec{\Delta}_{\perp}^{2})^{2}}\nonumber\\ 
   &-&\int_{\phi}\mathcal{A}^{(\ast \ast)}\vert_{\rm BT}\bigg[\frac{(M_{0}^{\prime 2}+M_{0}^{2}+\vec{\Delta}_{\perp}^{2})^{2}}{(M_{0}^{\prime 2}-M_{0}^{2}+\vec{\Delta}_{\perp}^{2})^{2}}\nonumber\\
    &+& \frac{2(M_{0}^{\prime 2}-M_{0}^{2})(M_{0}^{\prime 2}+M_{0}^{2}+\vec{\Delta}_{\perp}^{2})}{\vec{\Delta}_{\perp}^{2}(M_{0}^{\prime 2}-M_{0}^{2}+\vec{\Delta}_{\perp}^{2})}\bigg]~.\label{Apbt22}   
\end{eqnarray}

In Fig.~\ref{fig:AtermBT}, we show the different extraction of the $A$-term using Eqs.~\eqref{Apbt9}--\eqref{Apbt11} (shown by scattered points) and Eqs.~\eqref{Apbt16}--\eqref{Apbt18} (shown in continuous lines). Interestingly, we observe that the results of the five extractions merge, while one extraction, $A^{(ij)}$, shows a divergent trend at low momentum transfer. This demonstrates that the $(ij)$ component should be avoided in extracting the $A$-term without a proper implementation of the BT construction in LFQM. Similarly, we show different extractions of the $D$-term sector-wise (we are identifying the superscripts of $D$ in Eqs.~\eqref{Apbt12}--\eqref{Apbt15} or Eqs.~\eqref{Apbt19}--\eqref{Apbt22} as sectors) in Figs~\ref{fig:DpmBT}, \ref{fig:DmiBT}, and \ref{fig:DiiBT}, for $(+-)$, $(-i)$, and $(ii)$ sectors, respectively. In all these plots, we observe that after BT implementation, the $D$-term results for quarkonia with two different wavefunctions remain the same. An improper application of BT construction, i.e., only for the matrix elements, leads to different magnitudes for $D$-terms, which not only depend upon the sector chosen but also on the $A$ used to extract them. For the sector $(--)$ (not shown in the paper), we verified that even BT construction cannot cure the zero-mode contribution, and the extracted $D$ differs. This extends the conclusion of Ref.~\cite{Choi:2025rto} in reference to the pion for the quarkonia.
\begin{table*}[t]
\caption{The modifications of the Lorentz structure in the presence of BT construction.}
\label{tab:BTlor}
\begin{ruledtabular}
\begin{tabular}{c c c}
$(\mu\nu)$
&
$\Gamma_{A}^{\mu\nu}\vert_{\rm BT}$
&
$\Gamma_{D}^{\mu\nu}\vert_{\rm BT}$
\\
\hline
$(++)$
&
$\displaystyle 2 (P^{+})^{2}$
&
$0$
\\[3ex]
$(+-)$
&
$\displaystyle
2M_{0}^{\prime 2}+\vec{\Delta}_{\perp}^{2}
$
&
$\displaystyle
\vec{\Delta}_{\perp}^{2}
$
\\[3ex]
$(+i)$
&
$\displaystyle
P^{+}\Delta_{\perp}^{i}\left(1+\frac{M_{0}^{\prime 2}-M_{0}^{2}}{\vec{\Delta}_{\perp}^{2}}\right)
$
&
$0$

\\[3ex]
$(ij)$
&
$\displaystyle
\frac{1}{2}\left[\Delta_{\perp}^{i}\Delta_{\perp}^{j}\left(1+\frac{2(M_{0}^{\prime 2}-M_{0}^{2})}{\vec{\Delta}_{\perp}^{2}}\right)+\frac{(M_{0}^{\prime 2}-M_{0}^{2})^{2}}{\vec{\Delta}_{\perp}^{2}}\delta^{ij}\right]
$
&
$\displaystyle
\frac{1}{2}(\Delta_{\perp}^{i}\Delta_{\perp}^{j}-\delta^{ij}\vec{\Delta}_{\perp}^{2})
$
\\[3ex]

$(-i)$
&
$\displaystyle
\frac{(2M_{0}^{\prime 2}+\vec{\Delta}_{\perp}^{2})(M_{0}^{\prime 2}-M_{0}^{2}+\vec{\Delta}_{\perp}^{2})\Delta_{\perp}^{i}}{2\vec{\Delta}_{\perp}^{2}P^{+}}
$
&
$\displaystyle
\frac{M_{0}^{\prime 2}-M_{0}^{2}+\vec{\Delta}_{\perp}^{2}}{2P^{+}}\Delta_{\perp}^{i}
$

\\[3ex]
$(--)$
&
$\displaystyle
\frac{(M_{0}^{\prime 2}+M_{0}^{2}+\vec{\Delta}_{\perp}^{2})\left(\vec{\Delta}_{\perp}^{2}(M_{0}^{\prime 2}+M_{0}^{2}+\vec{\Delta}_{\perp}^{2})+2(M_{0}^{\prime 2}-M_{0}^{2})(M_{0}^{\prime 2}-M_{0}^{2}+\vec{\Delta}_{\perp}^{2})\right)}{2(P^{+})^{2}\vec{\Delta}_{\perp}^{2}}
$
&
$\displaystyle
\frac{(M_{0}^{\prime 2}-M_{0}^{2}+\vec{\Delta}_{\perp}^{2})^{2}}{2(P^{+})^{2}}$
\end{tabular}
\end{ruledtabular}
\end{table*}  

\section{Different spatial densities from EMT}\label{apedensity}
Here we collect some of the formulae that are discussed in several references~\cite{Polyakov:2018zvc,Freese:2021czn} for the sake of completeness. Before doing so, we will briefly explain how the lightfront densities are defined with the help of wavepackets closely following Ref.~\cite{Freese:2021czn}. Let us start with the definition of the spatial densities as provided in Ref.~\cite{Freese:2021czn},
\be
&&\int dz^{-} \langle \psi \vert \hat{T}^{\mu\nu} (z^{+}=0,z^{-},\vec{z}_{\perp}) \vert \psi \rangle=2\pi (2\sigma)^{2}\int \frac{d^{2}\vec{\bar{P}}_{\perp}}{(2\pi)^{2}}\nonumber\\
&& e^{-2\sigma^{2}\vec{\bar{P}}_{\perp}^{2}} \int \frac{d^{2}\vec{\Delta}_{\perp}}{(2\pi)^{2}2P^{+}} \langle \hat{T}^{\mu\nu}(0)\rangle~ e^{-i\vec{\Delta}_{\perp}\cdot\vec{z}_{\perp}}~ e^{-\sigma^{2}\vec{\Delta}_{\perp}^{2}/2}~.\nonumber\\\label{apd1}
\ee
The left-hand side of Eq.~\eqref{apd1} gives the definition of the spatial density associated with an operator (in our case, the EMT operator) in a localized meson state $\vert \psi \rangle$. Note that the $\vert \psi \rangle$ here is different from the $\Psi$ defined in Sec.~\ref{Sec:LCQM}; $\Psi$ was used for the distribution of the quarks inside a plane wave state of the meson, whereas the $\vert \psi \rangle$ here is the meson wavepacket made up of the plane wave states. To get the right-hand side of Eq.~\eqref{apd1}, the wavepacket is chosen in the momentum space so that it has a given momentum in the longitudinal direction, i.e., $P^{+}$, but is localized (spatially) in the perpendicular direction with a finite width $\sigma$~\cite{Freese:2021czn}. Note that a spatially localized wave packet is needed to make the description of density meaningful. To get the density in an arbitrarily localized wavepacket in the transverse plane, one needs to take the limit $\sigma \rightarrow 0$ after all the integrals in the right-hand side of Eq.~\eqref{apd1} are performed. Since one has to get rid of the width of the wave packet to have meaningful internal density, the matrix elements between the plane wave states $\langle \hat{T}^{\mu\nu}(0)\rangle$ should not have $\vec{\bar{P}}_{\perp}$ dependence~\cite{Freese:2021czn}. From Eq.~\eqref{G9} we notice that only the matrix element $\langle \hat{T}^{++}(0)\rangle$ and $\langle \hat{T}^{ij}(0)\rangle$ has no $\vec{\bar{P}}$ dependence. For such matrix elements the $\vec{\bar{P}}_{\perp}$ integral in Eq.~\eqref{apd1} can be done resulting,
\be
&&\int dz^{-} \langle \psi \vert \hat{T}^{\mu\nu} (z^{+}=0,z^{-},\vec{z}_{\perp}) \vert \psi \rangle\nonumber\\
&& =\int \frac{d^{2}\vec{\Delta}_{\perp}}{(2\pi)^{2}2P^{+}} \langle \hat{T}^{\mu\nu}(0)\rangle~ e^{-i\vec{\Delta}_{\perp}\cdot\vec{z}_{\perp}}~ e^{-\sigma^{2}\vec{\Delta}_{\perp}^{2}/2}.\label{apd2}
\ee
Now it is trivially seen that the shear tensor defined in Eq.~\eqref{G10} and the LF momentum density defined in Eq.~\eqref{G14} are actually the density associated with $\langle \hat{T}^{ij}(0)\rangle$ and $\langle \hat{T}^{++}(0)\rangle$, respectively, in the limit $\sigma \rightarrow 0$. We proceed from here to prove Eqs.~\eqref{ex5} and \eqref{ex4}, which define the pressure and shear function. The LF pure stress tensor is given by
\begin{eqnarray}
  {S}^{ij}&=& \frac{1}{4P^{+}}\int \frac{d^{2}\vec{\Delta}_{\perp}}{(2\pi)^{2}}~e^{-i\vec{\Delta}_{\perp}\cdot \vec{z}_{\perp}}~(\Delta_{\perp}^{i}\Delta_{\perp}^{j}-\vec{\Delta}_{\perp}^{2}~\delta^{ij})D\nonumber\\
  &=& \frac{1}{4P^{+}}\int \frac{d^{2}\vec{\Delta}_{\perp}}{(2\pi)^{2}}~D (-\partial_{i}\partial_{j}+\delta^{ij}\partial^{2}) e^{-i\vec{\Delta}_{\perp}\cdot \vec{z}_{\perp}}\nonumber\\
  &=& (-\partial_{i}\partial_{j}+\delta^{ij}\partial^{2}) \tilde{D}\label{apd3},
  \end{eqnarray}
where we define $\partial_{i}=\partial/\partial z_{\perp}^{i}$, $\partial^{2}=\partial_{i}\partial_{i}$ and $\delta^{ij}$ is the usual Kronecker delta. Note that we only assume that all the repeated indices are summed over and not following the known covariant-contravariant index prescription. 
It is useful to notice that the $\tilde{D}$ is only a function of $z_{\perp}=|\vec{z}_{\perp}|$ since $D$-term is isotropic, i.e., $D=D(-\vec{\Delta}_{\perp}^{2})$. In fact, any Fourier transform of this type can be cast as an integration over the Bessel function of the first kind. To see this, let us recall that the generating function of the Bessel function of the first kind is given by $e^{\frac{x}{2}(t-\frac{1}{t})}=\sum_{n=-\infty}^{n=\infty} J_{n}(x)t^{n}$ and identifying $x=\Delta_{\perp}z_{\perp}$ and $t=-ie^{i\phi}$, where $\vec{\Delta}_{\perp}\cdot \vec{z}_{\perp}=\Delta_{\perp} z_{\perp} \cos\phi$, we have,
\be
\tilde{D}=\frac{1}{4P^{+}}\times \frac{1}{2\pi} \int_{0}^{\infty} D~  J_{0}(\Delta_{\perp} z_{\perp})  ~\Delta_{\perp}d\Delta_{\perp}~.\label{apd4}
\ee
By exploiting the isotropic nature of $\tilde{D}$, Eq.~\eqref{apd3} can be further simplified by converting the transverse plane derivatives to radial derivatives in the transverse plane,
\be
\partial_{i}\partial_{j}&=&-\left(\frac{z_{\perp}^{i}z_{\perp}^{j}}{z_{\perp}^{2}}-\delta^{ij}\right)\frac{1}{z_{\perp}}\frac{d }{d z_{\perp}} + \frac{z_{\perp}^{i}z_{\perp}^{j}}{z_{\perp}^{2}}\frac{d^{2}}{d z_{\perp}^{2}},\label{apd5}\\
\partial^{2}&=&\frac{1}{z_{\perp}}\frac{d }{d z_{\perp}} + \frac{d^{2}}{d z_{\perp}^{2}}\label{apd6}~.
\ee
Substituting the Eqs.~\eqref{apd5} and \eqref{apd6} in Eq.~\eqref{apd3} with a bit of rearrangement we have,
\be
S^{ij}&=&\bigg[\left(\frac{z_{\perp}^{i}z_{\perp}^{j}}{z_{\perp}^{2}}-\frac{1}{2}\delta^{ij}\right)\left(\frac{1}{z_{\perp}}\frac{d}{d z_{\perp}}-\frac{d^{2}}{d z_{\perp}^{2}}\right) \nonumber \\
&+& \delta^{ij} \left(\frac{1}{2z_{\perp}}\frac{d}{d z_{\perp}}+\frac{1}{2}\frac{d^{2}}{d z_{\perp}^{2}}\right) \bigg] \tilde{D}\label{apd7}.
\ee
Comparing Eq.~\eqref{apd7} with Eq.~\eqref{G11}, we obtain the pressure and shear function as defined in Eqs.~\eqref{ex5} and \eqref{ex4}. Since pressure and shear function are obtained from the same function $\tilde{D}$, they are not independent. One can straightforwardly show from the definition~\eqref{apd3} of the shear stress that it is conserved, i.e.,
$\partial_{i}S^{ij}=\partial_{i}(-\partial_{i}\partial_{j}+\delta^{ij}\partial^{2})\tilde{D}=0$.
Eq.~\eqref{ex6} can be shown to hold directly from the definitions of pressure and shear functions provided in Eq.~\eqref{apd7} or from the shear stress conservation. From the conservation of the shear stress, we have
\begin{eqnarray}
    &&\partial_{i} \bigg[\left(\frac{z_{\perp}^{i}z_{\perp}^{j}}{z_{\perp}^{2}}-\frac{\delta^{ij}}{2}\right)~s(z_{\perp})+ \delta^{ij}~p(z_{\perp})\bigg]=0\nonumber\\
&& \text{or,} \left(\frac{z_{\perp}^{i}z_{\perp}^{j}}{z_{\perp}^{2}}-\frac{\delta^{ij}}{2}\right) \partial_{i} s(z_{\perp}) + s(z_{\perp})\partial_{i} \left(\frac{z_{\perp}^{i}z_{\perp}^{j}}{z_{\perp}^{2}}-\frac{\delta^{ij}}{2}\right)\nonumber\\
&&+ \delta^{ij} \partial_{i}p(z_{\perp})=0\nonumber\\
&& \text{or, }s^{\prime}\frac{z_{\perp}^{i}}{z_{\perp}}\left(\frac{z_{\perp}^{i}z_{\perp}^{j}}{z_{\perp}^{2}}-\frac{\delta^{ij}}{2}\right) +s(z_{\perp})\frac{z_{\perp}^{j}}{z_{\perp}^{2}}+ p^{\prime}\frac{z_{\perp}^{j}}{z_{\perp}}=0\nonumber\\
&& \text{or, }\frac{1}{2}s^{\prime}\frac{z_{\perp}^{j}}{z_{\perp}}+s(z_{\perp})\frac{z_{\perp}^{j}}{z_{\perp}^{2}}+ p^{\prime}\frac{z_{\perp}^{j}}{z_{\perp}}=0\nonumber\\
&& \text{or, }\frac{1}{2}s^{\prime}+s(z_{\perp})\frac{1}{z_{\perp}}+ p^{\prime}=0~,\label{ap8}
\end{eqnarray}
where we used $\frac{\partial }{\partial z_{i}}=\frac{z_{\perp}^{i}}{z_{\perp}}\frac{d}{dz_{\perp}}$, $p^{\prime}\equiv \frac{dp}{dz_{\perp}}$ and $s^{\prime}\equiv\frac{ds}{dz_{\perp}}$. Eq.~\eqref{ap8} is in agreement with Eq.~\eqref{ex6}. To obtain the second equality of Eq.~\eqref{ap8}, which is the von Laue stability criterion, we first rewrite the pressure in a different way using Eq.~\eqref{apd3},
\be
p(z_{\perp})=\frac{1}{2}S^{ij}\delta^{ij}=\frac{-1}{8P^{+}} \int \frac{d^{2}\vec{\Delta}_{\perp}}{(2\pi)^{2}}~\Delta^{2}_{\perp} D ~e^{-i\vec{\Delta}_{\perp}\cdot \vec{z}_{\perp}}.\label{apd9}
\ee
From Eq.~\eqref{apd9}, we get the von Laue stability condition by directly integrating in the transverse space as follows 
\be
\int d^{2}\vec{z}_{\perp}~p(z_{\perp})&=&\frac{-1}{8P^{+}} \int d^{2}\vec{\Delta}_{\perp}~\Delta^{2}_{\perp} D \int \frac{d^{2}\vec{z}_{\perp}}{(2\pi)^{2}} e^{-i\vec{\Delta}_{\perp}\cdot \vec{z}_{\perp}}\nonumber\\
&=&\frac{-1}{8P^{+}} \int d^{2}\vec{\Delta}_{\perp}~\Delta^{2}_{\perp} D ~\delta^{2}(\vec{\Delta}_{\perp})\nonumber\\
&=&0~,\label{apd10}
\ee
where we use $\lim_{\vec{\Delta}_{\perp} \rightarrow 0}\Delta^{2}_{\perp} D(-\Delta^{2}_{\perp})=0$. We can also find out the moment of the pressure and shear function by using their definitions in Eq.~\eqref{G12} and repeatedly applying integration by parts, 
\be
\int z_{\perp}^{2} p(z_{\perp}) d^{2}\vec{z}_{\perp}&=&2\pi\int \frac{z_{\perp}^{2}}{2}\frac{d}{dz_{\perp}}\left(z_{\perp}\frac{d\tilde{D}}{dz_{\perp}}\right)  dz_{\perp}\nonumber\\
&=&- 2\pi\int z_{\perp}^{2}\frac{d\tilde{D}}{dz_{\perp}}dz_{\perp}\nonumber\\
&=& 2\int \tilde{D}~d^{2}\vec{z}_{\perp}\nonumber\\
&=&\int \frac{d^{2}\vec{\Delta}_{\perp}}{2P^{+}} D(-\vec{\Delta}_{\perp}^{2}) \int \frac{d^{2}\vec{z}_{\perp}}{(2\pi)^{2}}  e^{-i\vec{\Delta}_{\perp}\cdot \vec{z}_{\perp}}\nonumber\\
&=& \frac{D(0)}{2P^{+}}~,\label{apd11}
\ee
where we assume that owing to the property of the $\tilde{D}$, the integrated part always vanishes at spatial infinity. Eq.~\eqref{apd11} suggest that for any system having $D(0)<0$, the $z_{\perp}^{2}$-weighted moment of pressure is negative. Similarly, it can be easily seen following the similar steps that lead to Eq.~\eqref{apd11},
\be
\int z_{\perp}^{2} s(z_{\perp}) d^{2}\vec{z}_{\perp}=\frac{-2D(0)}{P^{+}}~.\label{apd12}
\ee
Eq.~\eqref{apd12} states that for any system having $D(0)<0$, the $z_{\perp}^{2}$-weighted moment of shear function is positive. Similarly, one can obtain the shear function integrated over the space,
\be
\int s(z_{\perp})~d^{2}\vec{z}_{\perp}=\frac{-1}{4P^{+}}\int_{-\infty}^{0} D(t)dt~,\label{apd13}
\ee
where $t\equiv -\vec{\Delta}_{\perp}^{2}$. Combining Eqs.~\eqref{apd10}, \eqref{apd11}, \eqref{apd12}, and \eqref{apd13} the squared average mechanical radius defined in Eq.~\eqref{G17},
\be
\langle z_{\perp}^{2}\rangle_{\mathcal{F}}&=& \frac{\int d^{2}\vec{z}_{\perp}~ z_{\perp}^{2}~\left(p(z_{\perp})+\frac{1}{2}s(z_{\perp})\right)}{\int d^{2}\vec{z}_{\perp} ~\left(p(z_{\perp})+\frac{1}{2}s(z_{\perp})\right)}\nonumber\\
&=& \frac{4 D(0)}{\int_{-\infty}^{0} D(t)dt}~.\label{apd14}
\ee
The squared average momentum radius defined in Eq.~\eqref{G16} is,
 \be
 \langle z_{\perp}^{2}\rangle_{\mathcal{P}}&=&\frac{1}{P^{+}}\int d^{2}\vec{z}_{\perp} ~z_{\perp}^{2} \mathcal{P}(z_{\perp})\nonumber\\
 &=& \int \frac{d^{2}\vec{\Delta}_{\perp}}{(2\pi)^{2}} A\int d^{2}\vec{z}_{\perp} ~z_{\perp}^{2} e^{-i\vec{\Delta}_{\perp}\cdot \vec{z}_{\perp}}\nonumber\\
 &=&- \int d^{2}\vec{\Delta}_{\perp} A~\delta^{ij}\frac{\partial^{2}}{\partial \Delta_{\perp}^{i}\partial\Delta_{\perp}^{j}}\int \frac{d^{2}\vec{z}_{\perp}}{(2\pi)^{2}} e^{-i\vec{\Delta}_{\perp}\cdot \vec{z}_{\perp}}\nonumber\\
 &=&- \int d^{2}\vec{\Delta}_{\perp} A~\delta^{ij}\frac{\partial^{2} \delta^{2}(\vec{\Delta}_{\perp})}{\partial \Delta_{\perp}^{i}\partial\Delta_{\perp}^{j}}\nonumber\\
 &=&- \int d^{2}\vec{\Delta}_{\perp}\delta^{ij}\frac{\partial^{2} A(-\vec{\Delta}_{\perp}^{2})}{\partial \Delta_{\perp}^{i}\partial\Delta_{\perp}^{j}}  \delta^{2}(\vec{\Delta}_{\perp})\nonumber\\
 &=&-4\frac{dA}{d\vec{\Delta}_{\perp}^{2}}\Big|_{\vec{\Delta}_{\perp}^{2}=0}=4\frac{dA}{dt}\Big|_{t=0}\label{apd15}
 \ee
 where we integrated by parts to remove the derivatives from the delta function and used the identity $\delta^{ij}\frac{\partial^{2} A}{\partial\Delta_{\perp}^{i}\partial \Delta_{\perp}^{j}}=4\frac{d A}{d \vec{\Delta}_{\perp}^{2}}+4\vec{\Delta}_{\perp}^{2}\frac{d^{2} A}{d^{2}\vec{\Delta}_{\perp}^{2}}$. The LF longitudinal momentum density integrated over all space, we get, 
 \be
 \int \mathcal{P}~d^{2}\vec{z}_{\perp}&=& P^{+}\int d^{2}\vec{\Delta}_{\perp} A(-\vec{\Delta}_{\perp}^{2})\int  \frac{d^{2}\vec{z}_{\perp}}{(2\pi)^{2}} e^{-i\vec{\Delta}_{\perp}\cdot \vec{z}_{\perp}}\nonumber\\
 &=& P^{+}\int d^{2}\vec{\Delta}_{\perp} A(-\vec{\Delta}_{\perp}^{2})~\delta^{2}(\vec{\Delta}_{\perp})\nonumber\\
 &=& P^{+} A(0)\label{apd16}
 \ee
From which we can ascertain the sum rule $A(0)=1$. The energy density on the LF turns out to be a compound density consisting of two parts: the barycentric kinetic energy density and the internal energy density~\cite{Freese:2022fat}. Readers can see Ref.~\cite{Freese:2022fat} to learn about compound density and their derivation. In Eq.~\eqref{G15} we have provided only the internal energy density. One can also express the internal energy density as,
\be
\mathcal{E}&=& \frac{1}{2P^{+}}\left(M^{2}\tilde{A}-\frac{1}{4} \partial^{2}\tilde{A}-\frac{1}{2}\partial^{2}\tilde{D}\right)\nonumber\\
&=&\frac{1}{2P^{+}}\left[M^{2}\tilde{A}-\frac{1}{4}\left(\frac{1}{z_{\perp}}\frac{d(\tilde{A}+2\tilde{D})}{d z_{\perp}} + \frac{d^{2}(\tilde{A}+2\tilde{D})}{d z_{\perp}^{2}}\right)\right].\nonumber\\
\label{apd17}
\ee
We close the discussion by commenting on the fact that if one assumes a 3D density obtained from the Fourier transformation and an associated form factor $\mathscr{F} (-\vec{\Delta}^{2})$ following the similar steps as performed to obtain Eq.~\eqref{apd15}, the mean squared radius is given by, $\langle r^{2}\rangle_{\mathscr{F}}=-6\frac{d\mathscr{F}}{d\Delta^{2}}\vert_{\Delta^{2}=0}$.
\bibliography{ref_stress}

\end{document}